\documentclass[preprint]{aastex}
\usepackage{natbib}
\received{}
\accepted{}

\slugcomment{\apjs, accepted}

\lefthead{Richards}
\righthead{Intrinsic Absorption in Radio-Selected QSOs}

\begin{document}

\title{Intrinsic Absorption in Radio-Selected QSOs\altaffilmark{1}}

\author{Gordon T. Richards\altaffilmark{2,3}}
\affil{University of Chicago}
\affil{Department of Astronomy and Astrophysics, 5640 S. Ellis Avenue, Chicago, IL 60637}
\email{richards@oddjob.uchicago.edu}

\altaffiltext{1}{Based on observations obtained with the Apache Point
Observatory 3.5-meter telescope, which is owned and operated by the
Astrophysical Research Consortium.}
\altaffiltext{2}{Presented as part of a dissertation to the Department
of Astronomy and Astrophysics, The University of Chicago, in partial
fulfillment of the requirements for the Ph.D. degree.}
\altaffiltext{3}{Current address: Department of Astronomy and Astrophysics, The Pennsylvania State University, University Park, PA
16802}

\begin{abstract}

Moderate resolution spectra have been recorded for a sample of
radio-selected QSOs ($z \sim 2.5$) discovered in the ``Faint Images of
the Radio Sky at Twenty centimeters'' (FIRST) VLA survey.  This work
is motivated by the study of a heterogeneous set of QSO absorption
line spectra that showed that as many as 36\% of the absorbers
normally thought to be intergalactic are correlated with physical
properties of the background QSO, including the radio source spectral
index.  Spectra were taken for 24 newly discovered quasars in order to
test this finding on a more homogeneous data set.  The spectra have
been searched for the most frequently observed absorption lines.  The
absorption line properties are compared to the radio properties of the
quasars.  The primary results are summarized as follows: 1) an excess
of \ion{C}{4} absorbers in flat-spectrum quasars as compared to
steep-spectrum quasars is confirmed (based on a total sample of 165
systems), 2) a small excess of high-velocity \ion{C}{4} absorbers is
seen in radio-quiet QSOs as compared to radio-loud quasars, and 3) the
Sloan Digital Sky Survey QSO spectra will allow for the detection of a
high-velocity, narrow, intrinsic population of absorbers that is as
small as 7\% of the \ion{C}{4} absorber population in QSOs at
$z\sim2.5$.

\end{abstract}

\keywords{quasars: absorption lines --- quasars: general --- radio continuum: galaxies}

\section{Introduction}

The physical location of the narrow absorption features found in the
spectra of QSOs has been of considerable interest since their
discovery over 30 years ago \citep{sl66,blb66}.  There is a
considerable body of evidence that QSO absorption lines are, in
general, the result of a galaxy along the line of sight to the QSO,
where the galaxy is blocking some of the QSO light
\citep{ysb82,ssb88,ste90,sdm+97}.  Whether the material causing the
absorption is caused by extended galactic halos \citep{bs69}, or by
dwarf galaxies in the vicinity of a larger galaxy \citep{ydg+86}, is
still a matter of debate.

Despite the fact that intervening galaxies are clearly {\em a} cause
of absorption lines seen in the spectra of background QSOs, it is also
well-known that QSOs are also subject to self-absorption.  The broad
troughs that are characteristic of so-called Broad Absorption Line
QSOs (BALQSOs) are clearly material that is associated with the QSOs
\citep[e.g.,][]{tur88,wey97}.  There is also an excess of narrow
\ion{C}{4} absorption lines seen in the spectra of steep-spectrum
radio QSOs, within $5000\,{\rm km\,s^{-1}}$ of the emission redshift
of the QSO.  This excess is also thought to be associated with the QSO
in some way \citep{fwp+86,awf+87,fcw+88}.

These facts come together to form the commonly accepted picture that
heavy-element QSO absorption lines are caused by intervening galaxies,
unless: 1) the lines are broad (typically $2000\,{\rm km\,s^{-1}}$ or
wider), or 2) narrow, with a relative velocity with respect to the QSO
rest frame of less than $5000\,{\rm km\,s^{-1}}$.  For these two cases
the lines are either ``intrinsic'' to the QSO (directly related to the
QSO) or ``associated'' with the QSO (either directly related to the
QSO or influenced by the radiation field of the QSO).

There is a growing body of evidence that some QSO absorption line
systems do not fit into any of the above categories.  These are lines
that display characteristics of intrinsic or associated lines, but
have relative velocities that are greater than $5000\,{\rm
km\,s^{-1}}$ (blueward of the QSO emission redshift)\footnote{$v =
\beta c = \frac{R^2-1}{R^2+1}$, where $R=\frac{1+z_{em}}{1+z_{abs}}$,
such that positive velocities are blueshifted with respect to the QSO
redshift.} and are not broad enough to satisfy the BAL classification
scheme set forth by \citet{wmf+91}.  Examples of such systems can be
found in \citet{hbc+97} and \citet{jhk+96}.

This study attempts to determine if there is any evidence for a
population of narrow, high-velocity intrinsic/associated absorbers
that have been masquerading as intervening galaxies.  In particular,
this work is motivated by a desire to use the absorption line data
from the catalog of \citet{yyc+91}, an updated version of which is in
preparation (Vanden Berk et al. 2001), as probes of galaxies at high
redshift.  For example, \citet{qvy96} have used the data in the
catalog to explore the clustering of absorbers out to scales of
hundreds of Mpc.  In another paper based on this catalog,
\citet{vqy+96} found that bright QSOs tend to have more \ion{C}{4}
systems along their line of sight than faint QSOs.  A statistical weak
lensing effect was suggested as the origin of this effect; however, it
was acknowledged that a population of high velocity intrinsic
absorbers might yield similar observational consequences.

Studies like these use heavy-metal QSO absorption line systems as
powerful cosmological probes that allow astronomers to study
structures at redshifts higher than would otherwise be observable.
Thus, it is important to understand the nature of QSO absorption lines
and to what extent they actually do represent galaxies along the line
of sight to QSOs.  If the heavy-metal QSO absorption line population
is heavily contaminated by intrinsic QSO material, it may not be
possible to use QSO absorption lines as probes of high-$z$ galaxies
unless it is possible to identify intrinsic systems on a case-by-case
basis.

In a recent paper \citep[hereafter Paper I]{ryy+99}, the existing
evidence for a population of high-velocity, but narrow
intrinsic/associated absorption lines was summarized and new evidence
that supports the reality of said population was presented.  The
primary result of Paper I was the discovery that there is an excess of
narrow \ion{C}{4} absorption with $v>5000\,{\rm km\,s^{-1}}$ in
flat-spectrum quasars as compared to steep-spectrum quasars.  Since
some radio properties of quasars are thought to be correlated with the
line-of-sight angle of inclination of quasars \citep{rr77,ob82,pu92},
it was postulated that this excess may be the result of an
orientation-dependent population of intrinsic/associated absorbers.
This excess can be accounted for by a 36\% contamination of the
intervening \ion{C}{4} absorption line sample by intrinsic absorbers
that are preferentially found in flat-spectrum quasars.

While statistical tests by \citet{ryy+99} apparently removed various
biases from the inhomogeneous catalog of absorbers \citep{yyc+91},
there is no way to prove that all biases were removed.  In particular,
there was concern that variability in the radio data (compromising the
derivation of the radio spectral index from multi-band observations
that were not coincident in time) and the varying quality of the
optical spectra (and thus the equivalent width detection limit) might
affect the conclusions of Paper I.  \citet[hereafter Paper
III]{rlb00}, addresses the inhomogeneity of the radio data.
Confirming that the results from Paper I hold in light of more
homogeneous optical spectra is left to be examined herein.  The
primary goals of this work are to 1) confirm that the excess of
\ion{C}{4} absorbers in flat-spectrum quasars observed in Paper I
persists for a sample of more homogeneous data, and 2) test whether
the observed excess may instead be due to evolution of the absorber
population as a function of redshift.

This paper is laid out as follows: \S 2 presents the target list and
the details of the observations.  An indepth account of the data
reduction is given in \S 3.  In \S 4, detailed notes on the individual
objects are given.  \S 5 presents an analysis and discussion of the
science, including the impact that the Sloan Digital Sky Survey will
have on this and other QSO absorption line studies.  Finally, \S 6
presents the conclusions.

\section{Observations}

The primary goal of this work is to determine if the results of Paper
I hold for a more homogeneous set of absorption line data.  Since that
work depended on knowing the radio properties of the QSOs, it was
necessary to select radio-detected QSOs for observation.  It was
decided to select targets from those quasars discovered (or recovered)
as part of the FIRST Bright Quasar Survey (FBQS; \citealt{gbw+96}) as
of January 1998.  All FBQS quasars with $2.27 < z_{em} < 4.0$ were
included as potential targets.  The redshift range was chosen such
that any \ion{C}{4} absorption lines with velocities relative to
$z_{em}$ between $-5000\,{\rm km\,s^{-1}}$ and $75000\,{\rm
km\,s^{-1}}$ would fall between the wavelength limits of the spectra
($3965.2$ -- $8235.0\,{\rm \AA}$).  The redshift range was further
restricted to $z_{em} \le 2.8$ shortly after the start of the project
when it became clear that the red spectra would not be suitable for
finding \ion{C}{4} absorption lines.

Not all of the objects in the initial target list were observed as a
result of time constraints on the telescope.  Targets were observed
without prior knowledge of absorption line properties.  Brighter
objects close to the meridian were chosen over fainter objects and
those at higher airmass.

The final analysis includes some of the data from Paper I in order to
form a sample that is large enough for statistical analysis of the
correlation of absorption with radio spectral index.  Included are all
QSOs from the revised York catalog \citep{yyc+91,v+00} that meet the
following requirements: 1) $z_{em} \ge 2.2$, 2) $z_{em} \le 2.8$, 3)
spectral resolution between $35\,{\rm km\,s^{-1}}$ and $450\,{\rm
km\,s^{-1}}$, 4) not known or suspected gravitational lenses, and 5)
not known or suspected Broad Absorption Line QSOs (BALQSOs).
Furthermore, Q0237-233 was removed since this object has an unusually
high number of \ion{C}{4} absorbers in a small redshift range.

The entire target list is given in Table~1.  In Table~1, the columns
are the IAU name for the object, the redshift, the right ascension and
declination in J2000 coordinates, the E magnitude, the O-E color, the
peak (i.e. core) flux density (mJy) from the FIRST maps, the
integrated FIRST flux density (mJy), and the signal-to-noise of both
the blue and red spectra.  The optical magnitudes and radio flux
densities are taken from \citet{gbw+96}.  The ``blue'' signal-to-noise
is the average from the region between Lyman-$\alpha$ and \ion{C}{4}
emission; the ``red'' signal-to-noise is taken at $6000\,{\rm \AA}$.
The redshifts in Table~1 are flux weighted redshifts measured from the
\ion{C}{4} emission lines.

\placetable{tab:tab1}

The signal-to-noise threshold was set so as to yield equivalent width
limits similar to or better than previously achieved for \ion{C}{4}
absorption: the goal was to obtain a $5\sigma$ rest equivalent width
limit of $0.1\,{\rm \AA}$ for \ion{C}{4} doublets at $z=2$.  Given a
resolution of $3.2\,{\rm\AA}\,{\rm pixel}^{-1}$ and a two-pixel
resolution element, the required signal-to-noise is 75.  For the
spectra that are only signal-to-noise of 50, the $5\sigma$ rest
equivalent width limit is $0.15\,{\rm \AA}$.

The observations were carried out with the Double Imaging Spectrograph
(DIS)\footnote{The Double Imaging Spectrograph (DIS) was built at
Princeton University by Jim Gunn, Michael Carr, Brian Elms, Ricardo
Lucinio, Robert Lupton, and George Pauls.  See
http://www.apo.nmsu.edu/Instruments/DIS/dis.html for details.} on the
ARC 3.5m telescope at Apache Point Observatory (APO).  The DIS
detectors are a thinned, uv-coated SITe 512x512 CCD with 27 micron
pixels (blue camera), and a thinned 800x800 TI CCD with 15 micron
pixels (red camera).  The platescale at the detectors is
$1.086\arcsec\,{\rm pixel}^{-1}$ in the blue and $0.610\arcsec\,{\rm
pixel}^{-1}$ in the red.  The gratings used for these observations
have 600 lines\,${\rm mm}^{-1}$, giving a dispersion of about
$3.2\,{\rm \AA}\,{\rm pixel}^{-1}$ (blue camera) and 300 lines\,${\rm
mm}^{-1}$ with a dispersion of about $3.5\,{\rm \AA}\,{\rm
pixel}^{-1}$ (red camera).  Throughout the course of the observations,
a $2\arcsec$ wide steel slit (which is about $6\arcmin$ long) was
used.  All observations were made at the parallactic angle
\citep{fil82}.
 
Though DIS is only a medium resolution spectrograph, it is ideal for
the study at hand.  DIS is able to obtain complete spectral coverage
from $3965.2$ to $8235.0\,{\rm \AA}$ at a resolution sufficient to
split (two pixels between minima) the \ion{C}{4} doublet at $z \ge
1.488$.  Though the resolution is good enough to split \ion{C}{4}, it
is not always good enough to formally resolve it.  This apparent
detriment is more than made up for by the wavelength coverage, which
allows for the discovery of other lines in the system.  The presence
of other lines in the system thus confirm the reality of \ion{C}{4}
even though it may not be properly resolved.

DIS performed quite well throughout the course of this project;
however, there are a few problems worth noting.  The first is that the
red CCD is not nearly as sensitive as the blue CCD and suffers from
some noise problems.  Second, there is, inevitably, flexure in the
spectrograph when the spectrograph rotates during long exposures.
Therefore, one must be careful to obtain wavelength calibration
exposures on a frequent basis.  Finally, when attempting to observe at
the parallactic angle, it was sometimes difficult to place the target
directly on the boresight (the point around which the field rotates).
As a result, there may be some light loses that affect the absolute
spectrophotometry of the spectra.  Accurate guiding negates this
problem, but it was difficult at times to keep the object fully in the
slit as a result of long time delays between guiding exposures for
faint targets.

Table~2 gives a log of all of the observations.  The first column
gives a truncated name of the QSOs.  The next two columns give the UT
date and time of the observations.  The exposure times were typically
1800 seconds (30 minutes): exposure times different from 1800 seconds
are indicated by notes to the table.  No exposures that resulted in
signal-to-noise less than 10 per pixel were used.  At least 4
exposures were taken of each target to allow for proper detection (and
removal) of cosmic rays.

\placetable{tab:tab2}

\section{Data Reduction}

The data were reduced with the Image Reduction and Analysis Facility
(IRAF)\footnote{IRAF is distributed by the National Optical Astronomy
Observatories, which are operated by the Association of Universities
for Research in Astronomy, Inc., under cooperative agreement with the
National Science Foundation.} using a package called DISTOOLS, which
was written by the author.  The DISTOOLS package was written to
facilitate the reduction of data from the Double Imaging Spectrograph
(DIS) at the Apache Point Observatory (APO).  This package is largely
a series of ``wrappers'' to existing IRAF routines (version
2.11.x)\footnote{See the IRAF ``Help Pages''
(http://iraf.noao.edu/iraf/web/iraf-help.html) for detailed
explanations of any IRAF tasks describe herein.}, with parameters that
are optimized for the camera used for this project.

\subsection{Converting 2-D Images into 1-D Spectra}

The conversion of the 2-D images into 1-D spectra proceeded in the
usual manner, following the optimal extraction process of
\citet{hor86}.  Wavelength calibration was accomplished through the
observation of helium, neon, argon (HeNeAr) comparison lamps, which
were taken before and after each standard star and object spectrum.
The wavelengths are all vacuum heliocentric.

Typically the blue spectra required a third order Chebyshev
polynomial, giving an average root-mean-square (RMS) residual of $0.30
{\rm \AA}$, while the red spectra were best fit by a third order cubic
spline with an average RMS residual of $0.14 {\rm \AA}$.  The RMS
residuals of the dispersion solution fits are quite good; however,
they should be interpreted with caution: there are always additional
contributions to both the statistical and systematic errors in the
wavelength that can be as large as, or larger than the RMS values.
For this particular project, the wavelength errors are vastly more
accurate than what is required.

Next, the spectra are extinction corrected and flux calibrated.  The
standard Kitt Peak National Observatory (KPNO) extinction spectrum was
used despite the fact that these observations were taken at APO.
Tests showed that the use of the KPNO extinction vector was better
than not applying a correction at all.

After extinction correction and flux calibration, the red spectra were
also corrected for telluric absorption.  A template correction
spectrum was made by taking a very high signal-to-noise spectrum of
Feige~110 at an airmass of approximately 1.5 on the night of 9 July
1999.  Each separate red spectrum of every object was corrected for
telluric absorption using the TELLURIC task (new to IRAF V2.11.1) with
the above telluric correction template; the same template was used to
correct every individual spectrum.

All the spectra for each object are variance coadded to produce the
final 1-D spectrum.  Before variance coaddition, each spectrum has bad
pixels masked out and is scaled to the the individual spectrum with
the largest flux.  Bad pixels (cosmic rays, etc.) are masked out under
the premise that one should reject bad data, not attempt to fix it.
After bad pixels have been masked, all the spectra of each object are
scaled.  It is assumed that the spectrum with the highest mean flux is
that which is closest to the intrinsic spectrum.  Failure to normalize
the spectra can cause spectra with equal signal-to-noise but lower
absolute fluxes (as a result of flux calibration errors, slit losses,
etc.) to be weighted improperly.  Upon completion of the scaling
process, the spectra are then variance coadded.  The final weighted
error spectrum is the square root of the variance spectrum.

\subsection{Continuum Fitting}

A continuum vector was fit to and removed from each composite spectrum
before searching for absorption lines.  Each composite spectrum was
first split into segments at the peaks of strong emission lines.  Then
each segment was fit with a cubic spline polynomial using the
CONTINUUM procedure in IRAF.  The order, high- and low-rejection
thresholds were adjusted until the fit reproduced the continuum plus
emission spectrum suitably.  These pieces were then merged together
and the entire continuum spectrum was examined again.

That this method produced reasonable continuum spectra can be
confirmed by looking at the distribution of residuals (spectrum -
continuum).  If the continuum was well fit, then the residual
distribution should be nearly Gaussian (centered at zero with a width
of one over the signal-to-noise of the spectrum) with a tail of
positive-going points corresponding to absorption line features.  For
each object, the distribution of residuals was examined in order to
make sure that the continuum residuals indeed were reasonable.

\subsection{Redshift Determination \label{sec:redshift}}

Generally, for each QSO emission line the redshift is determined by
fitting a Gaussian (or Gaussians), using the peak flux, or taking a
flux weighted average.  For the sake of this project a simple
determination of the redshift will suffice.  The redshifts were
calculated as a flux weighted average of the signal in pixels with
more than 50\% of the peak flux in each \ion{C}{4} emission line.
Pixels for which the actual flux deviated from the fitted continuum
value by more than 10\% were weighted according to the continuum flux
and not the raw flux of the spectrum, since to do otherwise would
allow for strong absorption lines to affect the determination of the
redshift.

The use of a flux weighted average of the peak of the \ion{C}{4}
emission line only should be more than adequate for this work.  The
measurement error should typically be much less than the resolution of
the spectrum ($\Delta v = 350\,{\rm km\,s^{-1}}$ or $\Delta z =
0.004$).  Other methods may change the redshift by as much as $\Delta
z = 0.01$ ($\Delta v = 850\,{\rm km\,s^{-1}}$) or more, but such a
shift will have negligible impact on this study.  For asymmetric lines
with steeper red wings, the flux weighted method will yield slightly
lower redshifts than using the peak.  A more proper determination of
the redshift would use not only the \ion{C}{4} emission line, but also
all of the other observed lines.  Low-ionization lines are closer to
the systemic redshift, which is typically larger than that measured by
higher ionization lines such as \ion{C}{4} \citep{tf92}.

\subsection{Finding Absorption Lines}

The identification of significant absorption lines in the composite
QSO spectra largely follows the algorithm of \citet{shj+93}.
The process consists of normalizing the spectrum, finding local
minima, subtracting a Gaussian profile from significant lines, and
looking for secondary lines.  The blue and red spectra were treated
separately.

A number of quantities are determined for each significant absorption
line as per \citet{shj+93}.  These include the observed equivalent
width (${\rm EW}$), given by
\begin{equation}
{\rm EW} = D \sum_{j} P_j \left( F_{j} - 1 \right)/\sum_{j} P_j^2,
\end{equation}
where $P_j$ is the gaussian weighting function of the slit spread
function (SSF) whose width is given by $\sigma_{\rm SSF} = {\rm
FWHM}/2.354$, $F$ is the observed (normalized) flux at a given pixel,
and $D$ is the dispersion.  At the edges of the spectrum, no attempt
is made to buffer the spectrum and the SSF is modified accordingly.

The corresponding observed equivalent width error is then given by
\begin{equation}
\sigma({\rm EW}) = D\left(\sum_{j} P_j^2 E^2_{j}\right)^{1/2}/\sum_{j} P_j^2,
\end{equation}
where $E$ is the normalized flux error.  The so-called
``interpolated'' equivalent width error spectrum is the same except
for the use of the interpolated flux error in place of the raw
(normalized) flux error.  The interpolated flux error is determined by
interpolating the error spectrum across strong absorption lines and is
a better estimate of the signal-to-noise when determining the
significance of a line.  However, the usual flux error is still used
to calculate the error in the equivalent width measurement after a
line has been deemed significant.  Lines with significance limits ($SL
= |W|/\overline{\sigma}_W$) less than $3$ are rejected.  The SL is
simply the signal-to-noise (S/N) except that the equivalent width
error is replaced by the interpolated equivalent width error.

Some additional quantities are also determined.  The wavelength error
for a particular line is
\begin{equation}
\sigma(\lambda) = \sqrt{2} D \sigma_{SSF} \left|\frac{\sigma_W}{W}\right|.
\end{equation}
An alternative measure of the equivalent width is also calculated using:
\begin{equation}
W = D \sum_{j} (1 - F_j),
\end{equation}
where the sum over $j$ is carried out over the range of pixels that
are less than unity in the vicinity of the absorption feature.  These
equivalent width values can be compared to those computed for single,
isolated lines to test how well the input FWHM reproduces the SSF.

Finally, the observed FWHM is determined by
\begin{equation}
FWHM = D (\lambda_{\rm hi} - \lambda_{\rm lo}),
\end{equation}
where $\lambda_{\rm hi(lo)}$ is the wavelength where $F = 1/2(1-F_{\rm
min})$, $\lambda_{\rm hi(lo)}$ is larger(smaller) than the central
wavelength, and $F_{\rm min}$ is the flux at the minimum of the
absorption trough.  This observed FWHM can be compared to the input
FWHM (for single lines) to test the appropriateness of the input FWHM
for a given spectrum.  In addition, this measure allows one to
identify blended and resolved lines, since this measure of the FWHM
will be significantly larger than the input FWHM for such lines.

\subsection{Identification of Absorption Line Systems}

The identification of absorption line {\em systems} from the list of
individual absorption lines is a difficult task.  The code used for
this step in the data analysis was adapted from that used by Daniel
Vanden Berk in his 1997 University of Chicago Ph.D. Thesis
\citep{van97,vls+99}.  The code is based on the ZSEARCH algorithm
developed by the {\em HST} Quasar Absorption Line Key Project Team
\citep[e.g.,][]{bbb+93,jbb+98} and was modified to fit the needs of
this project.

Identification of the absorption lines begins by comparing the
observed lines to those that are typically found in the spectra of
QSOs.  The template lines used to identify systems are taken from
\citet*{myj88}.  The lines that were searched for are given in
Table~3.  The first column is the ion, the second column is the vacuum
wavelength of the particular transition, the third column is the
strength of the line, and the fourth column is the ionization
potential for the ion given.  Ionization potentials are taken from
\citet{m71}.  In this table, lines with $\lambda > 1036\,{\rm \AA}$
are sorted in increasing wavelength order of the the largest $f$ value
from each ion.  Then, within each ion, the lines are sorted in order
of decreasing $f$.

\placetable{tab:tab3}

The first step in the system identification procedure is to try to
match each input line with the line transitions listed in Table~3.  A
redshift is calculated for each observed-line/template-line pairing.
These redshifts are used as the basis set for the system
identification process.  The rest of the process involves rejecting
unlikely identifications.  That is, 1) are other lines from the same
atom/ion observable and are their strengths consistent with the first
line, and 2) are there other lines from different ions at the same
redshift that are also consistent?  All \ion{C}{4} and \ion{Mg}{2}
doublets that are stronger than the $5\sigma$ equivalent width limit
should be properly identified.  In some cases, these systems consist
only of the doublet, in which case the identification should be
considered as possible, but not definite.

Table~4 lists all of the lines identified in the spectra presented
herein.  All lines with significance limit greater than $3\sigma$ are
marked in the spectra, but only lines stronger than $5\sigma$ are
listed in the tables for the sake of brevity.  The numbering scheme
includes the $3\sigma$ lines, since those are marked in the spectra.
Lines weaker than $5\sigma$ may be included in the table if they are
identified as part of an identified system.  The first column gives
the number of the line.  Column 2 is the observed wavelength of the
line center, and column 3 is the error in the wavelength.  Columns 4
and 5 are the observed equivalent width and its error in Angstroms.
The FWHM is given in column 6.  The significance limit is in column 7.
If the line was identified with a given system, then the line
identification and the system redshift are given in columns 8 and 9
respectively.  The alphabetic notes to columns 2 and 8 are as follows:
a --- line in Lyman $\alpha$ forest, b --- possibly a continuum
fitting artifact, c --- questionable identification, d --- possibly a
telluric residual, e --- possible night sky artifact.  These
designations were made without knowledge of the identification of the
lines (if any).

The number of significant lines in the red spectra ($\lambda >
5500\,{\rm \AA}$) is somewhat larger than would be expected for the
signal-to-noise of the spectra.  This excess is probably caused by
difficulty in properly measuring the errors in the red spectra.  The
noise characteristics of the red CCD are not as stable as one might
like and the distribution of bias values is noticeably non-Gaussian.
This means that the read noise measured for the red CCD is not a true
measure of the error.  This effect causes the equivalent width errors
in the red spectra to be underestimated and, as a result, there are
more significant lines in the red spectra than there should be.
However, nearly all of the \ion{C}{4} systems are in the blue spectra
and only $5\sigma$ (or stronger) lines have been used for analysis, so
this problem is of little significance to the work at hand.

\section{Notes on Individual Objects}

Figures~1 through 24 present spectra of 24 quasars.  With the
exception of FBQS1625+2646 (also known as KP~76), none of these
quasars has been previously analyzed for absorption.  For each of
these objects the blue spectra ($\lambda < 5000\,{\rm \AA}$) are
typically quite good.  Lines identified within these blue spectra
should generally be real.  The quality of the red spectra is somewhat
worse: all lines stronger than $3\sigma$ are marked, but many are
likely to be artifacts.  Lines that are thought to be artifacts are
noted as such in Table~4.

An important facet of the analysis of individual absorption lines is
the knowledge of not only what lines are actually observed, but also
of what lines could have been observed.  Since the spectra all have
exactly the same wavelength coverage and are comparable in
signal-to-noise, it suffices to give the redshift ranges for
interesting combinations.  Of most interest is whether or not
\ion{Mg}{2} is observed in \ion{C}{4} systems.  For the spectra
presented herein any reasonably strong ($W_{obs} \ge 0.6\,{\rm \AA}$)
\ion{Mg}{2} system will be detected to $3\sigma$ or better for
$z\le1.937$.  Similarly, at least three of the five strongest lines of
\ion{Fe}{2} could be detected up to $z=2.456$.

\subsection{FBQS0047-0156 ($z=2.479$)}

This is a Broad Absorption Line QSO (BALQSO) with a strong, complex
BAL trough near the \ion{C}{4} line ($\sim\!5325\,{\rm \AA}$) and two
smaller troughs at larger velocities ($\sim\!5050\,{\rm \AA}$ and
$\sim\!5200\,{\rm \AA}$).  The lower velocity trough looks as though
it might break up at higher resolution.  There is no \ion{Si}{4}
trough, but the \ion{N}{5} BAL is prominent ($\sim\!4275\,{\rm \AA}$).
Of particular interest is the partially resolved \ion{Si}{4} emission
line ($\sim\!4875\,{\rm \AA}$).

There are 42 absorption lines with significance greater than $5\sigma$
redward of Lyman-$\alpha$ emission; however this number includes the
narrow lines that were fit to the BAL profiles.  Removing the narrow
line fits to the BAL profiles, there are only 14 strong, narrow lines,
six of which remain unidentified (\#34,45,69,72,76,81).  All but one
of the unidentified lines are broad features, noted as possible
artifacts of the continuum definition, or longward of $5500\,{\rm
\AA}$, where the noise characteristics are suspect.

$z_{\rm abs} = 1.8103$. --- Other than the BALs, this is the only
readily identifiable system in this spectrum.  With 10 lines including
\ion{C}{4} $\lambda\lambda 1548,\!1550$ (\#28,29), \ion{Mg}{2}
$\lambda\lambda 2796,\!2803$ (\#89,90), five lines of \ion{Fe}{2}
(\#33,78,79,83,84) and \ion{Al}{2} $\lambda 1670$ (\#35) this system
is certain.

\subsection{FBQS0210-0152 ($z=2.372$)}

This object shows very strong \ion{N}{5} emission and rather weak
\ion{Si}{4} emission.  It is rich in absorption lines, having 46 lines
stronger than $5\sigma$.  These lines define six systems of which four
are certain, one is probable and one is possible. Redward of
Lyman-$\alpha$ emission there are 34 strong lines: five remain
unidentified (\#43,55,58,59,61).  All but one of the unidentified
lines are in the lower quality, red spectrum.

$z_{\rm abs} = 0.9962$. --- Only the \ion{Mg}{2} doublet (\#45,46) is
observed in this questionable system, which is coincident with a
strong night sky feature.

$z_{\rm abs} = 1.3097$. --- This system is defined by ten lines
including \ion{Mg}{2} $\lambda\lambda 2796,\!2803$ (\#53,54),
\ion{Mg}{1} $\lambda 2852$ (\#56), \ion{Al}{3} $\lambda\lambda
1854,\!1862$ (\#19,20), and five lines of \ion{Fe}{2}
(\#40,41,42,49,50).

$z_{\rm abs} = 2.1474$. --- Seven lines define this mixed ionization
system.  \ion{Al}{2} $\lambda 1670$ (\#39), \ion{C}{2} $\lambda 1334$
(\#18), and \ion{Si}{2} $\lambda 1526$ (\#30) are observed in addition
to \ion{C}{4} $\lambda\lambda 1548,\!1550$ (\#31,32) and \ion{Si}{4}
$\lambda\lambda 1393,\!1402$ (\#24,25).  The \ion{Si}{4} doublet is
quite strong as is \ion{C}{4}.

$z_{\rm abs} = 2.3216$. --- This redshift is suggested by a relatively
weak \ion{C}{4} doublet (\#33,34), a possible detection of \ion{N}{5}
$\lambda\lambda 1238,\!1242$ (\#13,14) and Lyman-$\alpha$ (\#6).  If
the system is real, then the \ion{N}{5} $\lambda 1242$ is blended with
\ion{N}{5} $\lambda 1238$ from the $z_{\rm abs} = 2.3342$ system.

$z_{\rm abs} = 2.3342$. --- This is a strong, high-ionization
\ion{C}{4} system with Ly-$\alpha$ (\#8) and very strong \ion{N}{5}
(\#14,15), but no \ion{Si}{4}.  The \ion{C}{4} absorption (\#35,36) is
in the blue wing of the \ion{C}{4} emission line and is $\sim\!
3400\,{\rm km\,s^{-1}}$ from the emission redshift of the quasar.  The
presence of strong \ion{N}{5} coupled with the location of the system
make it a nearly certain associated system.  The \ion{N}{5} $\lambda
1238$ line may be blended with \ion{N}{5} $\lambda 1242$ from the
$z_{\rm abs} = 2.3216$ system.

$z_{\rm abs} = 2.3709$. --- In this system the \ion{C}{4} doublet
(\#37,38) is found at essentially the same redshift as the quasar
itself.  The Lyman-$\alpha$ line (\#12) is quite strong.  \ion{Si}{4}
$\lambda\lambda 1393,\!1402$ (\#28,29) are readily apparent and
\ion{N}{5} $\lambda\lambda 1238,\!1242$ (\#16,17) may be detected;
however they are shifted redward relative to \ion{C}{4}.

\subsection{FBQS0256-0119 ($z=2.491$)}

This is another BALQSO.  There are two \ion{C}{4} troughs; the higher
velocity trough has two minima.  The lower velocity trough has an
obvious \ion{N}{5} counterpart and there is evidence for a weaker
\ion{Si}{4} trough.  There are no unambiguous narrow line systems in
this QSO; however we list all of the lines, including the narrow line
fits to the BAL profiles.  There are seven unidentified lines at
$5\sigma$ significance (\#30,32,33,34,36,68,69).

\subsection{FBQS0725+2819 ($z=2.662$)}

This QSO has relatively weak broad emission lines.  There are 33
significant absorption lines longward of Lyman-$\alpha$ emission.
Twenty-eight of these lines are identified with four absorption line
systems.  Of the remaining five unidentified lines (\#49,53,62,63,79),
none are marked as possible blemishes.

$z_{\rm abs} = 1.4968$. --- This is an unusual system in that
\ion{Cr}{2} $\lambda\lambda\lambda 2056,\!2062,\!2066$ (\#54,55,56)
and \ion{Mn}{2} $\lambda\lambda\lambda 2576,\!2594,\!2606$
(\#71,73,75) are observed.  In addition to these uncommon lines, the
more common lines of \ion{Mg}{2} $\lambda\lambda 2796,\!2803$
(\#83,84), \ion{Mg}{1} $\lambda 2852$ (\#85) and \ion{Al}{3}
$\lambda\lambda 1854,\!1862$ (\#44,45) are observed.  Finally, there
is evidence for \ion{Si}{2} $\lambda 1808$ (blended with \ion{C}{4}
$\lambda 1550$ from $z_{\rm abs} = 1.9095$), \ion{Zn}{2}
$\lambda\lambda 2026,\!2062$ and \ion{Mg}{1} $\lambda 2026$.

$z_{\rm abs} = 1.9095$. --- Eleven lines make this system unambiguous.
Six lines of \ion{Fe}{2} (\# 46,77,81,82,86,87) are observed in the
company of \ion{Mg}{2} $\lambda\lambda 2796,\!2803$ (\#89,90),
\ion{Al}{2} $\lambda 1670$ (possible) and \ion{Si}{2} $\lambda 1526$
(possible).  It is also quite probable that the \ion{C}{4} doublet is
seen (\#40,41); however the red line is corrupted and may be blended
with \ion{Si}{2} $\lambda 1526$ from $z_{\rm abs} = 1.4968$. 

$z_{\rm abs} = 2.2674$. --- Only the \ion{C}{4} $\lambda\lambda
1548,\!1550$ doublet (\#50,51) is observed in this system.  However,
this identification is considered very unlikely since both lines can
be accounted for by other systems.

$z_{\rm abs} = 2.2735$. --- A strong Lyman-$\alpha$ line (\#1) and a
weak \ion{C}{4} doublet (\#51,52) make this a probable system.  The
\ion{Si}{4} $\lambda\lambda 1393,\!1402$ doublet (\#42,43) may also be
seen, though at lower significance.

\subsection{FBQS0729+2524 ($z=2.300$)}

In this object there are 24 lines stronger than $5\sigma$, 18 of which
are identified in three likely systems and one possible system.
Included in the unidentified lines (\#3,4,7,24,31,32) are three lines
that may be part of the $z_{\rm abs}= 2.2871$ system, but are not
deblended.

$z_{\rm abs} = 0.5114$. --- Only the two lines of the \ion{Mg}{2}
doublet (\#10,11) are observed in this system; however the only other
observable line would be \ion{Mg}{1} $\lambda 2852$, so this system is
considered possible.  If real, the \ion{Mg}{2} $\lambda 2803$ line is
partially blended with \ion{C}{4} $\lambda 1548$ from $z_{\rm abs} =
1.7422$.

$z_{\rm abs} = 1.7422$. --- This systems contains \ion{C}{4}
$\lambda\lambda 1548,\!1550$ (\#12,13), \ion{Si}{2} $\lambda 1526$
(\#9), \ion{Al}{2} $\lambda 1670$ (\#18, which may be blended with
\ion{Si}{4} $\lambda 1393$ from $z_{\rm abs} = 2.2871$), the
\ion{Mg}{2} doublet (\#40,41) and possibly \ion{Mg}{1} $\lambda 2852$.
There is also a hint of the \ion{Al}{3} doublet, but it is blended
with \ion{C}{4} from $z_{\rm abs} = 2.2871$.

$z_{\rm abs} = 2.2295$. --- Both the \ion{C}{4} (\#22,23) and
\ion{Si}{4} doublets (\#16,17) are observed in this probable system.

$z_{\rm abs} = 2.2871$. --- The identification of this very high
ionization system is certain and may be classified as an associated
system.  Lyman-$\alpha$ (\#2) is observed and is very strong.  Both
the \ion{C}{4} (\#25,26) and \ion{N}{5} doublets (\#5,6) are observed
and are very strong compared to the \ion{Si}{4} doublet (\#18,19).  In
fact, the \ion{Si}{4} doublet may be an artifact, since the proposed
\ion{Si}{4} $\lambda 1393$ line is blended with \ion{Al}{2} $\lambda
1670$ from $z_{\rm abs} = 1.7422$ and \ion{Si}{4} $\lambda 1402$
coincides with the trough between the two (partially resolved?) lines
of \ion{Si}{4} seen in emission.  The line fitting code required extra
lines to fit both the \ion{N}{5} and \ion{C}{4} doublets.  This may
indicate blending with other systems or that these lines would be
resolved at higher resolution.

\subsection{FBQS0804+2516 ($z=2.290$)}

The emission spectrum of this QSO is unusual.  The \ion{C}{4} emission
line is somewhat asymmetric.  Emission lines of Lyman-$\alpha$,
\ion{N}{5}, {\ion{O}{1}, \ion{C}{2} and \ion{Si}{4} are also seen, but
they are all shifted to the red as compared to \ion{C}{4}.  In
addition, there are two ``extra'' bumps ($\sim\!4150\,{\rm \AA}$) of
emission just redward of \ion{N}{5} emission, which straddle the
predicted location of \ion{Si}{2} $\lambda 1260$.  The redder of the
two is probably \ion{Si}{2} $\lambda 1260$, since the redshift as
measured from \ion{C}{4} is smaller than that measured from other
lines.  The bluer of the bumps may be \ion{S}{2}.

In contrast with the emission spectrum, the absorption spectrum is
rather bare.  There are only 21 relatively strong lines (many of which
are noted as possible artifacts) that can be grouped into two systems,
one of which is uncertain.  There also appears to be a pair of weak
lines sitting on top of the \ion{C}{4} emission line, but these lines
remain unidentified.

$z_{\rm abs} = 0.7096$. --- This system contains \ion{Mg}{2}
$\lambda\lambda 2796,\!2803$ (\#15,16) in addition to two iron lines,
\ion{Fe}{2} $\lambda 2382$ (\#5) and possibly \ion{Fe}{2} $\lambda
2600$ (\#9).

$z_{\rm abs} = 1.6408$. --- This system is quite uncertain.  It only
has a relatively weak \ion{Mg}{2} doublet (\#34,35) and a possible
detection of \ion{Fe}{2} $\lambda 2382$ (\#27) that is coincident with
a strong night sky feature.

\subsection{FBQS0821+3107 ($z=2.604$)}

This QSO spectrum may harbor a damped Lyman-$\alpha$ absorption
system.  It also has strong \ion{N}{5} emission and partially resolved
\ion{Si}{4} emission.  There is a wealth of absorption lines blueward
of Lyman-$\alpha$ emission, and 23 strong lines redward of
Lyman-$\alpha$.  Twenty-one lines are identified with five different
systems.

$z_{\rm abs} = 0.7806$. --- This is a low ionization system with
\ion{Mg}{2} $\lambda\lambda 2796,\!2803$ (\#57,58), \ion{Mg}{1}
$\lambda 2852$ (\#59) and \ion{Fe}{2} $\lambda 2600$ (\#48).  Other
lines of \ion{Fe}{2} may also be evident, but they are either in the
forest or blended with other systems.

$z_{\rm abs} = 1.9440$. --- Only the \ion{C}{4} doublet (\#44,45) is
observed in this system, which defines it as possible.

$z_{\rm abs} = 2.4445$. --- Similar to the above system, this redshift
is indicated by the \ion{C}{4} doublet (\#60,61).  However,
Lyman-$\alpha$ (\#17) is also observable at this redshift and is
detected.

$z_{\rm abs} = 2.5152$. --- Again \ion{C}{4} $\lambda\lambda
1548,\!1550$ and Lyman-$\alpha$ are detected, though the significance
of the red line of \ion{C}{4} is less than $3\sigma$.  There is also a
possible detection of both \ion{Si}{4} $\lambda 1393$ and \ion{Si}{4}
$\lambda 1402$, but the red line is obscured by \ion{Si}{4} $\lambda
1393$ from $z_{\rm abs} = 2.5351$.

$z_{\rm abs} = 2.5351$. --- At least 13 lines are found in this
mixed-ionization system.  Both the \ion{C}{4} (\#64,65) and
\ion{Si}{4} doublets (\#54,56) are found.  Also seen are
Lyman-$\alpha$ (\#29), \ion{Si}{2} $\lambda 1260$ (\#40), \ion{C}{2}
$\lambda 1334$ (\#50), \ion{Si}{2} $\lambda 1526$ (\#62), \ion{Fe}{2}
$\lambda 1608$ (\#67), \ion{Al}{2} $\lambda 1670$ (\#68) and possibly
\ion{S}{2}.  Furthermore \ion{O}{1} $\lambda 1302$ and \ion{Si}{2}
$\lambda 1304$ are observed (\#46,47); this pair is often mistaken for
a \ion{C}{4} doublet.

\subsection{FBQS0857+3313 ($z=2.340$)}

The emission lines of this QSO are fairly narrow.  In fact, they are
so narrow that the \ion{Si}{4}+\ion{O}{4]} emission line is clearly
resolved into two components.  In this spectrum there are four
absorption line systems, which are drawn from the 33 $5\sigma$ lines
redder than Lyman-$\alpha$.

$z_{\rm abs} = 1.3838$. --- This system, which is defined by
\ion{Mg}{2} $\lambda\lambda 2796,\!2803$ (\#44,45) is unusual in that
it does not show any of the other lines normally seen with a
\ion{Mg}{2} doublet this strong, but it does show weak \ion{Al}{3}
$\lambda\lambda 1854,\!1862$ (\#16,17).

$z_{\rm abs} = 1.8888$. --- At this redshift there is a strong
\ion{C}{4} doublet (\#18,19).  \ion{Si}{4} (\#7,9) may also be
detected in absorption, but it is lost in the forest.  There are no
other significant lines identified with this system.

$z_{\rm abs} = 2.2691$. --- Fifteen lines leave little doubt about the
reality of this system.  Strong lines are seen in both low and high
ionization.  Both the \ion{C}{4} (\#26,27) and \ion{Si}{4} (\#21,22)
doublets are strong.  Lyman-$\alpha$ (\#2) is also strong.  Also seen
are four \ion{Fe}{2} lines (\#33,54,55,56), three \ion{Si}{2} lines
(\#11,14,25), \ion{C}{2} $\lambda 1334$ (\#15), \ion{O}{1} $\lambda
1302$ (\#13) and \ion{Al}{2} $\lambda 1670$ (\#34).  The \ion{Si}{2}
$\lambda 1260$ line may be blended with a weak \ion{N}{5} $\lambda
1242$ line from $z_{\rm abs} = 2.3167$. \ion{Mg}{2} is beyond the
range of the spectrum.

$z_{\rm abs} = 2.3167$. --- Only \ion{C}{4} $\lambda\lambda
1548,\!1550$ (\#30,31) are unambiguous in this system.  However, there
is absorption at the expected position of Lyman-$\alpha$ and there is
evidence for \ion{N}{5} $\lambda 1238$.  \ion{N}{5} $\lambda 1242$ may
also be present, but if so, it is blended with \ion{Si}{2} $\lambda
1260$ from $z_{\rm abs} = 2.2691$.

\subsection{FBQS0910+2539 ($z=2.744$)}

This is the highest redshift QSO in the sample.  The \ion{C}{4}
emission line is shifted into the red spectrum.  This significantly
reduces the S/N near the emission line, which, in turn, makes it more
difficult to detect \ion{C}{4} absorption lines near the redshift of
the QSO.  The \ion{Si}{4} emission line is resolved.  There are six
strong absorption lines past Lyman-$\alpha$ emission.  Among these
lines only two systems are found.  There is evidence for absorption in
the blue wing of \ion{C}{4} emission (\#52,53), but a better spectrum
is needed in order to identify it.

$z_{\rm abs} = 1.6060$ --- The only lines seen at this redshift are
\ion{Mg}{2} $\lambda\lambda 2796,\!2803$ (\#61,62).  However, 
the agreement between the lines is not very good -- making this system
uncertain.

$z_{\rm abs} = 2.0101$. --- Only the \ion{C}{4} doublet (\#47,48) is
observed in this system.  However, the lines are strong, so the system
is probably real.

\subsection{FBQS0934+3153 ($z=2.420$)}

This is an obvious BALQSO.  No attempt has been made to search for
narrow lines and this object is excluded from any analysis.  The broad
troughs near $\sim\!4400\,{\rm \AA}$ and $\sim\!4875\,{\rm \AA}$ are
BALs where the continuum is incorrectly following the troughs.

\subsection{FBQS0955+3335 ($z=2.503$)}

The spectrum for this object is very nice.  There is a weak \ion{N}{5}
emission line that is well separated from Lyman-$\alpha$ emission.
The \ion{C}{4} emission line is fairly symmetric.  As for \ion{Si}{4}
emission, it is not clear whether it is resolved, or if there is
absorption at the peak of the line.  This spectrum has 25 significant
lines redward of Lyman-$\alpha$ emission that are part of six
absorption line systems.

$z_{\rm abs} = 1.5364$. --- Given that there are nine lines in this
system, it can be considered as certain.  The \ion{Mg}{2} doublet
(\#61,62) is quite strong and five lines of \ion{Fe}{2} are observed
(\#54,55,56,58,59). \ion{Al}{3} $\lambda\lambda 1854,\!1862$ (\#31,33)
and \ion{Al}{2} $\lambda 1670$ (\#23) may also be detected; however,
the \ion{Al}{2} line is in the Lyman-$\alpha$ forest and the red
\ion{Al}{3} line may be blended with a possible detection of
\ion{Si}{2} $\lambda 1526$ from $z_{\rm abs} = 2.0953$.

$z_{\rm abs} = 2.0953$. --- There are many lines in this system, but
all are relatively weak.  \ion{C}{4} $\lambda\lambda 1548,\!1550$
(\#34,35) is clearly seen; there is also evidence for the \ion{Si}{4}
doublet (\#28,29).  Other possible lines are \ion{Al}{2} $\lambda
1670$, \ion{Si}{2} $\lambda 1526$ (though, if real, it would be
blended with \ion{Al}{3} $\lambda 1862$ from $z_{\rm abs} = 1.5364$)
and possibly some weak \ion{Fe}{2} lines.

$z_{\rm abs} = 2.2857,2.2923$. --- These
\ion{C}{4}(\#40,41,42)+Lyman-$\alpha$(\#3,4) systems are uncertain;
however if one is real, then the other one is needed to explain the
doublet ratio.

$z_{\rm abs} = 2.3188$. --- This is another system with only
\ion{C}{4} $\lambda\lambda 1548,\!1550$ (\#43,44) and Lyman-$\alpha$ (\#7).

$z_{\rm abs} = 2.5087$. --- The reality of this redshift is
questionable.  The proposed detection of \ion{C}{4} $\lambda 1548$
(\#45) is below the $5\sigma$ detection threshold; however there is a
weak feature at the expected position of \ion{C}{4} $\lambda 1550$ and
there is a Lyman-$\alpha$ (\#25) line to go with it.

\subsection{FBQS1045+3440 ($z=2.353$)}

This QSO is extremely rich in absorption lines with 55 $5\sigma$ lines
redward of Lyman-$\alpha$ emission.  This is another example of a QSO
where it is not clear if \ion{Si}{4} emission is partially resolved or
if the apparent separation is caused by an absorption feature.  There
are as many as 11 absorption line systems.

$z_{\rm abs} = 0.6232$. --- This system is very uncertain; however, if
it were real it would explain some of the residual features in the
absorption spectrum that are not accounted for by the other systems.
Possible lines are \ion{Mg}{2} $\lambda\lambda 2796,\!2803$ (\#29,30)
and \ion{Fe}{2} $\lambda 2600$ (\#19).  They are badly blended with
lines from other systems.

$z_{\rm abs} = 0.8811$. --- Only the \ion{Mg}{2} doublet (\#50,51) is
seen at this redshift, making it ``possible''.

$z_{\rm abs} = 1.7157$. --- The \ion{Mg}{2} doublet (\#74,76) is
contaminated by residual telluric absorption; however, with a dozen or
more lines, this system is unambiguous.  Also seen are six lines of
\ion{Fe}{2} (\#23,57,59,60,67,68), \ion{Al}{3} $\lambda\lambda
1854,\!1862$ (\#40,42), \ion{Al}{2} $\lambda 1670$ (\#29) and possibly
\ion{Si}{2} $\lambda 1526$ (\#13) though it is blended with the \ion{N}{5}
absorption from the complex of ``associated'' systems.  Finally, there
is strong evidence for \ion{C}{4} $\lambda\lambda 1548,\!1550$
(\#15,16), though the doublet is not resolved --- this could reflect
the fact that the lines are broader than usual or that there is
contamination from another system.

$z_{\rm abs} = 1.7270$. --- This system has all the same lines as the
previous system; however, none of them are unblended.  As such the
system is considered probable, but not definite.

$z_{\rm abs} = 1.9456$.  --- The lines in this system are all very
strong.  The \ion{C}{4} $\lambda\lambda 1548,\!1550$ (\#31,32) lines
are stronger than $1\,{\rm \AA\,REW}$.  Many \ion{Fe}{2} lines
(\#36,65,66,\linebreak[1]67,76,78) are seen as are \ion{Al}{3} $\lambda\lambda
1854,\!1862$ (\#52,53), \ion{Al}{2} $\lambda 1670$ (\#39) and
\ion{Si}{2} $\lambda 1526$ (\#27).  The blue line of the \ion{Si}{4}
doublet (\#9) is also strong; the red line is blended with \ion{N}{5}
from the associated absorption complex.  The spectrum does not quite
cover \ion{Mg}{2}, but there is an indication of the \ion{Mg}{2}
$\lambda 2796$.

$z_{\rm abs} = 2.0750$. --- Only \ion{C}{4} $\lambda\lambda
1548,\!1550$ (\#37,38) and \ion{Si}{4} $\lambda\lambda 1393,\!1402$
(\#21,22) are found, but they are not blended with any other lines, so
this redshift is fairly certain.

$z_{\rm abs} = 2.2525$. --- This system is suggested by a likely
detection of the \ion{C}{4} doublet (\#40,41) and possible detection
of the \ion{Si}{4} doublet (\#29,32), though \ion{Si}{4} is badly
blended.

$z_{\rm abs} = 2.3325,2.3375,2.3432,2.3496$. --- This is a set of four
overlapping systems of \ion{C}{4} $\lambda\lambda 1548,\!1550$
(\#44,45,46,47,48), \ion{N}{5} $\lambda\lambda 1238,\!1242$
(\#10,11,12,13,14) and Lyman-$\alpha$ (\#5,6,7,8).  It is unclear if
all four systems are real; however, it is difficult to explain the
line ratios with fewer systems.

\subsection{FBQS1253+2905 ($z=2.565$)} 

This object is the host of yet another example of a complex
\ion{Si}{4} emission line.  Either there is significant absorption at
the redshift of the QSO, or the emission line doublet is partially
resolved.  \ion{N}{5} emission is fairly strong and is well-separated
from Lyman-$\alpha$ emission.  There is a complex of associated
absorption lines that are of particular interest.  Twenty-four
absorption lines with significance greater than $5\sigma$ are found to
the red of the Lyman-$\alpha$ emission line.  These lines constitute
six different systems.

$z_{\rm abs} = 1.3186$. --- \ion{Mg}{2} $\lambda\lambda 2796,\!2803$
(\#66,67) and \ion{Fe}{2} $\lambda 2382$ (\#57) are the only lines
observed at this redshift: as such it is considered ``possible''.

$z_{\rm abs} = 1.9127$. --- Only \ion{C}{4} $\lambda\lambda
1548,\!1550$ (\#38,39) is detected.

$z_{\rm abs} = 2.0654$. --- The \ion{C}{4} doublet (\#41,42),
\ion{Si}{2} $\lambda 1526$ (\#40) and \ion{Al}{2} $\lambda 1670$
(\#49) define this mixed ionization system.

$z_{\rm abs} = 2.1083$. --- \ion{C}{4} $\lambda\lambda 1548,\!1550$
(\#43,44) are observed and there is a possible detection of
\ion{Si}{4} $\lambda\lambda 1393,\!1402$ (\#30,32).

$z_{\rm abs} = 2.5411,2.5497$ --- Both of these systems exhibit
\ion{C}{4} $\lambda\lambda 1548,\!1550$ (\#52,53,\linebreak[1]54,55), \ion{N}{5}
$\lambda\lambda 1238,\!1242$ (\#33,34,35,36) and Lyman-$\alpha$
(\#27,28,29).  The $z_{\rm abs} = 2.5497$ system is quite strong and
fairly unambiguous.  The $z_{\rm abs} = 2.5411$ system is necessary to
account for the unusual profile of the complex and is less certain.

\subsection{FBQS1348+2840 ($z=2.466$)}

The \ion{C}{4} emission line of this QSO is somewhat asymmetric.
Otherwise it is a fairly ordinary QSO emission spectrum.  However,
there is a noticeable lack of absorption.

$z_{\rm abs} = 1.0179$. --- This redshift is suggested by a possible
\ion{Mg}{2} doublet (\#38,39) and \ion{Fe}{2} $\lambda 2382$ (\#27);
however, it is equally likely that the \ion{Mg}{2} absorption is a
continuum fitting feature.  As such this system is considered
unlikely.

$z_{\rm abs} = 1.6610$. --- There is little more than a possible
detection of \ion{Al}{3} $\lambda\lambda 1854,\!1862$ (\#31,32) to
indicate this rather unlikely system.

\subsection{FBQS1416+2649 ($z=2.303$)}

The signal-to-noise of this spectrum is lower than the goal as a
result of inclement weather during observations at this right
ascension.  Many of the lines in the red spectrum are likely to be
spurious.  In the blue spectrum there are 15 significant lines to the
red of Lyman-$\alpha$ emission.  These lines are found in five
systems.

$z_{\rm abs} = 1.6975$. --- The \ion{C}{4} doublet (\#8,9) in this
system is somewhat suspect as as result of blending; however, with six
or more other lines, the system itself is certain.  \ion{Mg}{2}
$\lambda\lambda 2796,\!2803$ (\#56,57) are obvious even at the lower
signal-to-noise of the red spectrum.  Also found are two or more
\ion{Fe}{2} lines (\#41,43,48) along with \ion{Si}{2} $\lambda 1526$
(\#6) and \ion{Al}{2} $\lambda 1670$ (\#14).

$z_{\rm abs} = 2.0162$. --- Weak \ion{Si}{4} $\lambda\lambda
1393,\!1402$ (\#10,11) is seen in this \ion{C}{4} (\#16,17) absorption
line system.  There is also a hint of \ion{Fe}{2} $\lambda 2382$
(\#49), but this may be an artifact.

$z_{\rm abs} = 2.2044$. --- Strong lines of \ion{C}{4} $\lambda\lambda
1548,\!1550$ (\#19,20) are the only lines found in this system.
\ion{Si}{4} would be observable but is not detected.

$z_{\rm abs} = 2.2328$. --- This is another \ion{C}{4} only system
(\#21,22).

$z_{\rm abs} = 2.2928$. --- This system is defined by a very strong
\ion{C}{4} doublet (\#23,24) found very close to the emission redshift
of the QSO.  \ion{N}{5} $\lambda\lambda 1238,\!1242$ (\#4,5) is also
clearly seen.  There is some evidence for \ion{Si}{4} (\#15), but the
red line is not detected.  In addition, there is a feature near the
expected position of \ion{C}{2} $\lambda 1334$ (\#13).

\subsection{FBQS1457+2707 ($z=2.531$)}

This QSO is yet another example from this sample of QSOs where the
\ion{Si}{4} emission line ($\sim\!4950\,{\rm \AA}$) appears resolved.
\ion{N}{5} emission is also clearly seen in the wing of
Lyman-$\alpha$.  Thirty lines stronger than $5\sigma$ are found to the
red of Lyman-$\alpha$ emission.  These lines constitute four systems.
The two lowest redshift systems are particularly interesting.  They
are low ionization systems separated by approximately the doublet
splitting of \ion{C}{4}.  This splitting made the identification
difficult, since the resulting features look like poorly resolved
\ion{C}{4} doublets with ratios that are not quite correct.  There is
also evidence for \ion{Mn}{2} absorption.

$z_{\rm abs} = 0.9854,0.9888$. --- This pair of low ionization systems
wreaked some havoc with the line identification code as mentioned
above.  However, from visual inspection it is clear that it is a pair
of systems with \ion{Mg}{2} $\lambda\lambda 2796,\!2803$ (\#43,44),
\ion{Fe}{2} (\#32,33,34,35,37,38,39,40) and probably \ion{Mn}{2}.  The
higher redshift line is similar in strength to the blue line in
\ion{Fe}{2}, but is considerably weaker in \ion{Mg}{2}.

$z_{\rm abs} = 1.5701$. --- The only evidence for this redshift is a
possible detection of the \ion{Mg}{2} doublet (\#60,61) that is near
a telluric feature.

$z_{\rm abs} = 1.8346$. --- \ion{C}{4} $\lambda\lambda 1548,\!1550$
(\#28,29) is clearly detected and there is evidence for many other
lines.  These include \ion{Mg}{2} $\lambda\lambda 2796,\!2803$
(\#76,77), two lines of \ion{Fe}{2} (\#51,52), \ion{Si}{2} $\lambda
1526$ (\#27) and possibly \ion{Al}{3} $\lambda 1854$ (\#41).  There is
also a line at the locations of \ion{Al}{2} $\lambda 1670$ (\#35);
however, this line overlaps the \ion{Fe}{2} $\lambda 2382$ lines from
the $z_{\rm abs} = 0.9854,0.9888$ complex and may not be real.

\subsection{FBQS1537+2716 ($z=2.445$)}

This QSO has a rather unusual emission spectrum; the emission lines
are {\em very} weak.  In particular, the \ion{C}{4} emission line
merges smoothly with the continuum on the blue side and is discernible
only by the red wing of the line.  Redward of Lyman-$\alpha$ emission
there are 9 strong absorption lines, of which half are identified.
Many of the unidentified lines are in the red spectra and may be
continuum fitting artifacts resulting from lower S/N in the red
spectra.

$z_{\rm abs} = 0.8595$. --- Only \ion{Mg}{2} $\lambda\lambda
2796,\!2803$ (\#32,33) is found.

$z_{\rm abs} = 1.7341$. --- The \ion{Mg}{2} doublet (\#53,54) and
\ion{Fe}{2} $\lambda 2382$ (\#44) lines are seen; however, they are all
questionable identifications.

\subsection{FBQS1540+4138 ($z=2.512$)}

The emission spectrum of this QSO has \ion{C}{4} much stronger than
\ion{Si}{4}, and the \ion{N}{5} line is clearly seen in the red wing
of Lyman-$\alpha$.  Thirty-five significant lines are detected
longward of Lyman-$\alpha$ emission.  These lines are distributed
among six systems.

$z_{\rm abs} = 0.8697$. --- This system is identified by a saturated
\ion{Mg}{2} doublet (\#50,51) and a weak \ion{Mg}{1} $\lambda 2852$
(\#53) line. \ion{Fe}{2} lines are also found at rest wavelengths of
$\lambda 2344, 2382, 2586\,{\rm and}\,2600$ (\#30,31,38,39).  The \ion{Fe}{2}
$\lambda 2586$ line is blended with \ion{Si}{2} $\lambda 1526$ from
$z_{\rm abs} = 2.1675$.

$z_{\rm abs} = 1.9219$. --- The \ion{C}{4} $\lambda\lambda
1548,\!1550$ doublet (\#32,33) suggests the possibility of this
redshift.

$z_{\rm abs} = 2.0265$. --- At this redshift a nearly saturated
\ion{C}{4} doublet (\#36,37) is seen as are \ion{Si}{2} $\lambda 1526$
(\#35) and \ion{Al}{2} $\lambda 1670$ (\#46).  \ion{Si}{4}
$\lambda\lambda 1393,\!1402$ and \ion{C}{2} $\lambda 1334$ may also be
present, but they are in the forest.

$z_{\rm abs} = 2.1675$ --- A strong \ion{C}{4} doublet (\#41,42) is
found in the midst of the \ion{Si}{4} emission line.  Other lines
include \ion{Si}{4} $\lambda\lambda 1393,\!1402$ (\#29,30),
\ion{Al}{2} $\lambda 1670$ (\#52) and \ion{Si}{2} $\lambda 1526$ (which is
blended with \ion{Fe}{2} from $z_{\rm abs} = 0.8697$).  Other low
ionization lines also may be detected, but are lost in the noise of
the forest.

$z_{\rm abs} = 2.2291$ --- A weak \ion{C}{4} doublet (\#44,45) is the
only thing observed at this redshift.

$z_{\rm abs} = 2.2791$ --- Again only \ion{C}{4} $\lambda\lambda
1548,\!1550$ (\#47,48) is seen.

\subsection{FBQS1625+2646 ($z=2.507$)}

This object, otherwise known as KP~76, was previously studied by
\citet{cro89}.  The Lyman-$\alpha$, \ion{Si}{4} and \ion{C}{3]}
emission lines are shifted significantly redward from the redshift
indicated by \ion{C}{4}.  There are 40 lines of significant strength
not including those lines found in the Lyman-$\alpha$ forest.  As many
as 12 systems may be found in this spectrum, though most are
questionable.  There is one strong line on top of the \ion{C}{4}
emission line that remains unidentified.

$z_{\rm abs} = 0.7503$ --- It is possible that there might be
absorption by \ion{Mg}{2} $\lambda\lambda 2796,\!2803$ (\#35,37) at
this redshift.

$z_{\rm abs} = 0.8873$ --- A \ion{Mg}{2} doublet (\#42,43),
\ion{Mg}{1} $\lambda 2852$ (\#47) and three \ion{Fe}{2} lines
(\#25,28,37) define this system.  The blue line in the \ion{Mg}{2}
doublet is blended with one or more other lines.

$z_{\rm abs} = 1.0388$ --- \ion{Mg}{2} $\lambda\lambda 2796,\!2803$
(\#54,55) is evident.  \ion{Fe}{2} $\lambda 2382, \lambda 2344$ and
$\lambda 2600$ (\#32,34,44) are also present.

$z_{\rm abs} = 1.1346$ --- While it is unlikely, a doublet of
\ion{Mg}{2} (\#58,59) may be detected at this redshift.

$z_{\rm abs} = 1.4240$ --- There appears to be an absorption feature
at the wavelength indicated by a \ion{Mg}{2} doublet (\#63,64) at this
redshift.  The reality of the identification is highly questionable.

$z_{\rm abs} = 1.4805$ --- The feature near $\lambda 6940$ may be a
continuum fitting residual, but may also be \ion{Mg}{2}
$\lambda\lambda 2796,\!2803$ (\#66,67).

$z_{\rm abs} = 1.8799$ --- A doublet of \ion{C}{4} (\#26,27) may
suffice to explain this absorption feature near $\lambda 4460$.

$z_{\rm abs} = 2.0520$ --- Strong \ion{C}{4} $\lambda\lambda
1548,\!1550$ (\#30,31) absorption defines this system.  This redshift
may be confirmed by possible identification of \ion{Si}{4}
$\lambda\lambda 1393,\!1402$ and \ion{C}{2} $\lambda 1334$; however
those lines are all in the forest.

$z_{\rm abs} = 2.2453$ --- A weak \ion{C}{4} doublet (\#38,39)
suggests this redshift.

$z_{\rm abs} = 2.4010$ --- This may be a system with \ion{C}{4}
$\lambda\lambda 1548,\!1550$ (\#41,42).  If real the blue line is
clearly observed, but the red line is blended with \ion{Mg}{2}
$\lambda 2796$ from $z_{\rm abs} = 0.8873$.  However, a strong line is
seen at the expected position of Lyman-$\alpha$ (\#11) so this system
is considered probable.

$z_{\rm abs} = 2.4441$ --- This is another Lyman-$\alpha$+\ion{C}{4}
system.  The \ion{C}{4} doublet (\#45,46) is weaker in this system
than in the previous one.

$z_{\rm abs} = 2.5288$ --- There is strong evidence for both
Lyman-$\alpha$ (\#23) and \ion{C}{4} $\lambda 1548$ (\#49) absorption
at this redshift; however, \ion{C}{4} $\lambda 1550$ is not detected.
As such this system is questionable.

\subsection{FBQS1634+3203 ($z=2.336$)} 

The \ion{C}{4} emission line is very broad as is the
Lyman-$\alpha$+\ion{N}{5} blend.  Thirty-four significant lines are
found in the six systems redward of Lyman-$\alpha$ emission.

$z_{\rm abs} = 0.7293$ --- This system is defined by a weak
\ion{Mg}{2} doublet (\#25,26).

$z_{\rm abs} = 1.2839$ --- There is evidence for \ion{Mg}{2}
$\lambda\lambda 2796,\!2803$ (\#49,50) and some \ion{Fe}{2} lines at
this redshift; however they are all very likely to be spurious.

$z_{\rm abs} = 1.4942$ --- The detection of eight lines leaves little
doubt regarding the reality of this systems.  Doublets of \ion{Mg}{2}
(\#56,57) and \ion{Al}{3} (\#19,20) are found along with \ion{Al}{2}
$\lambda 1670$ (\#11).  Three or more \ion{Fe}{2} lines are also seen.

$z_{\rm abs} = 2.0921$ --- This system is somewhat unusual:
\ion{Al}{2} $\lambda 1670$ (\#31), \ion{Si}{2} $\lambda 1526$ (\#23)
and \ion{C}{2} $\lambda 1334$ (\#9) are the strongest lines in the
system.  The \ion{Fe}{2} lines are somewhat weaker and the \ion{C}{4}
and \ion{Si}{4} doublets are even weaker.  In addition, there is
absorption at the expected position of \ion{Ni}{2} $\lambda 1709$
(\#35).  \ion{Mg}{2} is beyond the range of the spectrum.

$z_{\rm abs} = 2.2582$ --- \ion{C}{4} $\lambda\lambda 1548,\!1550$
(\#29,30) are seen along with \ion{Si}{4} $\lambda\lambda 1393,\!1402$
(\#16,17) and possibly \ion{Si}{2} $\lambda 1526$ (\#23); however,
\ion{Al}{2} $\lambda 1670$ is not detected.

$z_{\rm abs} = 2.3504$ --- This is clearly an ``associated'' system
with a redshift slightly higher than that of the QSO.  The \ion{C}{4}
(\#32,33), \ion{Si}{4} (\#21,22) and \ion{N}{5} (\#10,11) doublets are
all clearly detected.  A very strong Lyman-$\alpha$ line (\#7) is also
observed.

\subsection{FBQS1645+2244 ($z=2.723$)} 

The redshift of this QSO pushes the \ion{C}{4} emission line into the
red spectrum.  Both \ion{C}{3]} and \ion{Al}{3} are also observed in
emission in the red part of the spectrum.  The blue spectrum shows a
broad Lyman-$\alpha$ emission line along with relatively strong lines
of \ion{C}{2} ($\sim\!4975\,{\rm \AA}$) and \ion{O}{1} ($\sim\!4850\,{\rm
\AA}$), with weaker \ion{Si}{4}.  There are 21 $5\sigma$ lines
longward of Lyman-$\alpha$ emission.  Only three systems are
identified with these lines.

$z_{\rm abs} = 1.9648$ --- This is a \ion{C}{4} doublet (\#49,50) of
moderate strength.  \ion{Si}{4} may also be detected, but it is in the
forest.

$z_{\rm abs} = 2.3992$ --- Both \ion{C}{4} $\lambda\lambda
1548,\!1550$ (\#59,60) and \ion{Si}{4} $\lambda\lambda 1393,\!1402$
(\#52,53) are fairly strong in this absorber.  Lower ionization lines
of \ion{Si}{2} $\lambda 1526$ (\#58) and \ion{C}{2} $\lambda 1334$
(\#46) may also be detected.  In the forest, the Lyman-$\alpha$ line
(\#10) is strong and stands out from the forest.  \ion{N}{5} may also
be present, but it is difficult to determine if the lines are real or
simply Lyman-$\alpha$ forest lines.

$z_{\rm abs} = 2.5738$ --- There is a possible detection of \ion{C}{4}
(\#62,63) at this redshift.  The reality of this system is further
supported by detections of \ion{Si}{4} (\#54,55) and Lyman-$\alpha$
(\#27).

\subsection{FBQS1651+4002 ($z=2.339$)}

The emission lines in this QSO all have very nearly the same redshift.
Redward of Lyman-$\alpha$ emission, there are 29 absorption lines with
greater than $5\sigma$ significance.  These lines constitute six
systems.

$z_{\rm abs} = 0.4991$ --- This is a very low redshift \ion{Mg}{2}
doublet (\#10,11).  The presence of \ion{Mg}{1} $\lambda 2852$
(\#12) supports the reality of this redshift.

$z_{\rm abs} = 1.6380$ --- A strong doublet of \ion{C}{4} (\#7,8) in
the red wing of Lyman-$\alpha$ emission defines this system.
\ion{Mg}{2} may also be detected, but it is uncertain.

$z_{\rm abs} = 1.7992$ --- Only \ion{C}{4} $\lambda\lambda
1548,\!1550$ (\#14,15) serves to define this system.

$z_{\rm abs} = 1.8912$ --- This system is certain; it has at least six
strong lines with many more weaker lines.  Both the \ion{C}{4}
(\#19,20) and \ion{Mg}{2} (\#67,68) doublets are found.  \ion{Si}{2}
$\lambda 1526$ (\#16) and \ion{Al}{2} $\lambda 1670$ (\#24) also are
clearly detected and there is evidence for a few of the \ion{Fe}{2}
transitions as well.

$z_{\rm abs} = 1.9920$ --- Weak \ion{C}{4} $\lambda\lambda
1548,\!1550$ (\#21,22) is seen in the blue wing of \ion{Si}{4}
emission.  Lyman-$\alpha$ is not in the range of the spectrum.

$z_{\rm abs} = 2.1905$ --- This is another \ion{C}{4} only system
(\#27,28).  If real, it is very weak.

\subsection{FBQS2216-0057 ($z=2.392$)}

This QSO has an asymmetric \ion{C}{4} emission line ($\sim\!5250\,{\rm
\AA}$) that is skewed towards the red.  \ion{N}{5} and \ion{O}{1} are
also clearly observable.  Twenty-one lines that are significant are
observed longward of Lyman-$\alpha$ emission.  There are five
absorption line systems.

$z_{\rm abs} = 0.5372$ --- This redshift is suggested by weak
\ion{Mg}{2} $\lambda\lambda 2796,\!2803$ (\#17,18) absorption.  It is
considered questionable.

$z_{\rm abs} = 0.6791$ --- Strong lines of \ion{Mg}{2} (\#23,24) and
\ion{Mg}{1} $\lambda 2852$ (\#27) are seen as are \ion{Fe}{2} $\lambda 2586$
and $\lambda 2600$ (\#19,20).

$z_{\rm abs} = 1.5724$ --- This is another strong \ion{Mg}{2}
$\lambda\lambda 2796,\!2803$ (\#50,51) system.  \ion{Mg}{1} $\lambda
2852$ is not detected but six \ion{Fe}{2} lines are
(\#14,42,43,44,45,46).  The \ion{Al}{3} $\lambda\lambda 1854,\!1862$
doublet (\#25,27), \ion{Al}{2} $\lambda 1670$ (\#17) and \ion{Si}{2}
$\lambda 1808$ (\#22) are also seen.  \ion{C}{4} is lost in the
forest.

$z_{\rm abs} = 2.1674$ --- Only \ion{C}{4} (\#29,30) serves to
indicate this redshift.  Lyman-$\alpha$ is out of range.

$z_{\rm abs} = 2.3927$ --- Strong Lyman-$\alpha$ (\#13) is seen in
this relatively weak \ion{C}{4} system (\#36,37).  It is just blueward
of the emission redshift.

\subsection{FBQS2233-0838 ($z=2.339$)}

This QSO has a fairly innocuous emission spectrum.  \ion{N}{5} is seen
in the red wing of Lyman-$\alpha$.  Longward of Lyman-$\alpha$
emission, there are 24 strong lines in nine systems.  Six lines remain
unidentified.

$z_{\rm abs} = 0.5860$ --- Strong \ion{Mg}{2} (\#20,21) is detected at
this redshift.  Also seen are \ion{Mg}{1} $\lambda 2852$ (\#22) and
\ion{Fe}{2} $\lambda 2600$ (\#11).  \ion{Fe}{2} $\lambda 2586$ would
also be observable, but it is blended with \ion{C}{4} from $z_{\rm
abs} = 1.6465$.

$z_{\rm abs} = 1.0674$ --- There is a hint of \ion{Mg}{2} (\#38,39) to
suggest this redshift, which is very uncertain.

$z_{\rm abs} = 1.1123$ --- Another uncertain \ion{Mg}{2}
$\lambda\lambda 2796,\!2803$ system (\#40,41).

$z_{\rm abs} = 1.5140$ --- No match to the pair of lines on top of
\ion{Si}{4} emission could be found other than the \ion{Al}{3} doublet
(\#24,25).  The agreement is not perfect, but there is absorption at
the expected locations of \ion{Mg}{2} (\#52,53) and \ion{Al}{2}
$\lambda 1670$ (\#15).  As such, this redshift is considered probable.

$z_{\rm abs} = 1.6465$ --- Strong \ion{C}{4} $\lambda\lambda
1548,\!1550$ (\#9,10) is detected at this redshift.  It is partially
contaminated by \ion{Fe}{2} $\lambda 2586$ from $z_{\rm abs} =
0.6791$.

$z_{\rm abs} = 1.7527$ --- This is a questionable \ion{C}{4} (\#16,17)
only system.  Lyman-$\alpha$ is not observable.

$z_{\rm abs} = 1.7809$ --- Same as for $z_{\rm abs} = 1.7527$
(\#18,19).

$z_{\rm abs} = 2.3226$ --- Weak \ion{C}{4} (\#29,30) absorption is
seen is the blue wing of \ion{C}{4} emission.  A strong Lyman-$\alpha$
(\#5) line bolsters the reality of this system.

$z_{\rm abs} = 2.3398$ --- This is an associated absorption system
that is at nearly the same redshift of the QSO.  Lyman-$\alpha$ (\#7)
is seen along with \ion{N}{5} (\#12,13) and \ion{C}{4} (\#31,32).
\ion{Si}{4} is not detected.

\placefigure{fig:fig1}
\placefigure{fig:fig2}
\placefigure{fig:fig3}
\placefigure{fig:fig4}
\placefigure{fig:fig5}
\placefigure{fig:fig6}
\placefigure{fig:fig7}
\placefigure{fig:fig8}
\placefigure{fig:fig9}
\placefigure{fig:fig10}
\placefigure{fig:fig11}
\placefigure{fig:fig12}
\placefigure{fig:fig13}
\placefigure{fig:fig14}
\placefigure{fig:fig15}
\placefigure{fig:fig16}
\placefigure{fig:fig17}
\placefigure{fig:fig18}
\placefigure{fig:fig19}
\placefigure{fig:fig20}
\placefigure{fig:fig21}
\placefigure{fig:fig22}
\placefigure{fig:fig23}
\placefigure{fig:fig24}

\section{Analysis and Discussion}

Prior to discussing the evidence for and against high-velocity,
intrinsic \ion{C}{4} absorption, it is necessary to have a basis for
comparison.  As such, the distribution of all \ion{C}{4} and
\ion{Mg}{2} absorbers both as a function of redshift and velocity are
presented here.  Some preliminary comments are necessary.  When
discussing the redshift distribution of absorbers, we make the
assumption that {\em all} of the absorbers have redshifts that are
indicative of their cosmological distribution and that there are no
significant relativistic velocity components to their redshifts.  Yet,
it is well-known that there are absorbers that have large relativistic
velocity components to their redshifts, for example, BALs and
associated absorption lines.  Fortunately, both of these species are
easy to recognize and they can be removed from the sample when and if
called for.

Similarly, when discussing the velocity distribution of absorbers, it
is assumed that {\em all} the absorbers have cosmological redshifts
that are very close to the redshift of the QSO and that any
differences are due to relative velocities.  However, many, if not
most, QSO absorption lines are cosmological in origin, i.e. they are
caused by material along the line of sight, but not physically
associated with the QSO.  Unlike the case described in the previous
paragraph, there is almost no way to distinguish these systems a
priori so that the intrinsic/associated systems can be studied by
themselves.

This fundamental inability to separate the two classes of absorbers is
the basis for this study.  The ability to separate these populations
is crucial to the science that comes from the analysis of QSO
absorption lines systems.  Should it be confirmed that there is indeed
a significant population of apparently high-velocity absorbers that
also have narrow profiles, it will have a strong impact on previous
and future absorption lines studies.  The ability to discern the
differences in the classes of absorption lines systems using low
resolution spectra, as opposed to more time consuming high resolution
spectra as discussed by \citet{bs97}, would be of considerable value.

The primary diagnostics used throughout this work are plots of the
number of absorbers per unit redshift ($dN/dz \equiv N(z)$) and the
number of absorbers per unit velocity ($dN/d\beta \equiv N(\beta)$).
Both of these quantities are normalized by the number of times a given
absorber of a given strength could have been observed in the other
QSOs in the sample, and should be independent of the velocity and
redshift distribution of the parent objects (unless there is a physical
meaning to such correlations).

Table~5 gives the different samples that are studied herein.  Sample
S1 includes all objects from the \citet{yyc+91} absorption line
catalog, plus some systems from the update to this catalog
\citep{v+00}.  Sample S3 includes only the absorbers discovered in the
24 spectra presented herein.  Sample S2 includes all of Sample S3 in
addition to absorbers from objects in Sample S1 that met the selection
criteria of Sample S3.  The other sample types restrict either the
intrinsic QSO properties or the element studied.  Each of these
samples can be combined with a bitwise AND to create a new, more
restricted sample.  For example, Sample S1C4L contains only \ion{C}{4}
systems from the radio-loud quasars in Sample S1.

\placetable{tab:tab5}

\subsection{Redshift distribution of absorbers}

Presented here are the normalized redshift distributions for
\ion{C}{4} and \ion{Mg}{2}.  The redshift distribution of absorbers as
a function of atom/ion will be discussed in detail by \citet{v+00};
however, a brief discussion is necessary here.  Figure~\ref{fig:fig25}
shows the normalized redshift distribution of all the \ion{C}{4}
absorbers from Sample S1C4.  The solid line in the bottom half of the
plot gives the number of absorbers found in a given redshift bin (left
axis).  The dashed line in the bottom plot gives the number of quasars
searched (right axis).  The top plot is essentially the ratio of the
two bottom plots and gives the normalized redshift distribution after
correcting for selection effects; error bars are Poisson.  In the top
plot, thick lines are for bins with more than ten absorbers and are
considered significant.  Bins delineated by thinner lines have fewer
than ten absorbers in a bin; these bins should be treated with caution
(or they could be combined together in a single, more significant
bin).  Lines within $5000\,{\rm km\,s^{-1}}$ of the QSO redshift have
been removed.

\placefigure{fig:fig25}

For comparison, the new data taken specifically for this project
(Sample S3), yield a value for the number of absorbers per unit
redshift (at $z\sim 2$) of $2.54\pm0.44$ for \ion{C}{4} absorbers with
rest equivalent widths stronger than $0.15\,{\rm \AA}$.  Note that
this value is somewhat higher than that of Sample S1 as shown in
Figure~\ref{fig:fig25}.  For absorbers stronger than $0.30\,{\rm
\AA}$, $dN/dz$ is $1.43\pm0.32$.  Both values are consistent with
those determined by \citet*{ssb88}.

Figure~\ref{fig:fig26} is similar to Figure~\ref{fig:fig25} except
that here \ion{Mg}{2} is plotted instead of \ion{C}{4} (i.e. Sample
S1M2).  The most obvious difference is that the \ion{Mg}{2} absorption
is found at much lower redshift, since \ion{Mg}{2} is at longer
wavelengths and moves out of the optical and into the IR sooner than
\ion{C}{4}.  Ideally, one would like to compare the two species as
tracers of high- and low-ionization lines, but the redshift
discrepancy makes such a comparison difficult.  Of particular interest
is the well-known fact that \ion{Mg}{2} absorption is seen less than
half as often as \ion{C}{4}.  This is true even in the small redshift
range where the samples overlap.

\placefigure{fig:fig26}

\subsection{Velocity Distribution}

As with the redshift distribution analysis, it is also possible to
assume that all of the absorbers are associated with the QSO and
observed wavelengths that represent outflow velocities (positive or
negative) with respect to the QSO redshift.  Figure~\ref{fig:fig27}
shows the velocity distribution of all of the \ion{C}{4} (and
\ion{Mg}{2}) absorbers in the entire sample.  The 375 \ion{C}{4}
absorbers with relative velocities in excess of $5000\,{\rm
km\,s^{-1}}$ are the same absorbers as in Figure~\ref{fig:fig25},
which emphasizes the different assumptions made in each type of
analysis.  The flatness of the distribution at large velocities
typically is taken as evidence for the intervening hypothesis
\citep{ssb88}.  That is, if the absorbers are {\em not} associated
with the QSOs but rather with intervening galaxies, then the erroneous
assumption that the absorbers are associated with the QSOs will yield
a random velocity distribution.  For these (and all of the following)
graphs of $dN/d\beta$, the first bin includes all systems within
$5000\,{\rm km\,s^{-1}}$ of the QSO redshift.  That most of these
systems (those within $5000\,{\rm km\,s^{-1}}$) are primarily
associated with QSOs and not intervening galaxies is generally
accepted.  Discussion of data in this velocity range will be reserved
for a future study.  For \ion{C}{4} absorbers with $v > 5000\,{\rm
km\,s^{-1}}$ in Sample S3, the \ion{C}{4} $dN/d\beta$ values are
$7.84\pm1.34$ and $4.44\pm0.99$ at $z\sim2$ for $W_{rest} > 0.15\,{\rm
\AA}$ and $W_{rest} > 0.30\,{\rm \AA}$, respectively.  These values
are somewhat larger than the average value for Sample S1.

\placefigure{fig:fig27}

Figure~\ref{fig:fig27} also shows the velocity distribution of
\ion{Mg}{2}.  Since the \ion{Mg}{2} emission line is at a much longer
wavelength than the Lyman-$\alpha$ forest (as compared to \ion{C}{4}),
\ion{Mg}{2} absorption is detected at apparent velocities much higher
than that of \ion{C}{4}.  Note that the distribution is quite flat and
that there is no peak at or near zero apparent velocity.  These facts
are consistent with \ion{Mg}{2} absorption being purely cosmological.
The distributions of \ion{C}{2}, \ion{Si}{2} and \ion{Fe}{2} are
similar.

\subsection{Absorption Lines as a Function of Radio Properties}

In this section the velocity distribution of QSO absorption lines is
studied as a function of radio properties.  In each case, two samples
are examined.  First, the entire data set from \citet{yyc+91} and
\citet{v+00} as of this writing (Sample S1), and second, Sample S2,
the sample that includes Sample S3 and that is supplemented with
data of similar quality from Sample S1.

\subsubsection{Radio-Loud versus Radio-Quiet}

The initial premise of this work was the comparison of the
distribution of absorption lines as a function of radio luminosity.
The observed dichotomy of radio-detected quasars provides a natural
starting point for such a comparison \citep{smw+92}.  If radio-loud
and radio-quiet quasars constitute separate species (whether as a
result of actual physical differences, or apparent physical
differences due to the viewing angle) and if QSO absorption line
systems are not associated with the parent QSOs, then the distribution
of these absorbers should be the same for both radio-loud and
radio-quiet sources.  We test this hypothesis for both \ion{C}{4} and
\ion{Mg}{2} in Samples S1 and S2.

Figure~\ref{fig:fig28} gives the velocity distribution of
absorbers for both radio-loud and radio-quiet quasars from Sample S1.
Plotted is the number of \ion{C}{4} absorbers per unit velocity
interval versus velocity.  The solid line represents \ion{C}{4}
absorbers seen in radio-loud quasars, while the dashed line represents
the lines seen in radio-quiet quasars.  In each case the number of
absorbers is given in the legend since that number cannot be derived
from looking at the plot (because it is a {\em normalized} number per
unit velocity and not simply the number in a velocity bin).  The
legend gives the number with $v>5000\,{\rm km\,s^{-1}}$ and the
expected number in the same velocity range as determined by the
average level for the entire sample.  This average level for the
entire sample (and its $1\sigma$ error) for $5000\,{\rm km\,s^{-1}} <
v < 55,000\,{\rm km\,s^{-1}}$ is represented by the dotted line.

\placefigure{fig:fig28}

Figure~\ref{fig:fig29} gives the velocity distribution of absorbers
for both radio-loud and radio-quiet quasars for Sample S2.  What is
interesting is that the radio-loud sample agrees well with the
predicted $dN/d\beta$ from the entire sample, but the radio-quiet
sample shows a small excess over the expected level.  Though the
excess doesn't appear all that significant, it is the way that these
excess absorbers are distributed that are interesting.  In particular,
for the whole sample, as depicted in Figure~\ref{fig:fig28}, the
excess is at low and high velocities and is not prevalent at all
velocities.  This effect could simply be statistical error, but it may
also be indicative of interesting physical processes.  For comparison,
the new data (Sample S3), give $dN/d\beta$ as $6.96\pm2.01$ and
$6.09\pm2.72$ for \ion{C}{4} absorbers at $z\sim2$ with $W_{rest} >
0.15\,{\rm \AA}$ in radio-quiet and radio-loud quasars, respectively.

\placefigure{fig:fig29}

Similar to \ion{C}{4}, the velocity distribution of \ion{Mg}{2} is
plotted in Figure~\ref{fig:fig30} for Samples S1M2L and S1M2Q.  The
first bin shows just those systems within $\pm 5000\,{\rm km\,s^{-1}}$
of the QSO redshift.  The other bins are $50,000\,{\rm km\,s^{-1}}$
wide.  A similar analysis using Sample S2 confirms that there is no
significant difference in the velocity distribution of \ion{Mg}{2}
between radio-quiet and radio-loud quasars.  This fact demonstrates
that the intervening galaxy hypothesis for \ion{Mg}{2} is probably
correct.  It is unclear why the last bin of Figure~\ref{fig:fig30}
shows a difference between the radio-loud and radio-quiet \ion{Mg}{2}
absorbers.  One possibility is that this difference is caused by
\ion{N}{5} doublets being mistaken for \ion{Mg}{2} doublets in
BAL-like radio-quiet QSOs.

\placefigure{fig:fig30}

\subsubsection{Steep-spectrum versus Flat-spectrum}

Though difference in the distribution of \ion{C}{4} in QSOs as a
function of radio luminosity is interesting, a comparison of the
distribution of \ion{C}{4} between flat-spectrum and steep-spectrum
sources is somewhat more intriguing.  Using the spectral indices of
the quasars might be expected to be a better way to discern if there
is any evidence for a population of high velocity, intrinsic narrow
absorption given that quasars are distinctly asymmetric and that
spectral indices are thought to correlate with the viewing angle of
quasars \citep{pu92}.

Figure~\ref{fig:fig31} shows the velocity distribution of \ion{C}{4}
in the steep- and flat-spectrum quasars from Sample S1.  At least
three things of interest can be discerned from this plot.  First,
there is a well-known excess of absorbers in steep-spectrum quasars at
low velocity \citep{fcw+88}.  Second, a small ``spike'' in
steep-spectrum absorbers is seen near $\beta \sim 0.2$, which also
will be commented on later.  Finally, there is the excess of absorbers
in flat-spectrum quasars as compared to steep-spectrum quasars that
was first reported in Paper I.  (However, we note that Sample S1 is
not independent of the sample as used in Paper I).

\placefigure{fig:fig31}

The primary goal of this work is to attempt to determine if this
result from Paper I is real.  It is important to determine if the use
of relatively inhomogeneous optical spectra have biased the result.
For example, if the steep-spectrum quasars were not searched for
absorption as thoroughly as the flat-spectrum quasars, such an effect
might be observed.  Since the levels have been corrected for the
number of times each line could have been observed in terms of the
wavelength coverage {\em and} the equivalent width detection
threshold, such an effect in unlikely.  Nevertheless, it is worth
investigating this result with a more homogeneous sample.

Sample S2 is just such a sample.  The systems from Sample S2 are all
either from Sample S3, or from those objects in \citet{v+00} that meet
the criteria of Sample S3.  As such, Sample S2 is intended to be a
sample that is relatively unaffected by inhomogeneity in the quality
of the optical spectra.  Figure~\ref{fig:fig32} presents the velocity
distribution of \ion{C}{4} absorption in a sample of steep- and
flat-spectrum quasars from Sample S2, which is more homogeneous in its
optical spectra than Sample S1 and includes only QSOs with $z_{em}
\sim 2.5$.  As was first reported in Paper I and as can be seen in
Figure~\ref{fig:fig31}, there is an excess of \ion{C}{4} absorption
from $5000$ to $55,000\,{\rm km\,s^{-1}}$ in the flat-spectrum sample
as compared to the steep-spectrum sample.  The dotted line is the
expected distribution (shown with one standard deviation errors) from
all of the data in Paper I.

\placefigure{fig:fig32}

The excess of \ion{C}{4} in flat-spectrum quasars in Sample S2, while
not terribly significant, is consistent with
Figure~\ref{fig:fig31} and the result from Paper I.  The
persistence of this excess in the more homogeneous sample is
interpreted as a confirmation of the excess of \ion{C}{4} in
flat-spectrum quasars and therefore of an orientation dependent
population of intrinsic absorbers.  This new result suggests that the
old result from Paper I is not an artifact of inhomogeneous optical
spectra.  In Paper I, we used the excess absorption in the
flat-spectrum sources to postulate that there is contaminated at the
36\% level of the \ion{C}{4} sample in flat-spectrum quasars by a
population of intrinsic absorbers.  The numbers are too small in this
sample for a similar analysis, but they are not inconsistent with the
value derived in Paper I.  The observed value of $dN/d\beta$ for
Sample S3 is $8.03\pm2.15$ \ion{C}{4} absorbers with $W_{rest} >
0.15\,{\rm \AA}$ in flat-spectrum quasars.  This value is consistent
with the claimed excess in the larger sample.  Unfortunately, the
number of steep-spectrum quasars in Sample S3 is too small to
determine a similar value for steep-spectrum quasars.

It is also significant that Sample S2 includes only those QSOs with
$z_{em} \sim 2.5$.  The result from Paper I could also have been
biased if there is strong evolution in the frequency of \ion{C}{4}
absorption as a function of redshift and if the flat- and
steep-spectrum QSOs sampled somewhat different redshift ranges (which
they do).  Therefore, the continued excess, albeit small, suggests
that the excess observed in Paper I (and as seen in
Figure~\ref{fig:fig31}) is not an artifact of redshift evolution.

For comparison of \ion{Mg}{2} with \ion{C}{4},
Figure~\ref{fig:fig33} gives the distribution of \ion{Mg}{2}
absorption line systems towards both steep- and flat-spectrum quasars
in Sample 1M2.  This plot serves to demonstrate that this excess of
absorbers in flat-spectrum quasars does not extend to lower ionization
species such as \ion{Mg}{2}.  Sample S2M2 has too few \ion{Mg}{2}
absorption line systems to make a similar comparison.

\placefigure{fig:fig33}

In addition to suffering from the use of inhomogeneous absorption line
data, Paper I also suffered from a lack of homogeneous radio data.  As
such, one must be concerned that the evidence that was found for an
intrinsic population of narrow, high-velocity \ion{C}{4} absorption
line systems was merely an artifact of the inhomogeneity of the data.
We address the problem of inhomogeneous radio spectral indices in
Paper III using new, simultaneous 3.5 and 20\,cm images (from the VLA
in the A configuration) of 144 of the quasars from Paper I.

Since the idea that narrow absorption lines are caused by intervening
galaxies is so entrenched and is backed up by considerable evidence,
it is worth discussing what might cause the observed effects that lead
to the conclusion that intrinsic absorption is not the only possible
explanation for the steep/flat dichotomy.  One obvious way to produce
an excess of absorbers in one sample over another is for there to be
systematic differences in the quality of the spectra or in the search
algorithms.  For instance, if the equivalent width limits for the
steep-spectrum quasars are worse than for the flat-spectrum quasars
because most of the steep-spectrum quasars come from a single study,
then this might explain the dearth of absorption in those objects.
While such a problem would certainly produce the observed effect, it
is unlikely simply because $N(\beta)$ and $N(z)$ have been normalized
not only by the number of quasars that the lines could have been
observed in, but also by the number of times that a line of that
equivalent width could have been measured in those quasars.  However,
more absorption line data from bright, steep-spectrum quasars are
needed to fully rule out selection effects.

\subsection{The Sloan Digital Sky Survey (SDSS)}

The amount of data used for this study is only just enough for any
sort of reasonable statistical analysis.  Even the larger, less
homogeneous sample (Sample 1) from which the primary sample studied
herein (Sample 2) is drawn from is not large enough for certain kinds
of analysis.  However, the Sloan Digital Sky Survey should resolve any
problems caused by the lack of data, let alone the lack of homogeneous
data.  The SDSS will produce spectra for approximately $10^5$ QSOs to
a limiting magnitude of $i'\sim19$.  All stellar sources that are
detected by FIRST will be selected as QSO candidates.  These spectra
will be of sufficient signal-to-noise and resolution that they will be
very valuable for QSO absorption line studies \citep{ybl+99}.

With a spectral resolution $(R=\lambda/\Delta\lambda)$ of 2000 and
wavelength coverage from 3900 to $9000\,{\rm \AA}$, the SDSS QSO
spectra will be well suited for QSO absorption line studies.  Between
redshift 2 and 3 alone, the SDSS will find about 7700 QSOs to a
magnitude limit of $i' = 18.5$ (D. P. Schneider 1999, private
communication).  Typical studies of \ion{C}{4} at moderate redshifts
obtain a $5\sigma$ rest equivalent width limit of $0.15\,{\rm \AA}$
\citep{ssb88}.  For a \ion{C}{4} absorption line system, the
signal-to-noise ratio needed to achieve a $5\sigma$ rest equivalent
width limit in an $R=2000$ spectrum is $25.8$ --- independent of
redshift.  If the average $i'=17.5$ magnitude SDSS QSO reaches this
signal-to-noise (which it should), then there will be approximately
1050 QSOs in the SDSS sample between redshift 2 and 3 with spectra
suitable for measuring \ion{C}{4} absorption to a rest equivalent
width limit of $0.15\,{\rm \AA}$.  Furthermore, the 7700 ($2 < z < 3$)
SDSS QSOs brighter than $i'=18.5$ and will be more than suitable for
studying \ion{C}{4} absorbers stronger than $W_{rest} > 0.30\,{\rm
\AA}$.

\citet{ssb88} found that at $z\sim2$ there are about $2.57$ \ion{C}{4}
absorbers per unit redshift with $W_{\rm o} > 0.15\,{\rm \AA}$; they
found that lines stronger than $W_{\rm o} > 0.30\,{\rm \AA}$ are found
at the rate of approximately 1.48 per unit redshift.  Given this
frequency of \ion{C}{4}, we can ask how small of an intrinsic
absorption population the SDSS sample could measure.  That is, if
there really are narrow, intrinsic absorption lines in QSOs at large
velocities, what fraction of \ion{C}{4} systems would need to be
intrinsic for them to be detected statistically?  If we assume that we
can separate the parent population with 100\% efficiency and that the
extra absorbers are all in one type of QSO, then we can determine the
minimum observable fraction.  Here we assume that the observed value
of $dN/d\beta$ is the average of all QSOs over a velocity range of
$5000 < \Delta v/c < 70000\,{\rm km\,s^{-1}}$.  Given these
assumptions, 16\% of the \ion{C}{4} absorption line systems must be
intrinsic in order to have a statistically significant detection in
the SDSS data.  For absorption stronger than $0.30\,{\rm \AA}$, an
intrinsic population as small as 7\% could be detected, largely as a
result of the greatly increased number of QSOs.

\section{Conclusions}

New spectra of 24 $z\sim2.5$ radio-detected quasars are presented.
These spectra were searched for the most commonly observed QSO
absorption lines.  The number of \ion{C}{4} absorbers per unit
redshift was found to be $2.54\pm0.44$ (at $z\sim2$, with both members
of the doublet stronger than $0.15\,{\rm \AA}$) --- consistent with
the finding of \citet{ssb88}.  In terms of velocity space, the number
of absorbers per unit velocity ($dN/d\beta$) is $7.84\pm1.34$.  For
radio-quiet, radio-loud, and flat-spectrum quasars, $dN/d\beta$ was
found to be $6.96\pm2.01$, $6.09\pm2.72$, and $8.03\pm2.15$,
respectively.

The absorption line systems found in these quasars were combined with
systems of similar quality from \citet{v+00} to form a statistical
sample.  This sample is restricted to redshifts between $2.2$ and
$2.8$.  The construction of this sample is such that it has more
homogeneous absorption line properties and a smaller range of
redshifts than the sample studied in Paper I.  As such, this sample
can be used to address potential biases in Paper I.  Analysis of this
sample and comparison to the less homogeneous parent sample yield the
following conclusions:

1) The excess of \ion{C}{4} absorbers in flat-spectrum quasars as
compared to steep-spectrum quasars is confirmed and is found to be
consistent with the 36\% excess as reported in Paper I.  This excess
does not seem to be the result of a bias in the quality of the QSO
spectra (and thus the detection limits of \ion{C}{4}) as a function of
radio spectral index.  Nor does the effect result from a redshift
evolution of the \ion{C}{4} distribution that is reflected by the
differences in the redshift distributions of the steep- and
flat-spectrum populations, since the effect remains when considering
only a small range in redshift ($z\sim2.5$).  No such excess is
observed in the \ion{Mg}{2} sample.

2) There is also a small excess of high velocity \ion{C}{4} absorbers
in radio-quiet QSOs as compared to radio-loud quasars.  This excess
appears to be a function of velocity and remains (at $2\sigma$
significance) in the more homogeneous sample.  Though the significance
is small, the effect deserves further consideration.  No such effect
is observed for \ion{Mg}{2} absorbers.

3) The QSO spectra from the Sloan Digital Sky Survey will allow for
the detection of a high velocity, narrow, intrinsic population of
absorbers that is as small as 7\% of the \ion{C}{4} absorber
population in QSOs between redshift 2 and 3.

\acknowledgements

I would like to acknowledge Don York, my advisor, for his support and
guidance in this project.  I thank Sally Laurent-Muehleisen and Bob
Becker for their assistance on this and related projects.  Dan Vanden
Berk, Jean Quashnock, Arieh K\"{o}nigl and John Kartje were additional
sources of guidance and inspiration.  Comments from an anonymous
referee were very helpful to the flow of the paper.

\clearpage
\begin{figure}[phtb]
\plotone{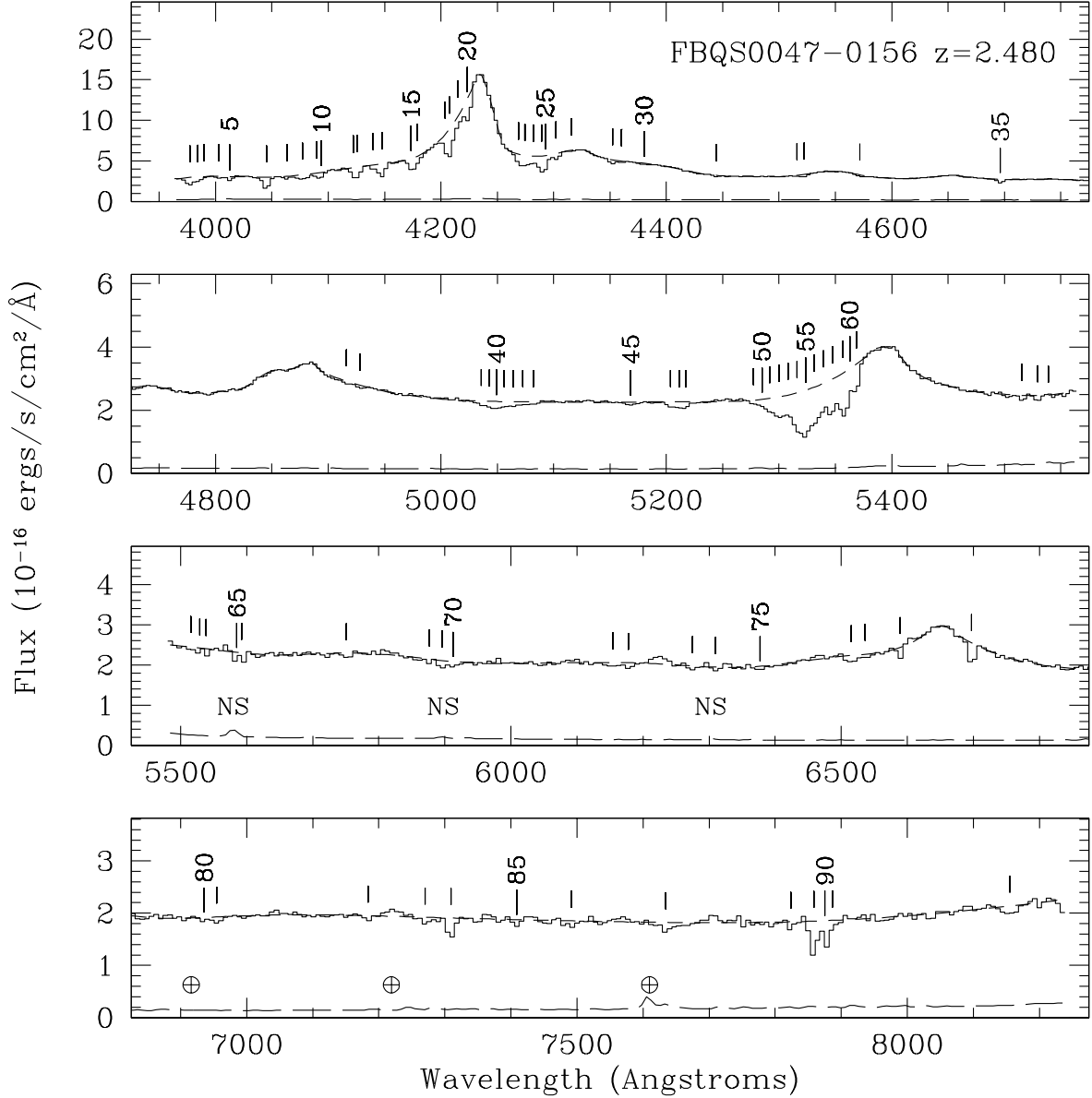}
\caption{FBQS0047-0156.  Coadded blue (upper panels) and red (lower
panels) spectra of FBQS0047-0156.  Solid line is the coadded spectrum,
short dashed line is the continuum fit, and long dashed line is a
$5\sigma$ error array.  Absorption features with $3\sigma$
significance or better are indicated by tick marks.  Regions effected
by strong night sky lines are labelled with a ``NS'', whereas regions
effected by telluric absorption are marked by a circled
cross. \label{fig:fig1}}
\end{figure}

\begin{figure}[phtb]
\plotone{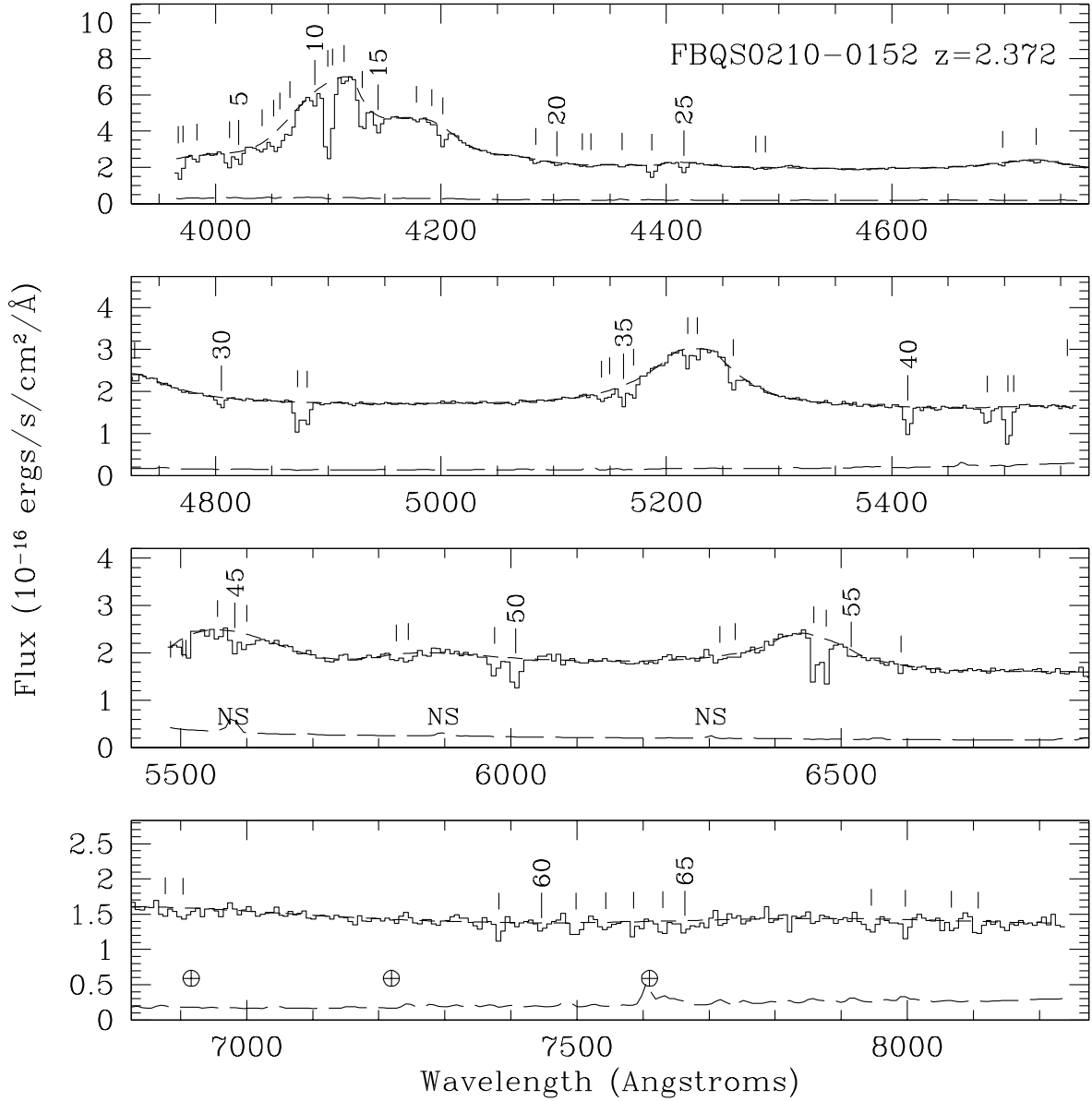}
\caption{FBQS0210-0152\label{fig:fig2}}
\end{figure}

\begin{figure}[phtb]
\plotone{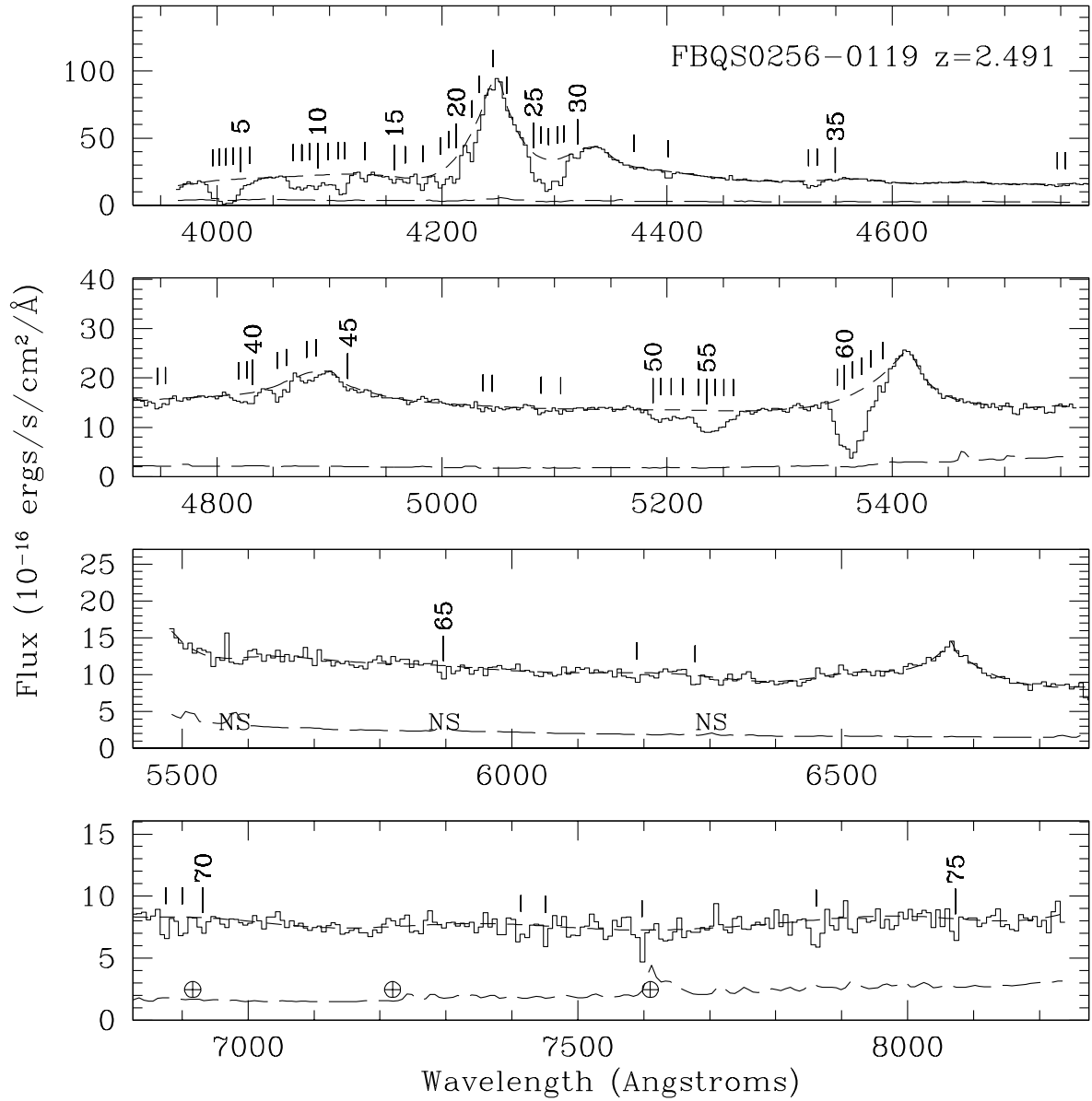}
\caption{FBQS0256-0119\label{fig:fig3}}
\end{figure}

\begin{figure}[phtb]
\plotone{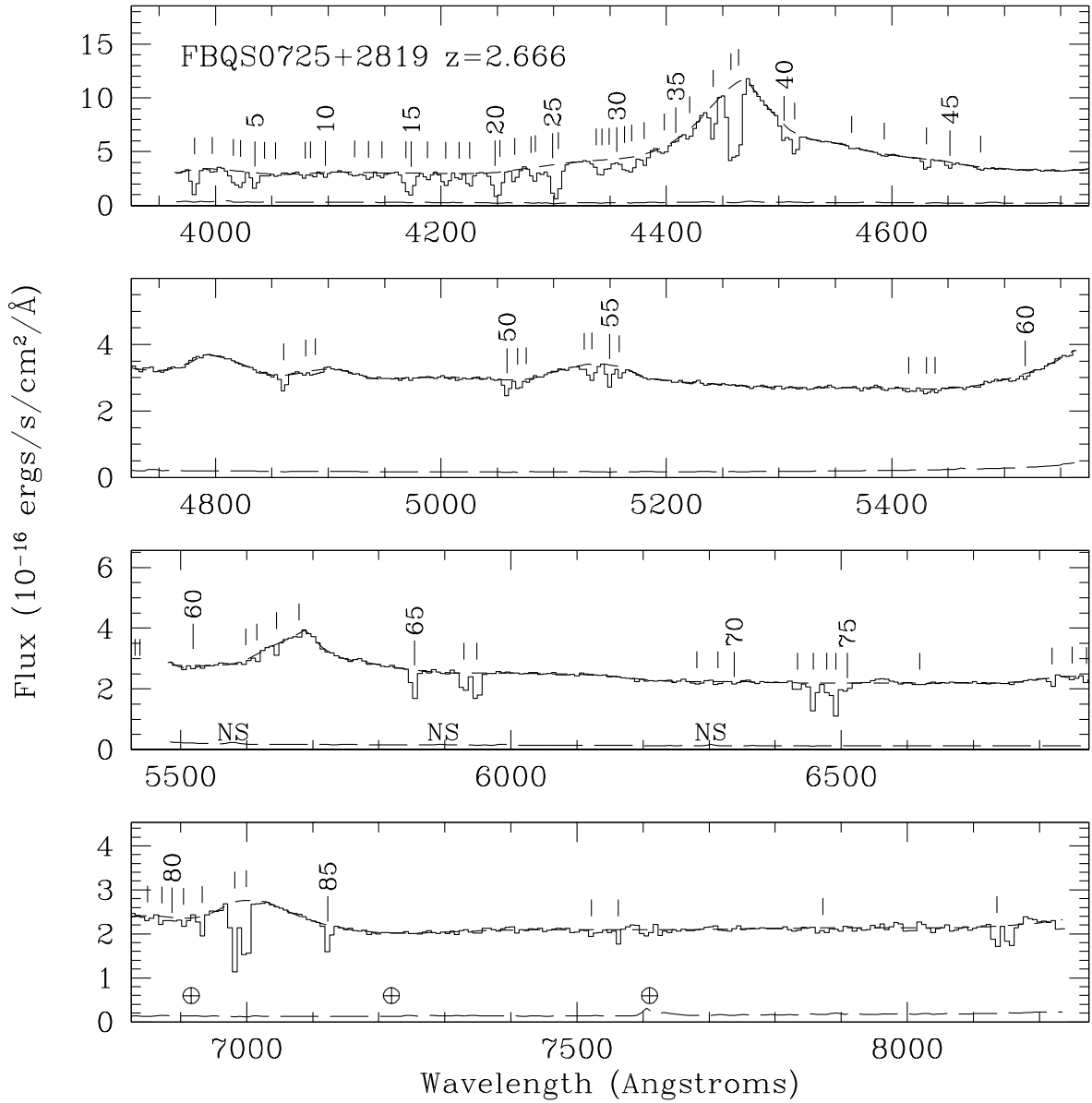}
\caption{FBQS0725+2819\label{fig:fig4}}
\end{figure}

\begin{figure}[phtb]
\plotone{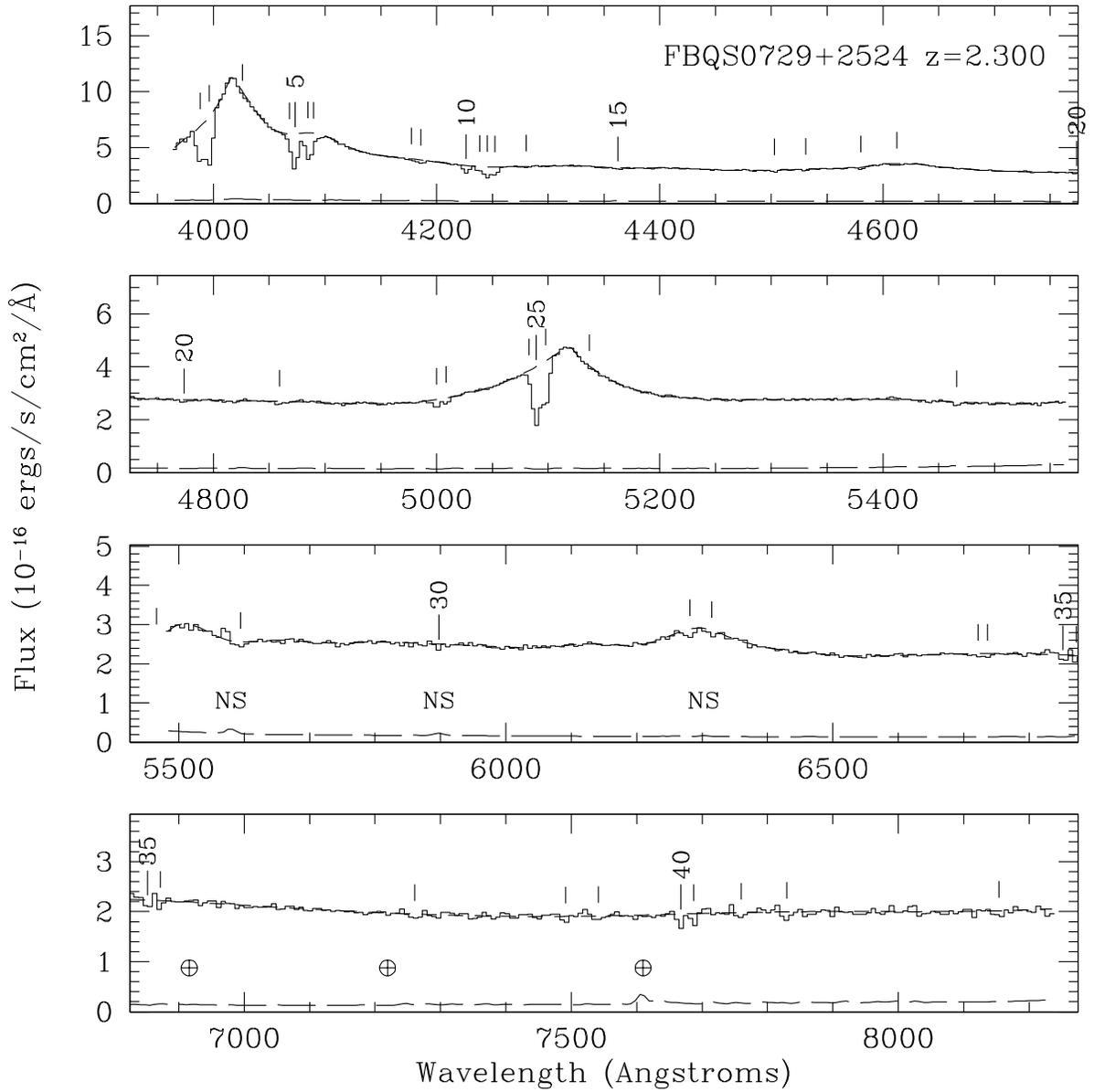}
\caption{FBQS0729+2524\label{fig:fig5}}
\end{figure}

\begin{figure}[phtb]
\plotone{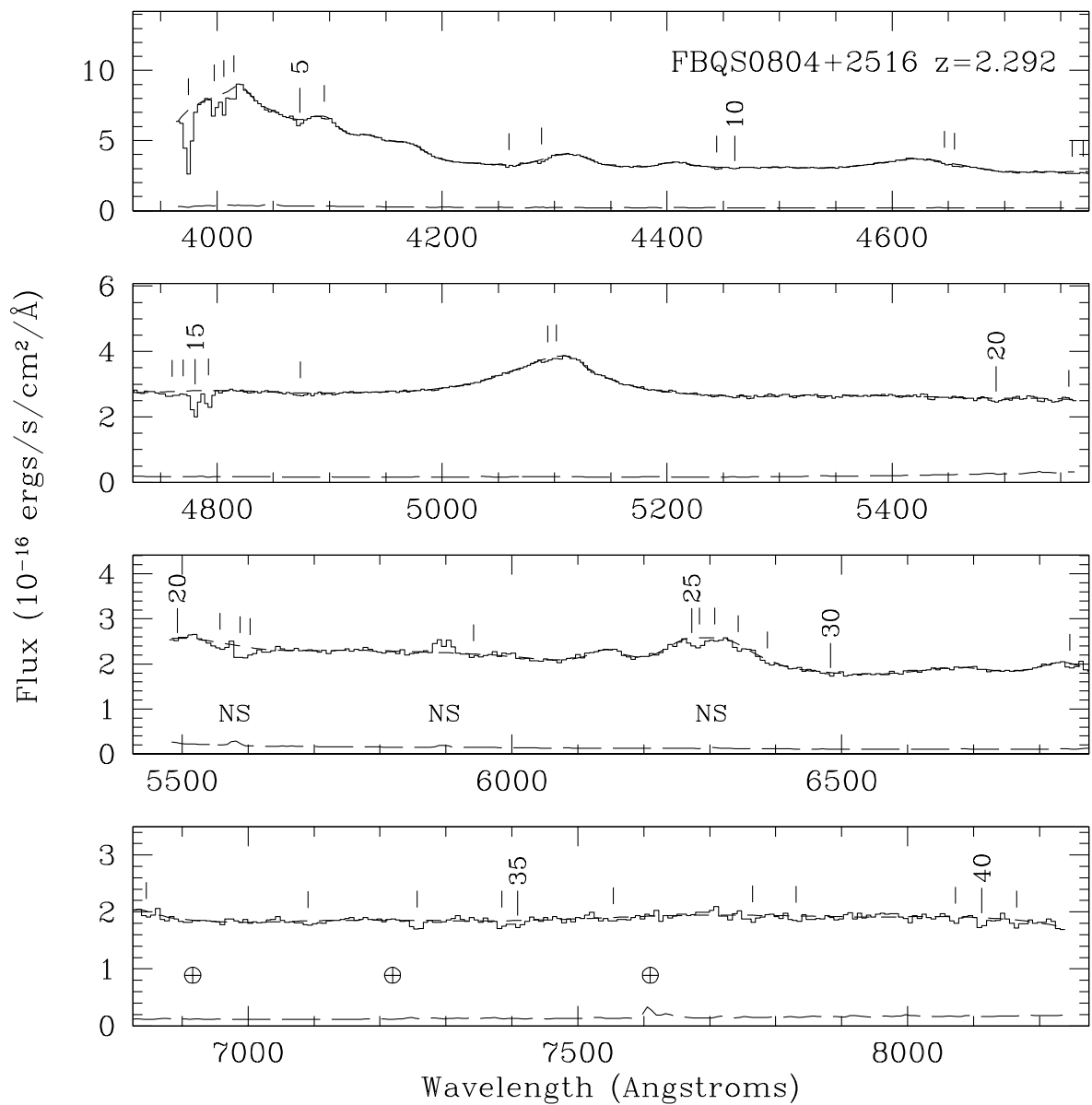}
\caption{FBQS0804+2516\label{fig:fig6}}
\end{figure}

\begin{figure}[phtb]
\plotone{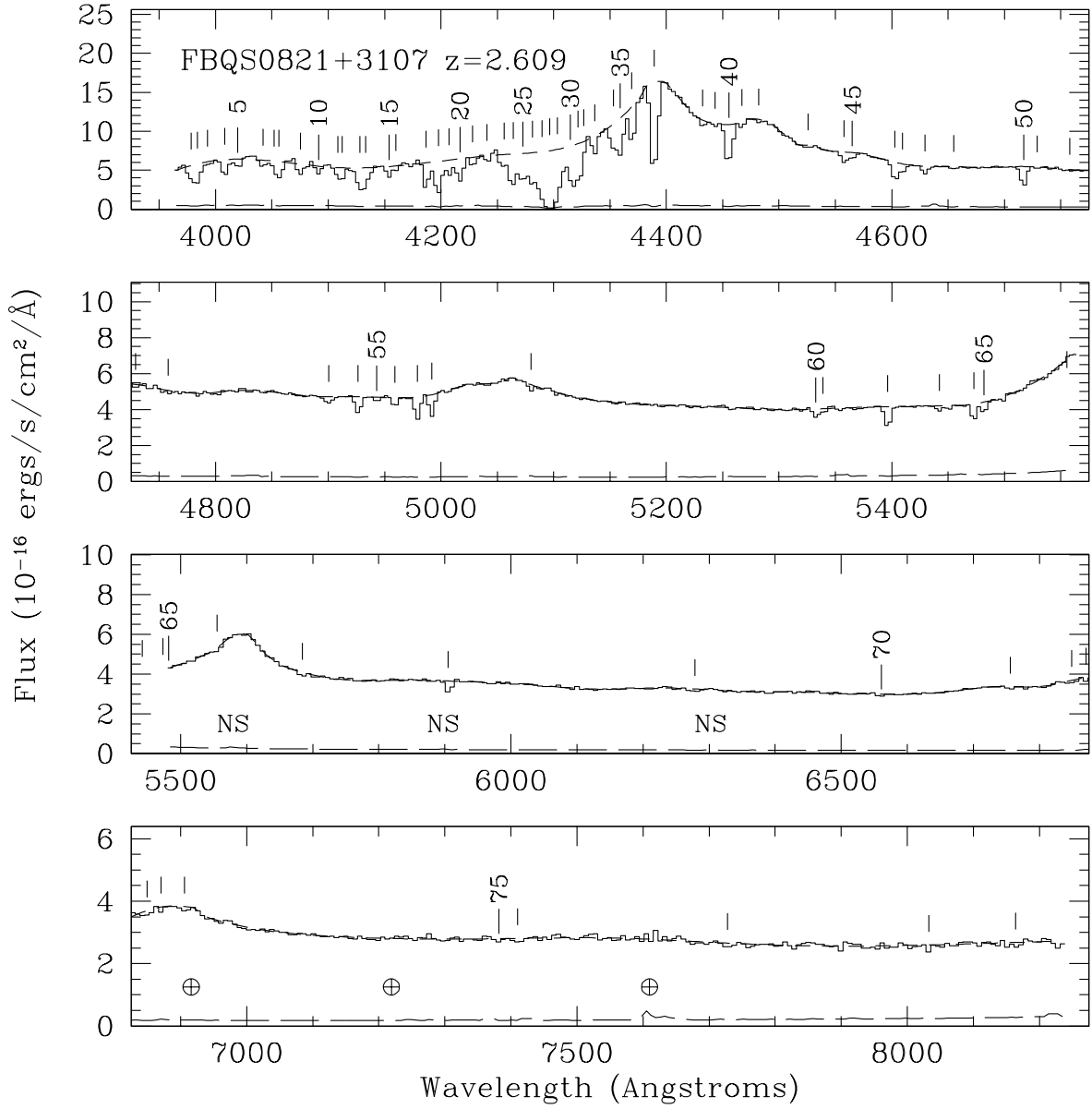}
\caption{FBQS0821+3107\label{fig:fig7}}
\end{figure}

\begin{figure}[phtb]
\plotone{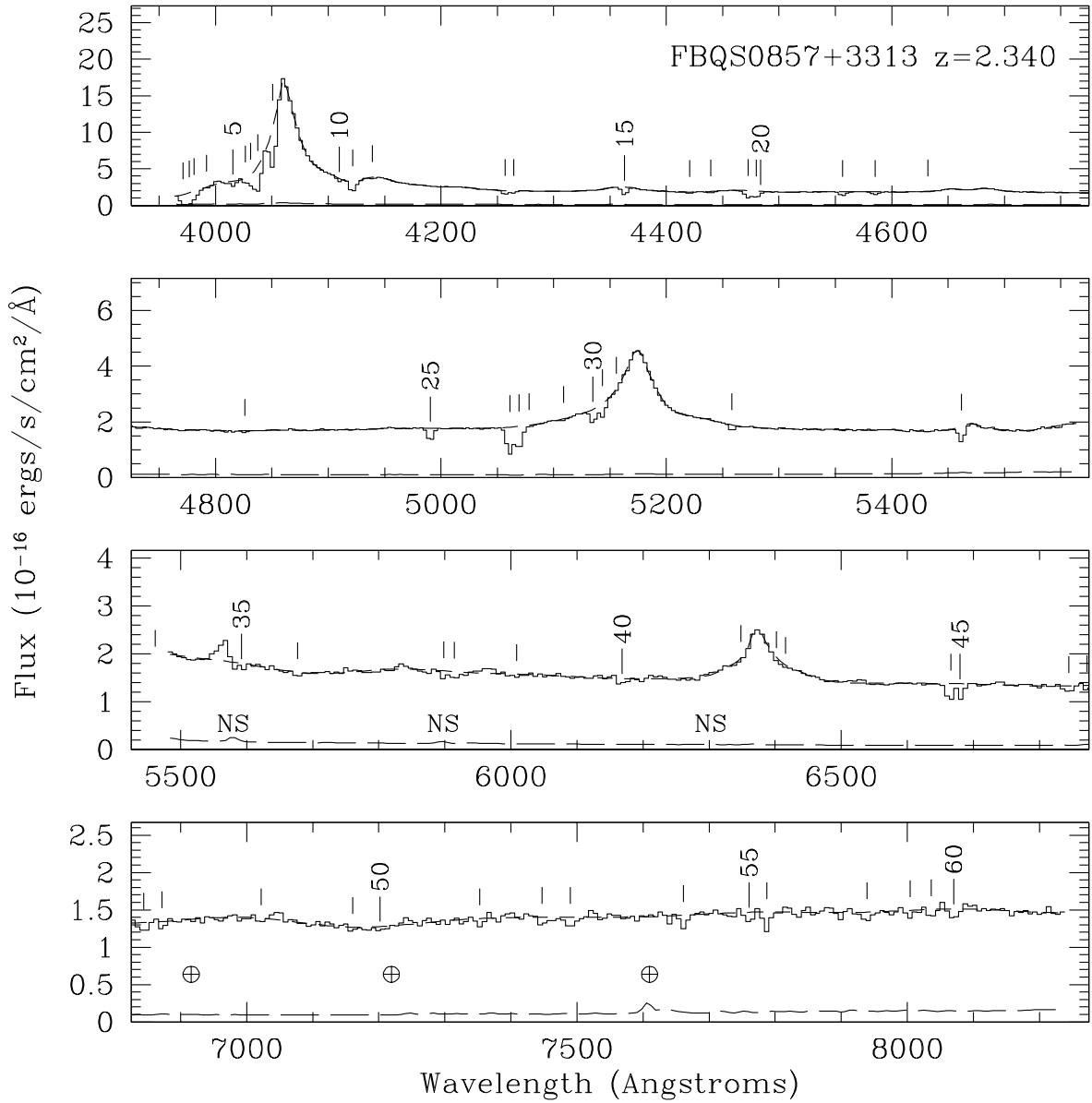}
\caption{FBQS0857+3313\label{fig:fig8}}
\end{figure}

\clearpage
\begin{figure}[phtb]
\plotone{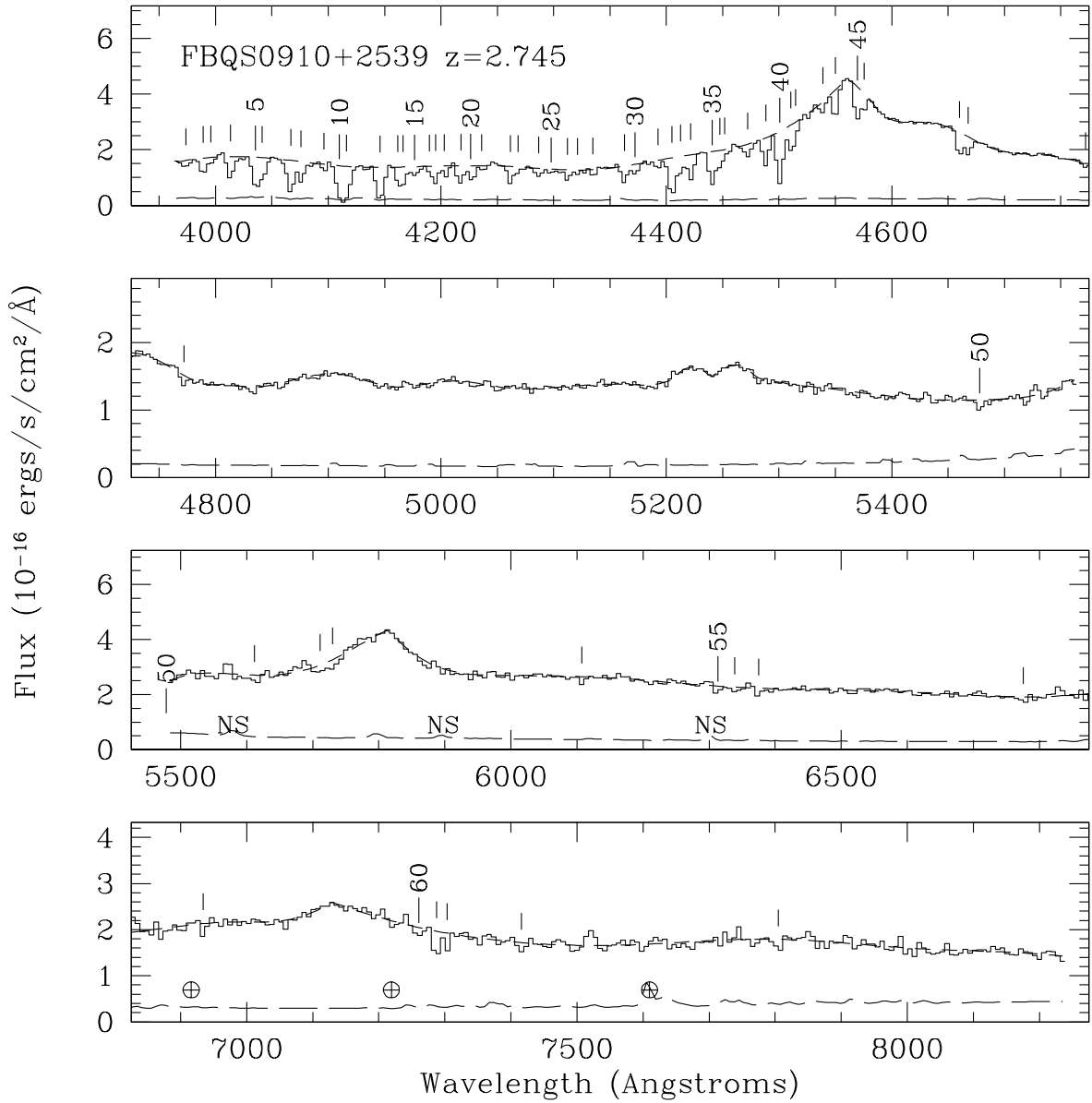}
\caption{FBQS0910+2539\label{fig:fig9}}
\end{figure}

\begin{figure}[phtb]
\plotone{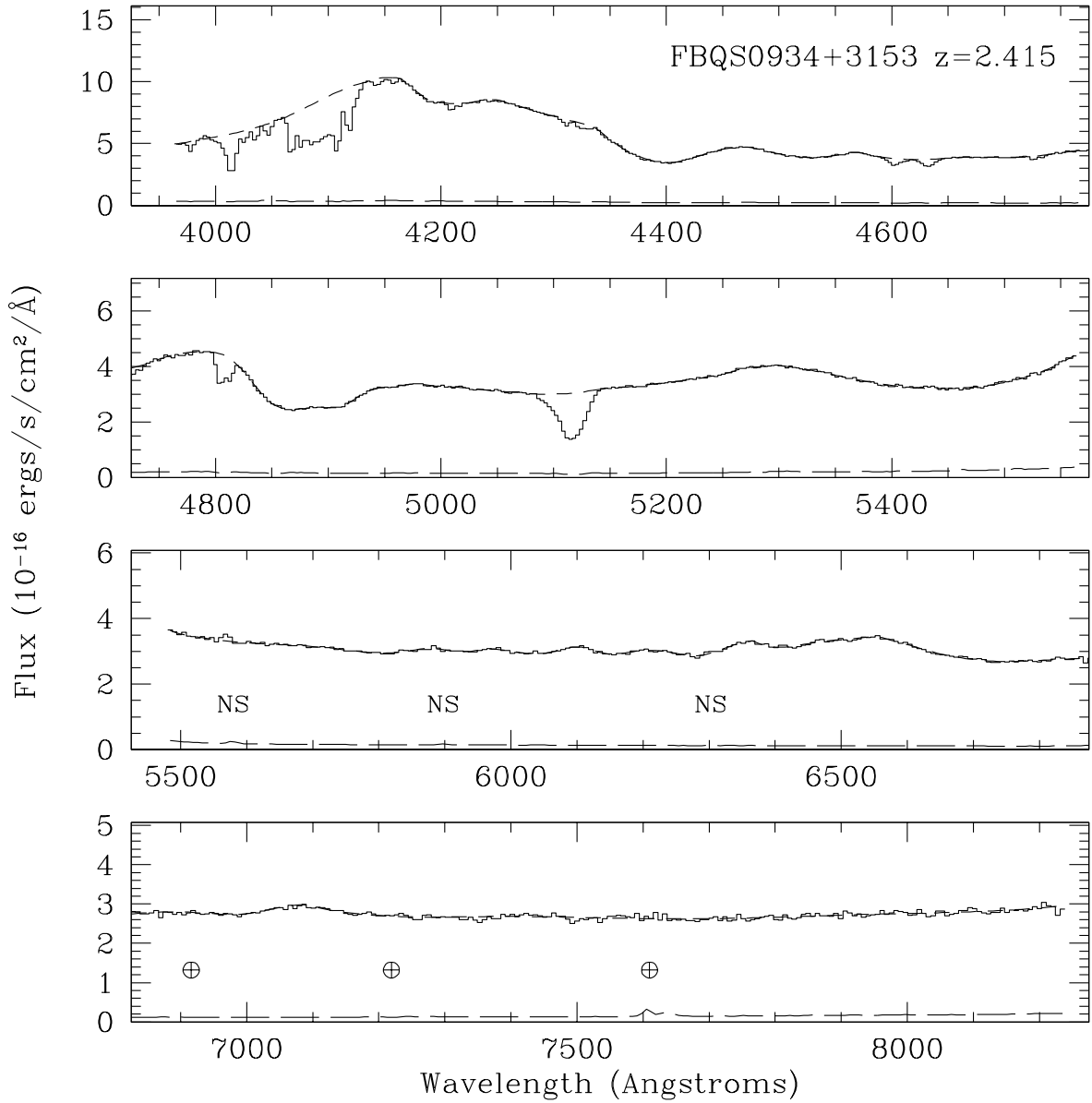}
\caption{FBQS0934+3153\label{fig:fig10}}
\end{figure}

\begin{figure}[phtb]
\plotone{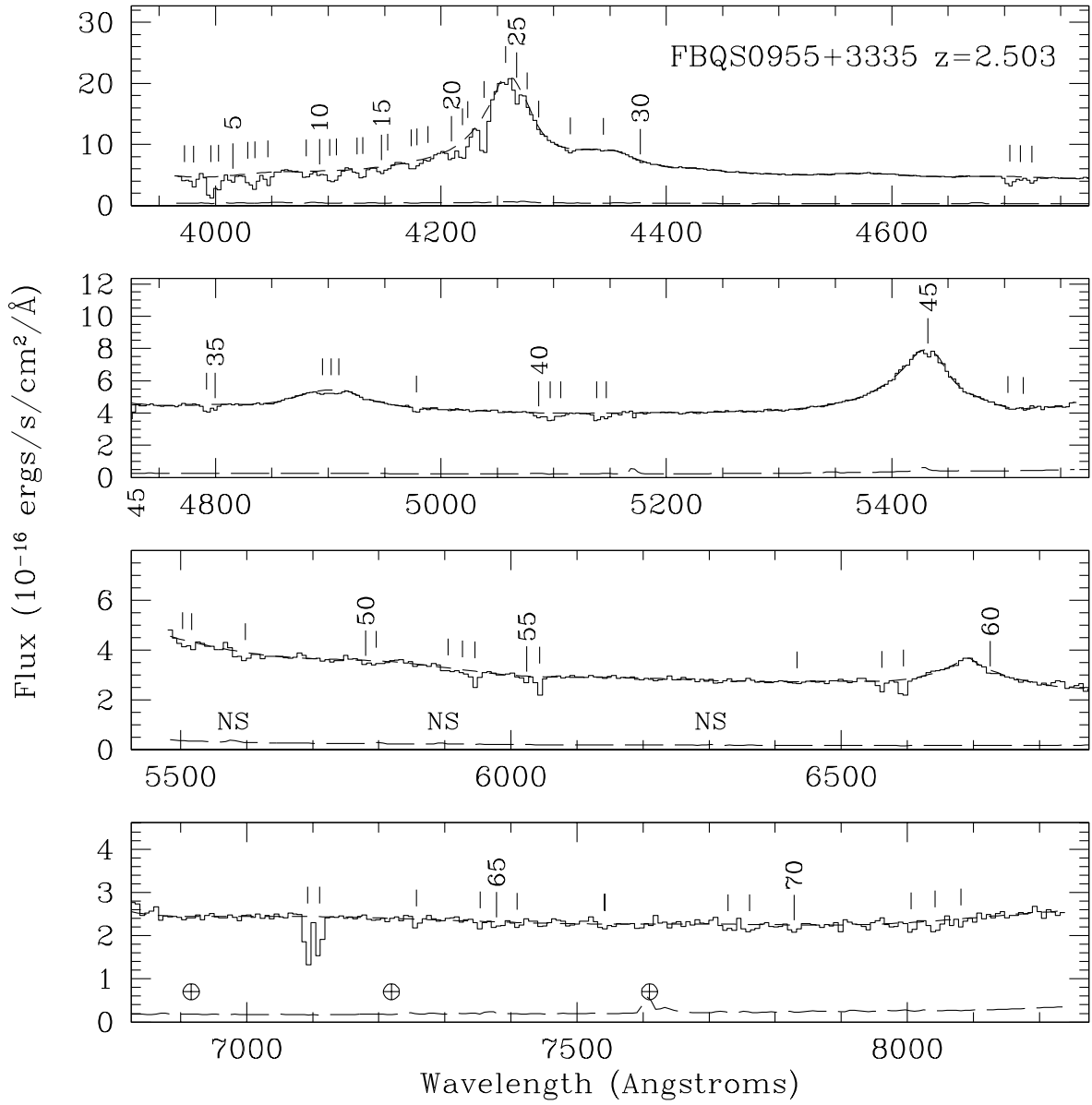}
\caption{FBQS0955+3335\label{fig:fig11}}
\end{figure}

\begin{figure}[phtb]
\plotone{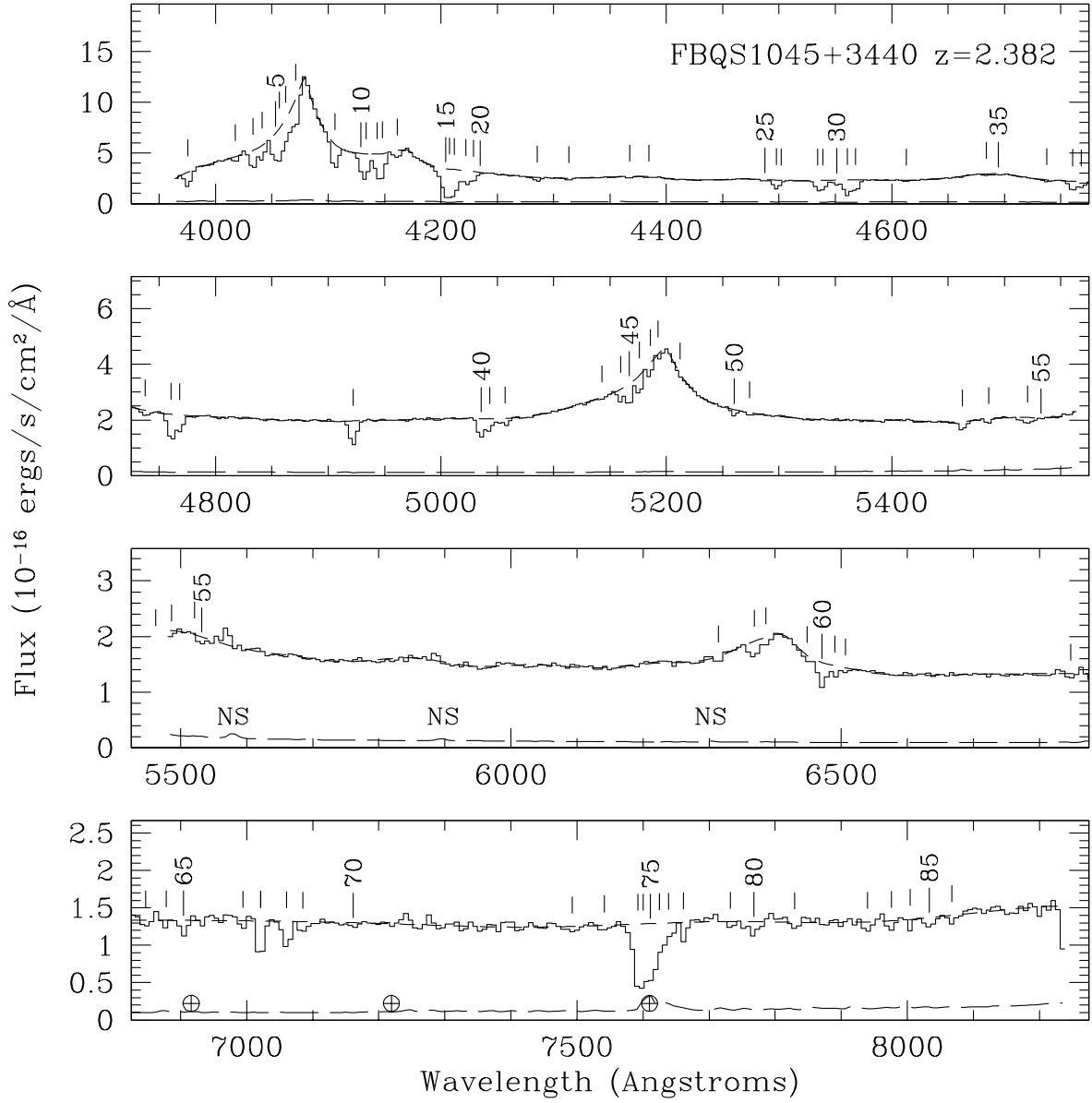}
\caption{FBQS1045+3440\label{fig:fig12}}
\end{figure}

\begin{figure}[phtb]
\plotone{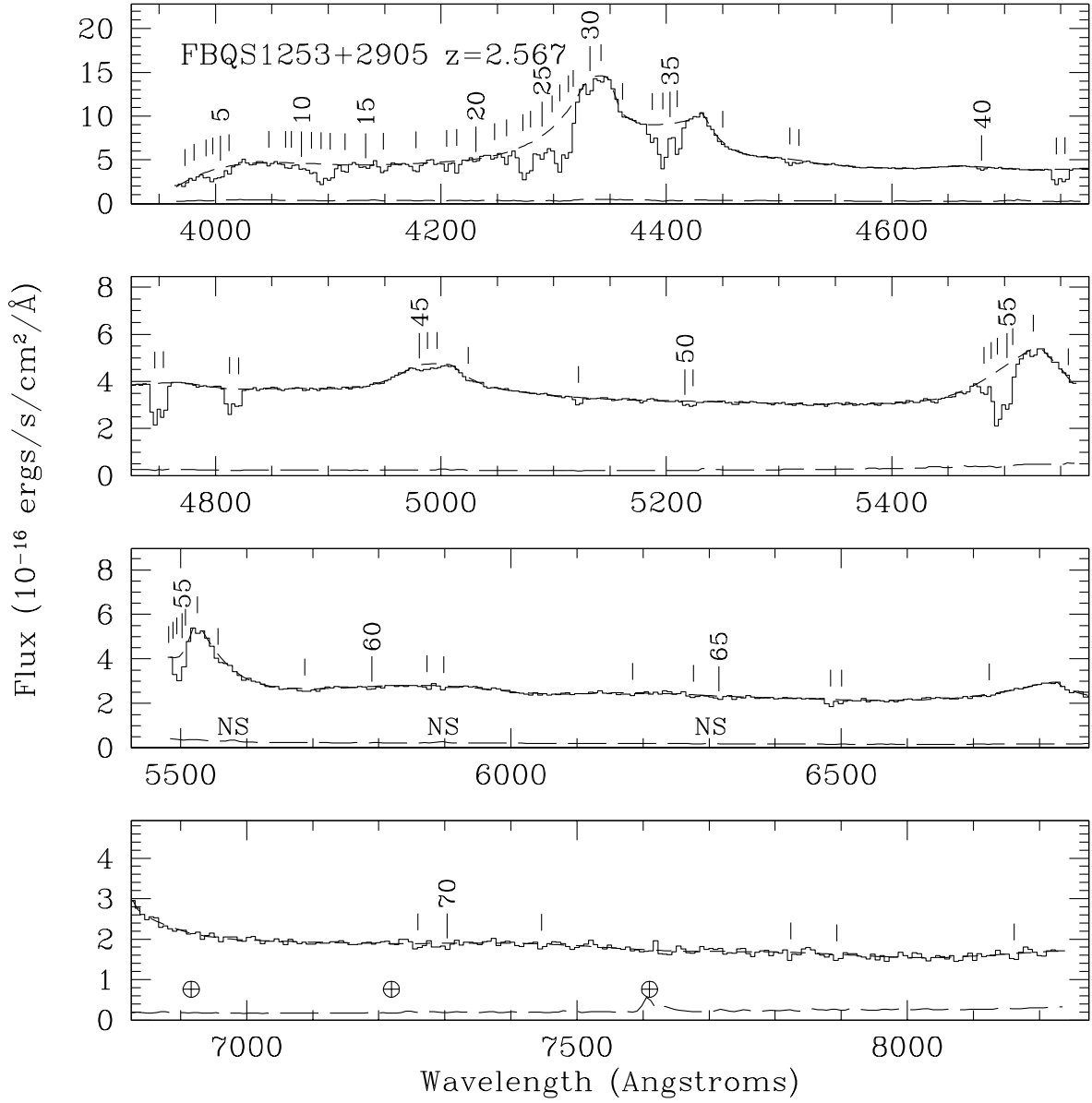}
\caption{FBQS1253+2905\label{fig:fig13}}
\end{figure}

\begin{figure}[phtb]
\plotone{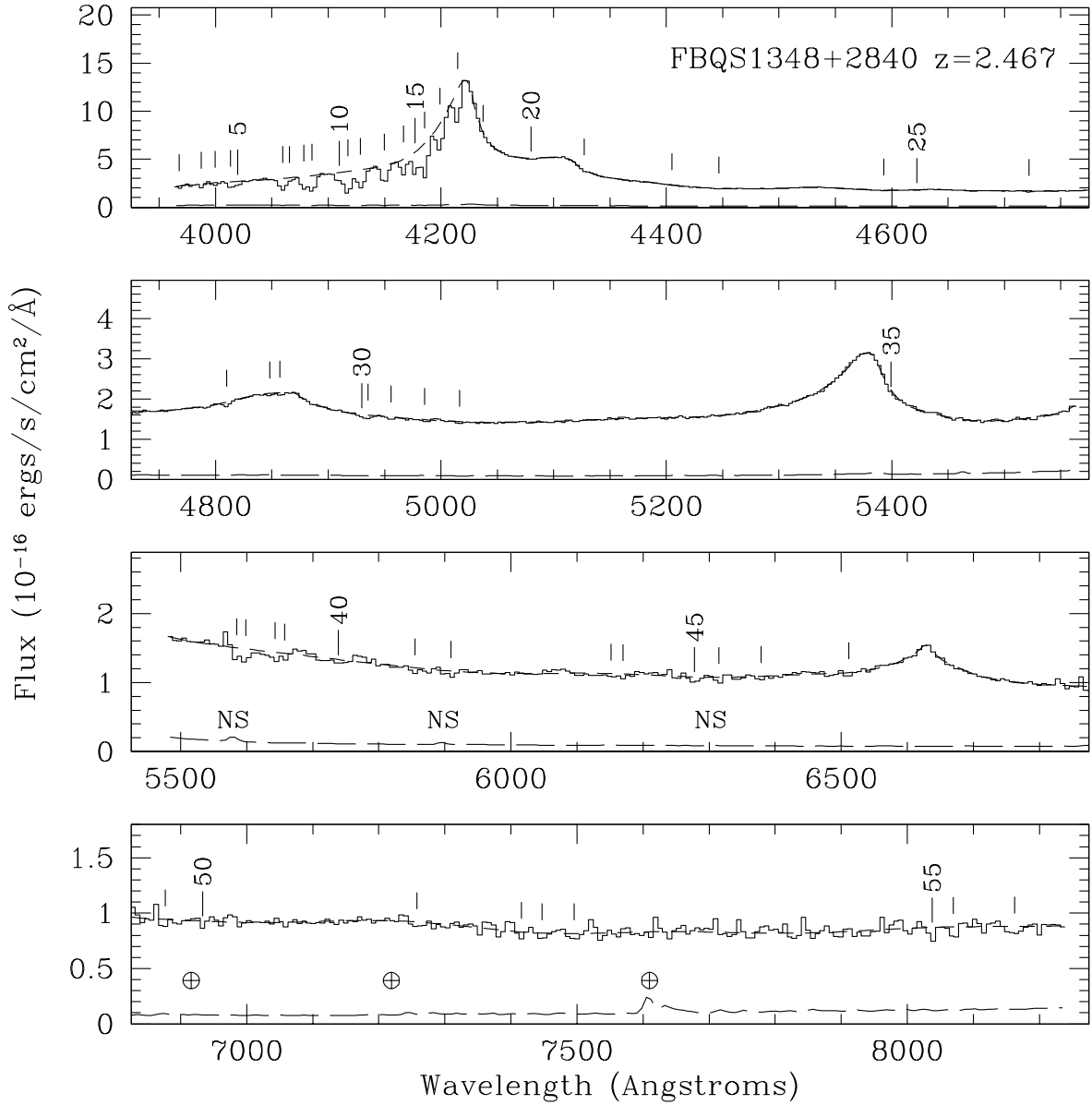}
\caption{FBQS1348+2840\label{fig:fig14}}
\end{figure}

\begin{figure}[phtb]
\plotone{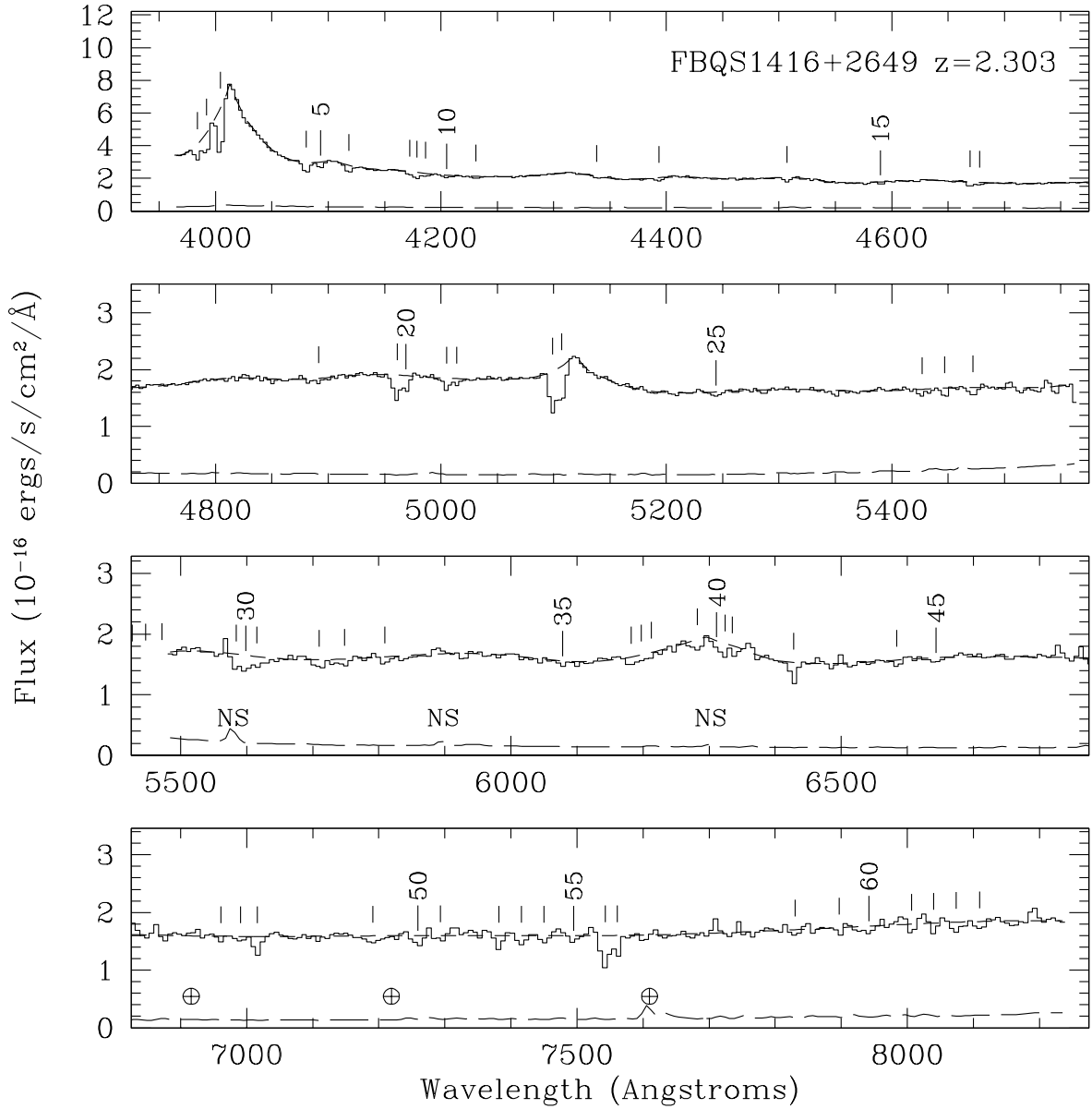}
\caption{FBQS1416+2649\label{fig:fig15}}
\end{figure}

\begin{figure}[phtb]
\plotone{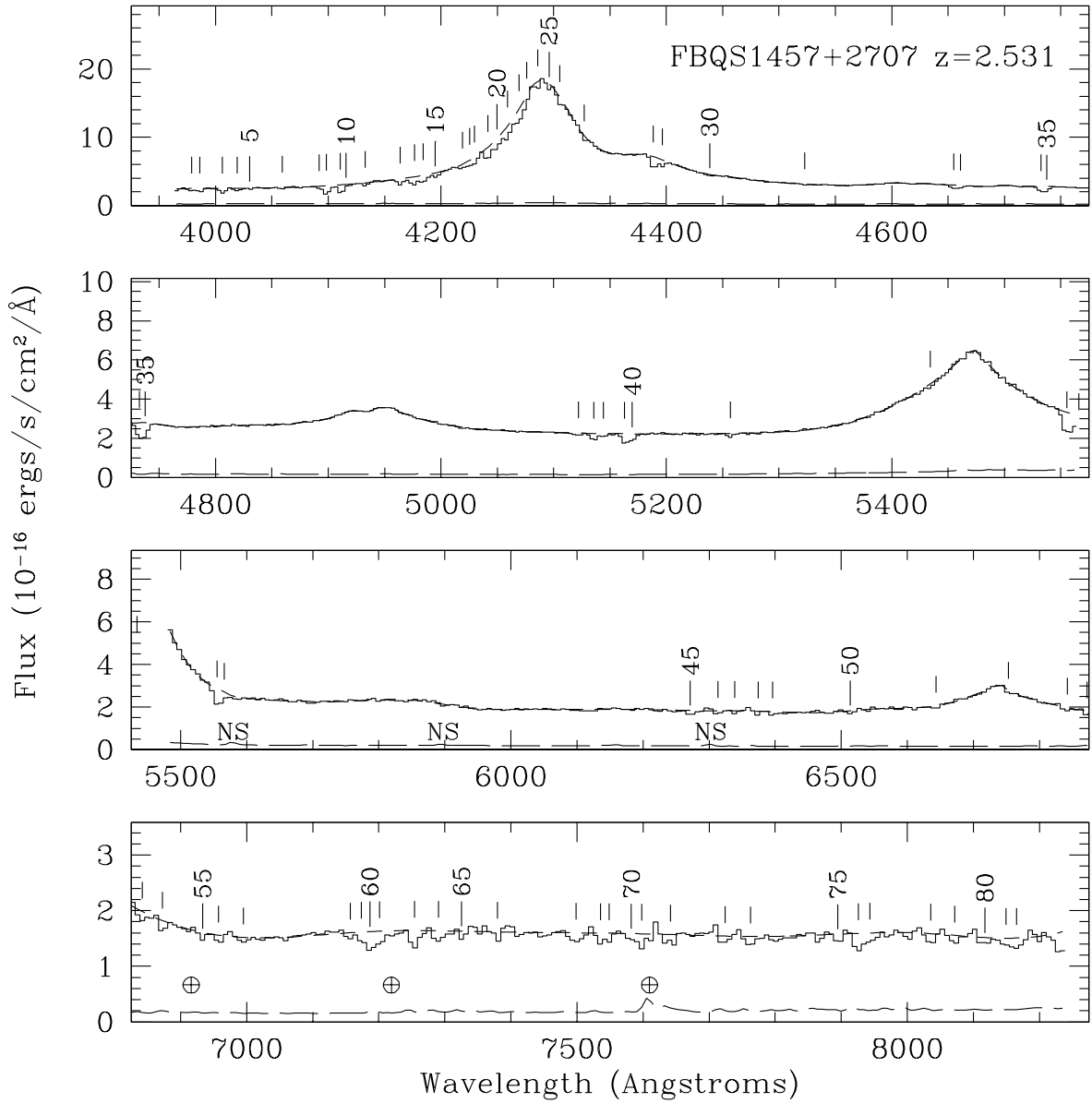}
\caption{FBQS1457+2707\label{fig:fig16}}
\end{figure}
\clearpage
\begin{figure}[phtb]
\plotone{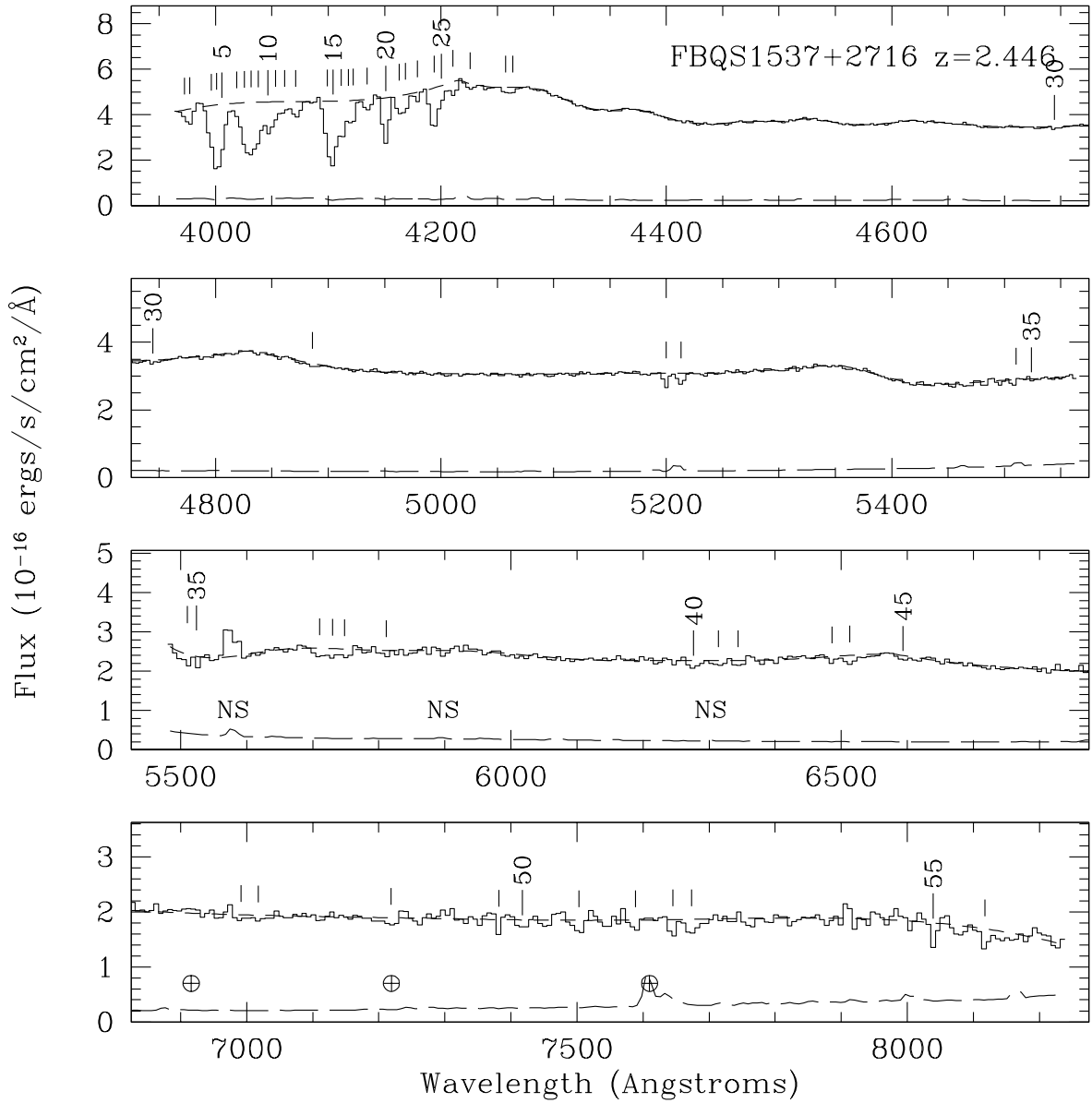}
\caption{FBQS1537+2716\label{fig:fig17}}
\end{figure}

\begin{figure}[phtb]
\plotone{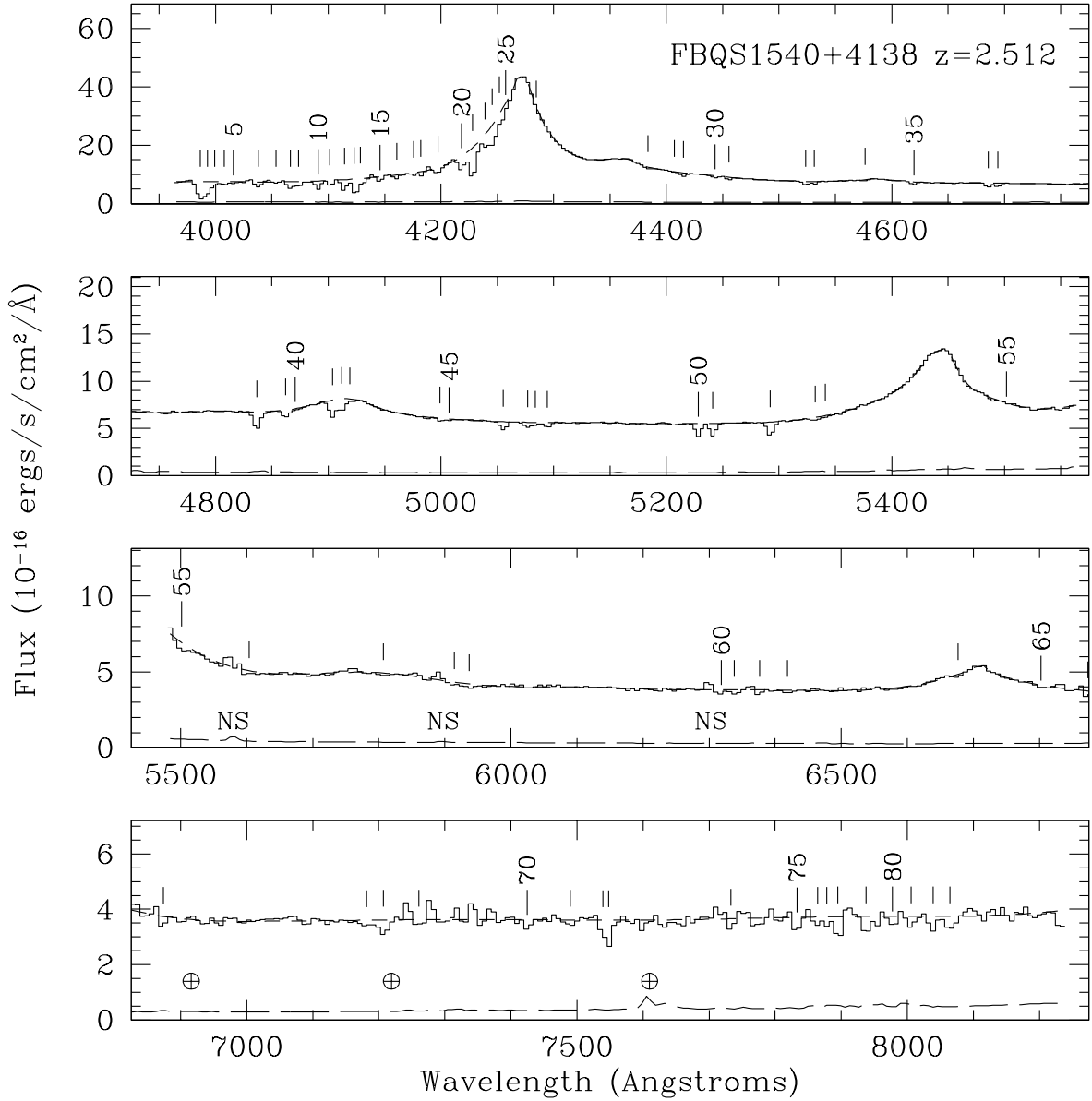}
\caption{FBQS1540+4138\label{fig:fig18}}
\end{figure}

\begin{figure}[phtb]
\plotone{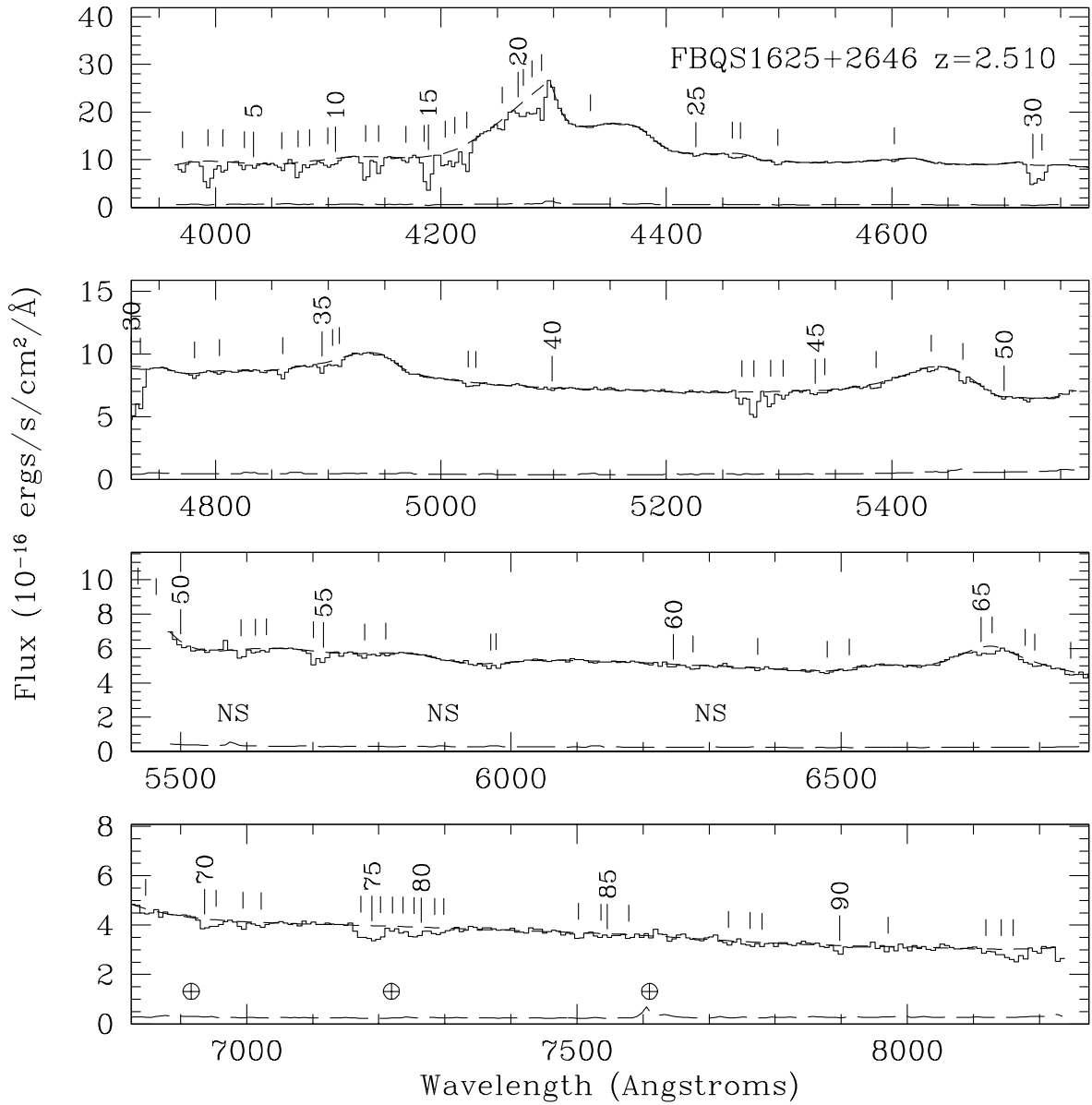}
\caption{FBQS1625+2646\label{fig:fig19}}
\end{figure}

\begin{figure}[phtb]
\plotone{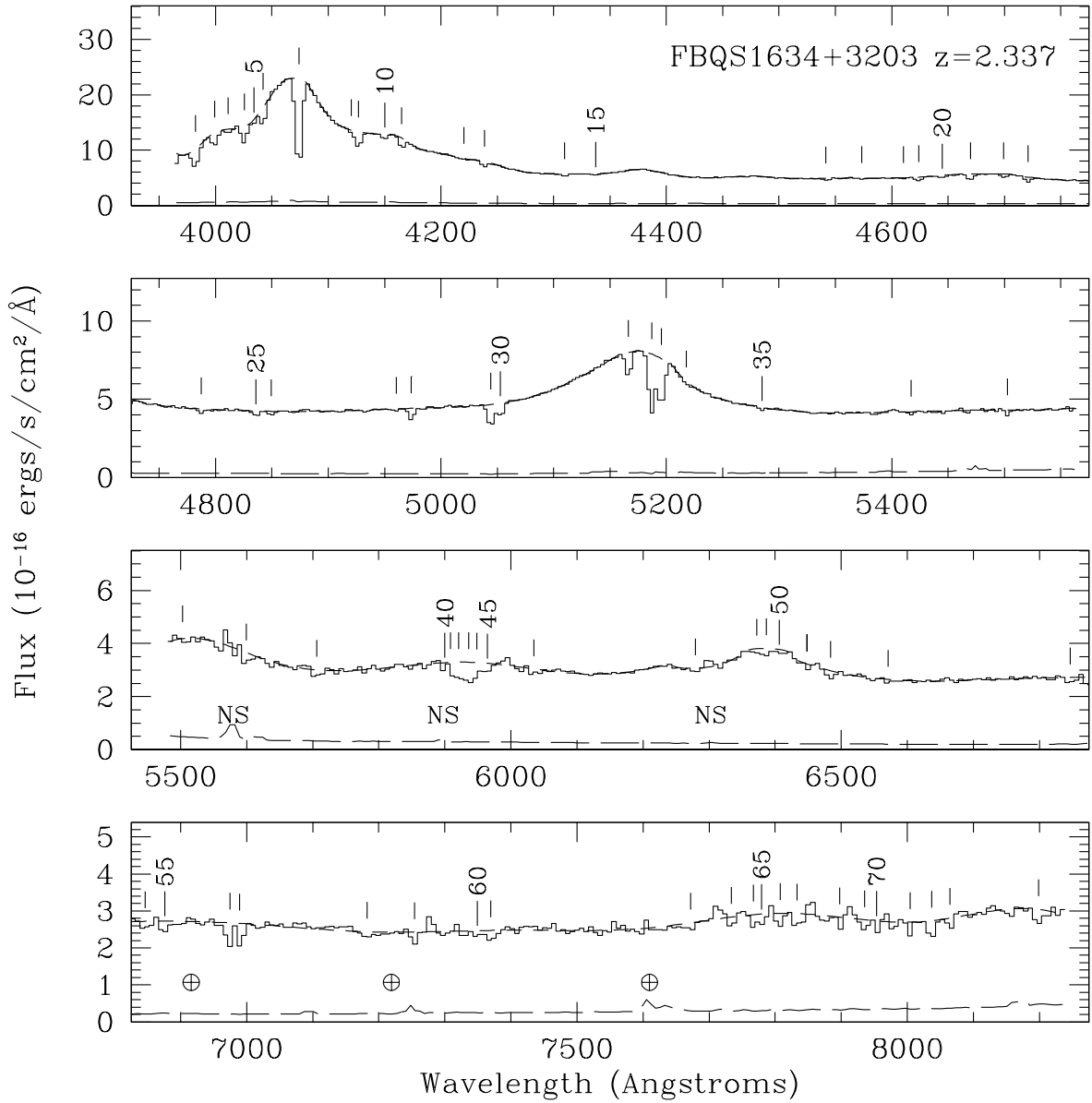}
\caption{FBQS1634+3203\label{fig:fig20}}
\end{figure}

\begin{figure}[phtb]
\plotone{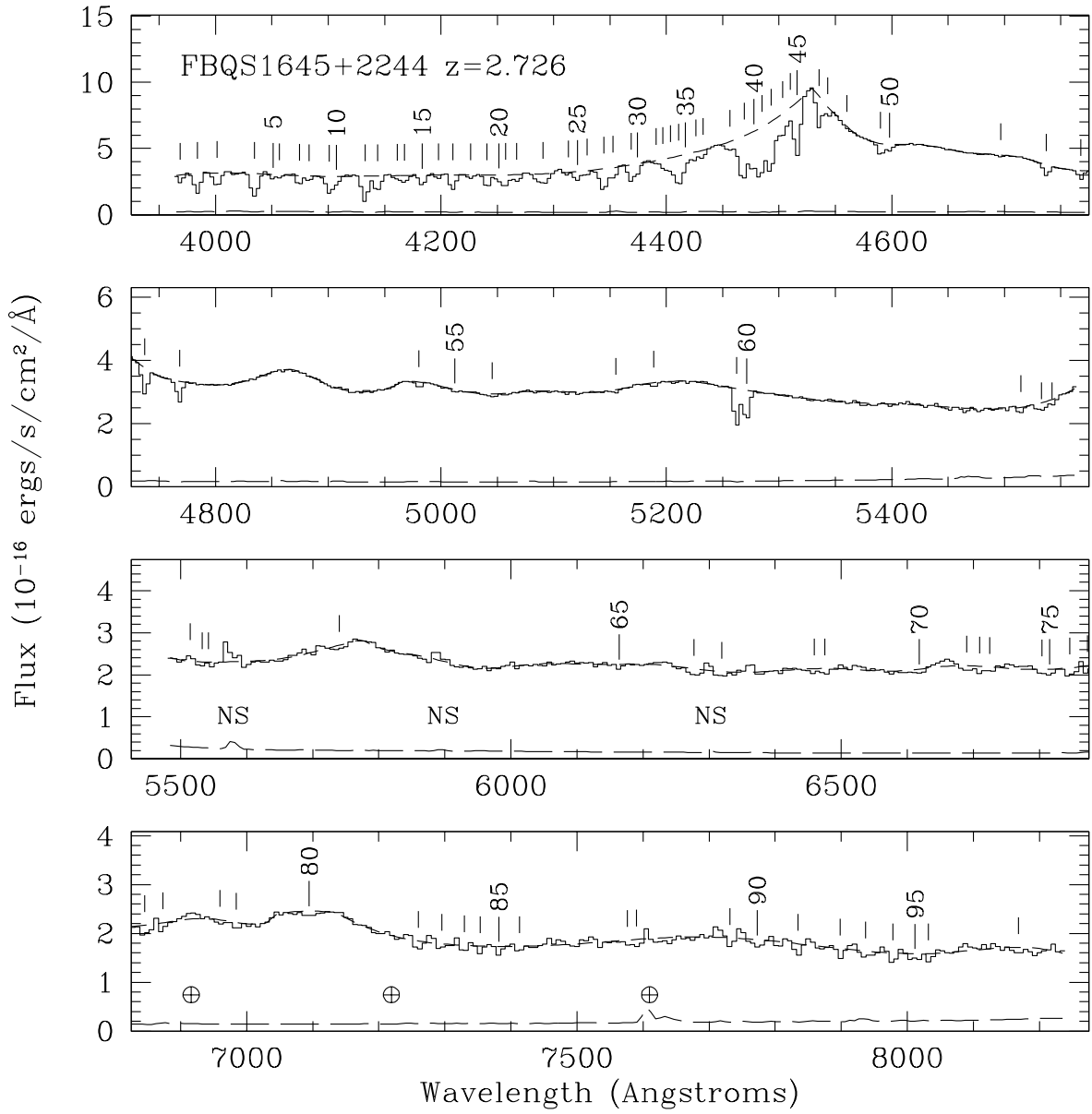}
\caption{FBQS1645+2244\label{fig:fig21}}
\end{figure}

\begin{figure}[phtb]
\plotone{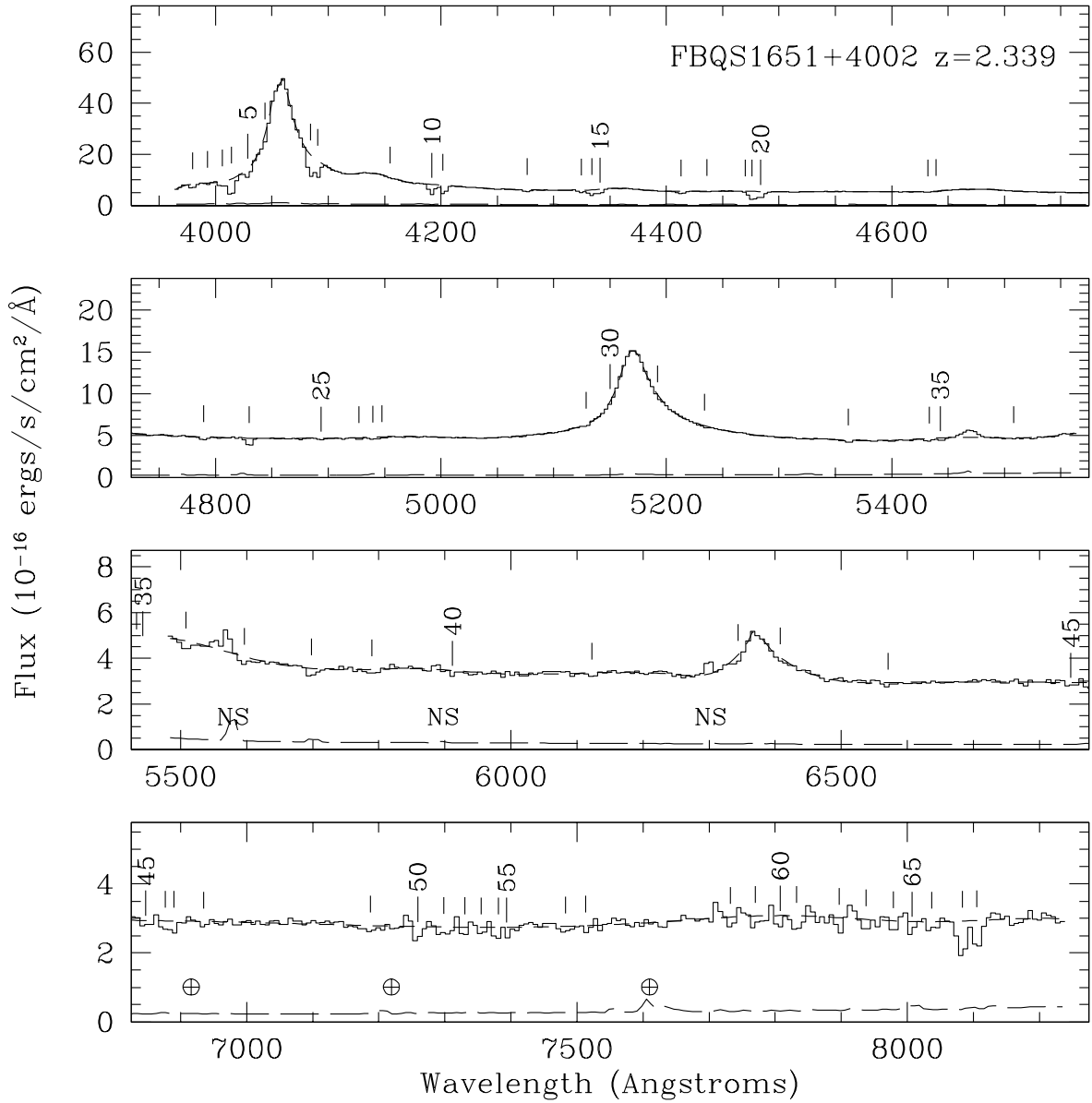}
\caption{FBQS1651+4002\label{fig:fig22}}
\end{figure}

\begin{figure}[phtb]
\plotone{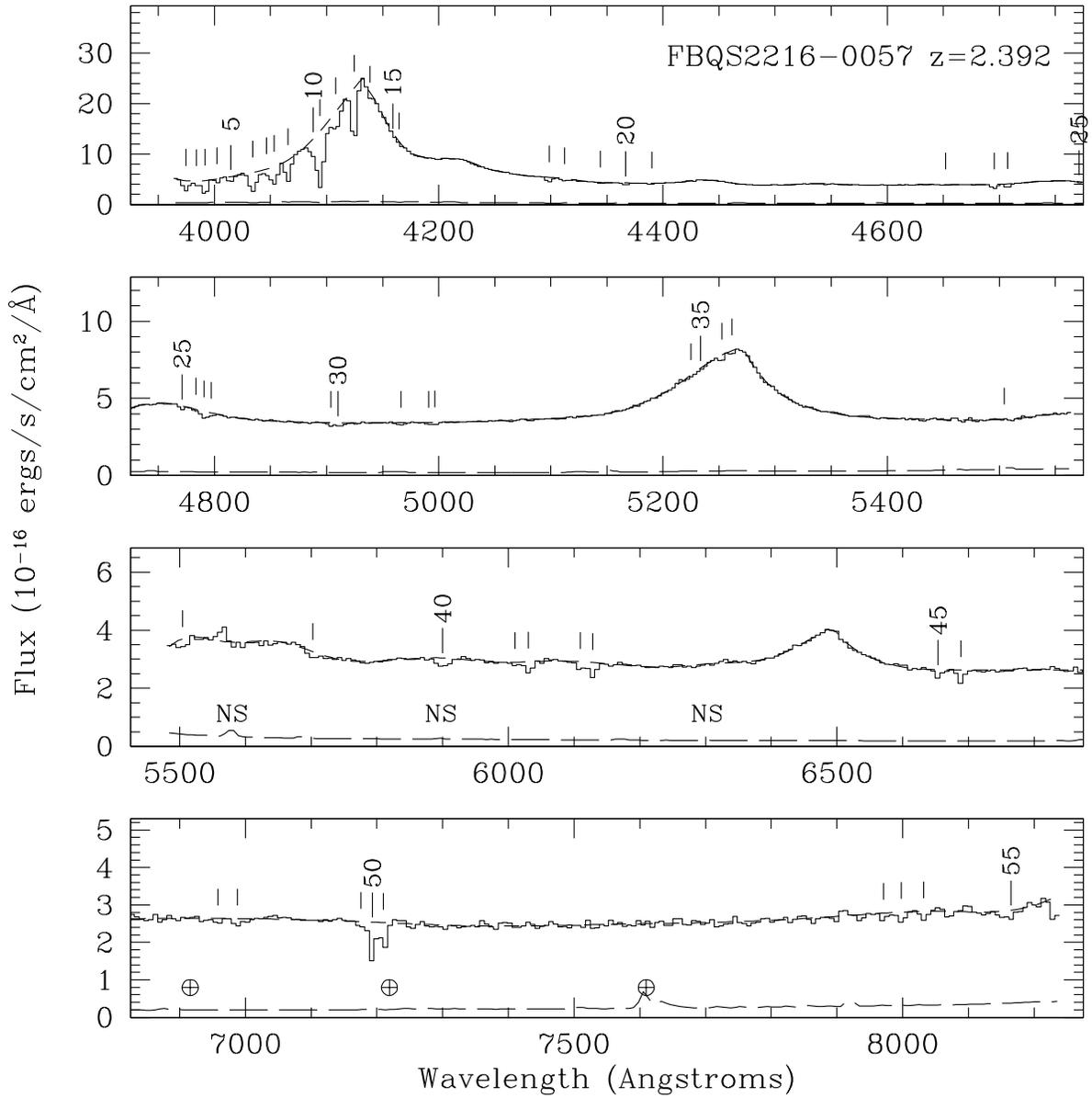}
\caption{FBQS2216-0057\label{fig:fig23}}
\end{figure}

\begin{figure}[phtb]
\plotone{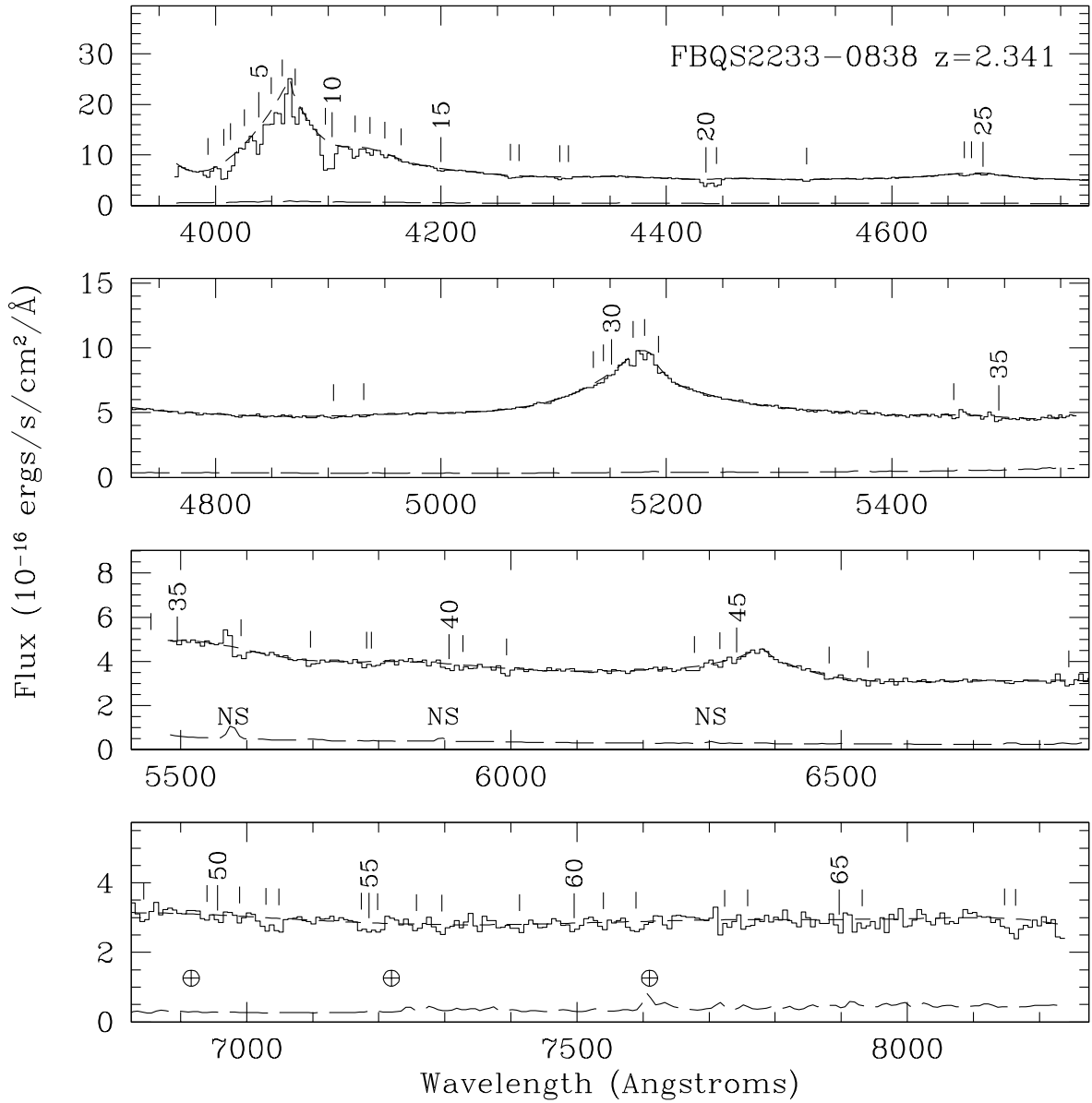}
\caption{FBQS2233-0838\label{fig:fig24}}
\end{figure}

\clearpage

\begin{figure}[htpb]
\plotone{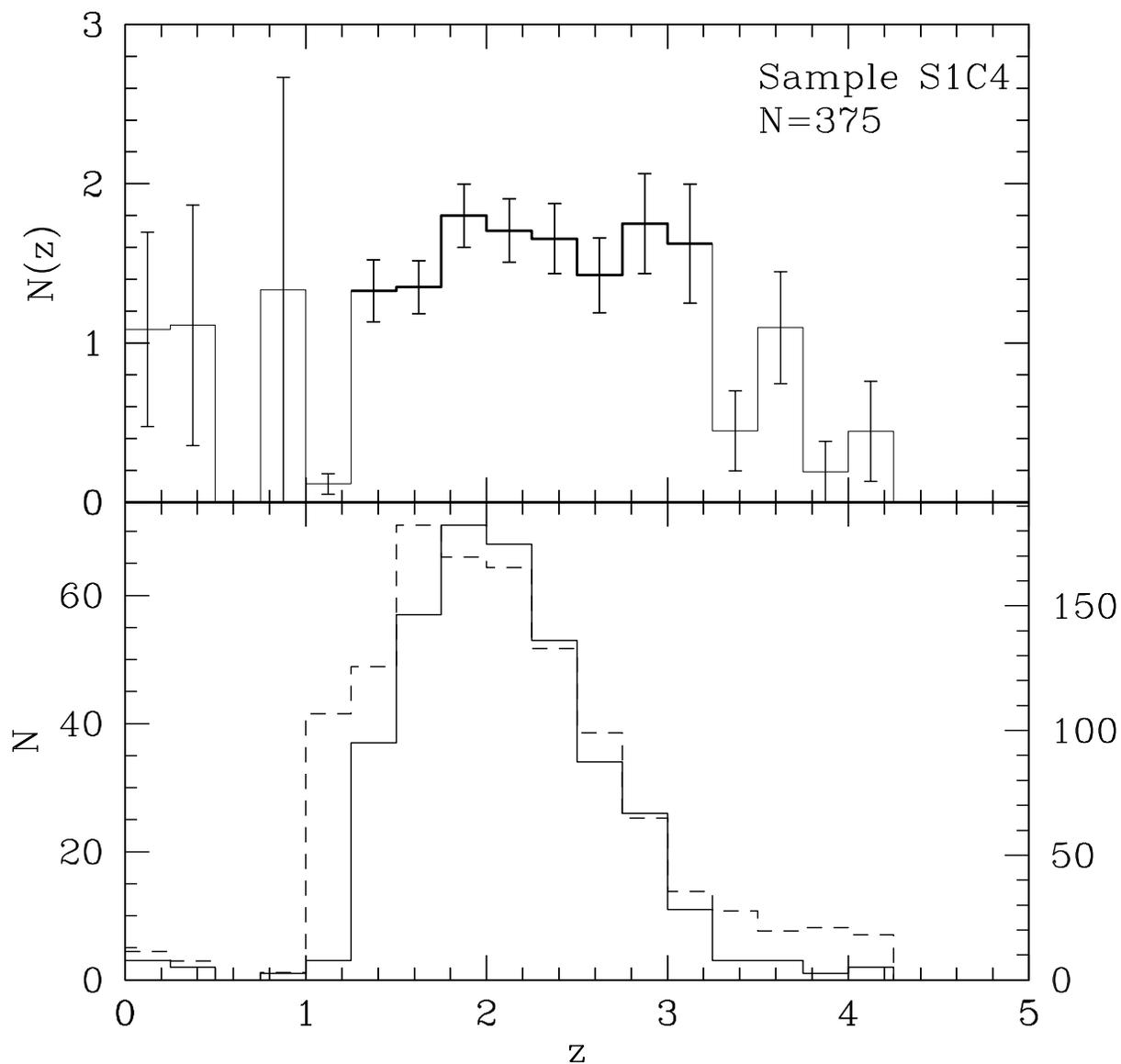}
\caption{Sample S1C4 $dN/dz$.  Normalized redshift distribution of
\ion{C}{4} from Sample S1C4.  ({\em Bottom}) Solid line shows the
number of absorbers found in each bin (left axis).  Dashed lines show
the number of QSOs searched (right axis).  ({\em Top}) Normalized
number of absorbers per unit redshift: essentially the ratio of the
two bottom plots.  Thicker lines symbolize bins with more than ten
absorbers: these bins are considered significant.  Error bars are
Poisson.  The total number of absorbers and the sample from which they
were drawn are given in the upper right hand corner.  Lines within
$5000\,{\rm km\,s^{-1}}$ of the QSO redshift have been removed.}
\label{fig:fig25}
\end{figure}

\begin{figure}[htpb]
\plotone{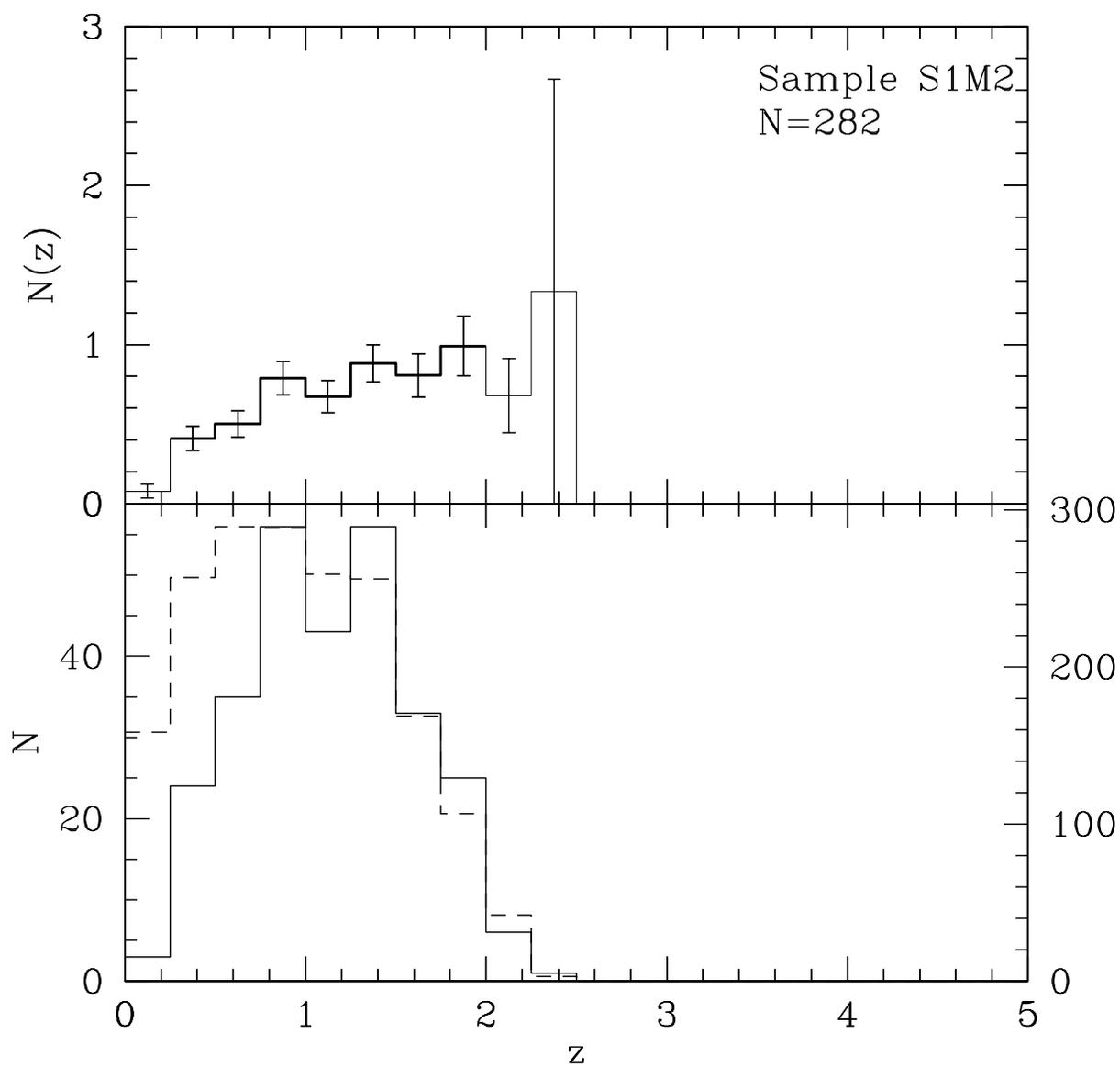}
\caption{Sample S1M2 $dN/dz$.  Normalized redshift distribution of
\ion{Mg}{2} in Sample S1M2.  See Figure~\ref{fig:fig25} for an
explanation of the lines.\label{fig:fig26}}
\end{figure}

\begin{figure}[htpb]
\plotone{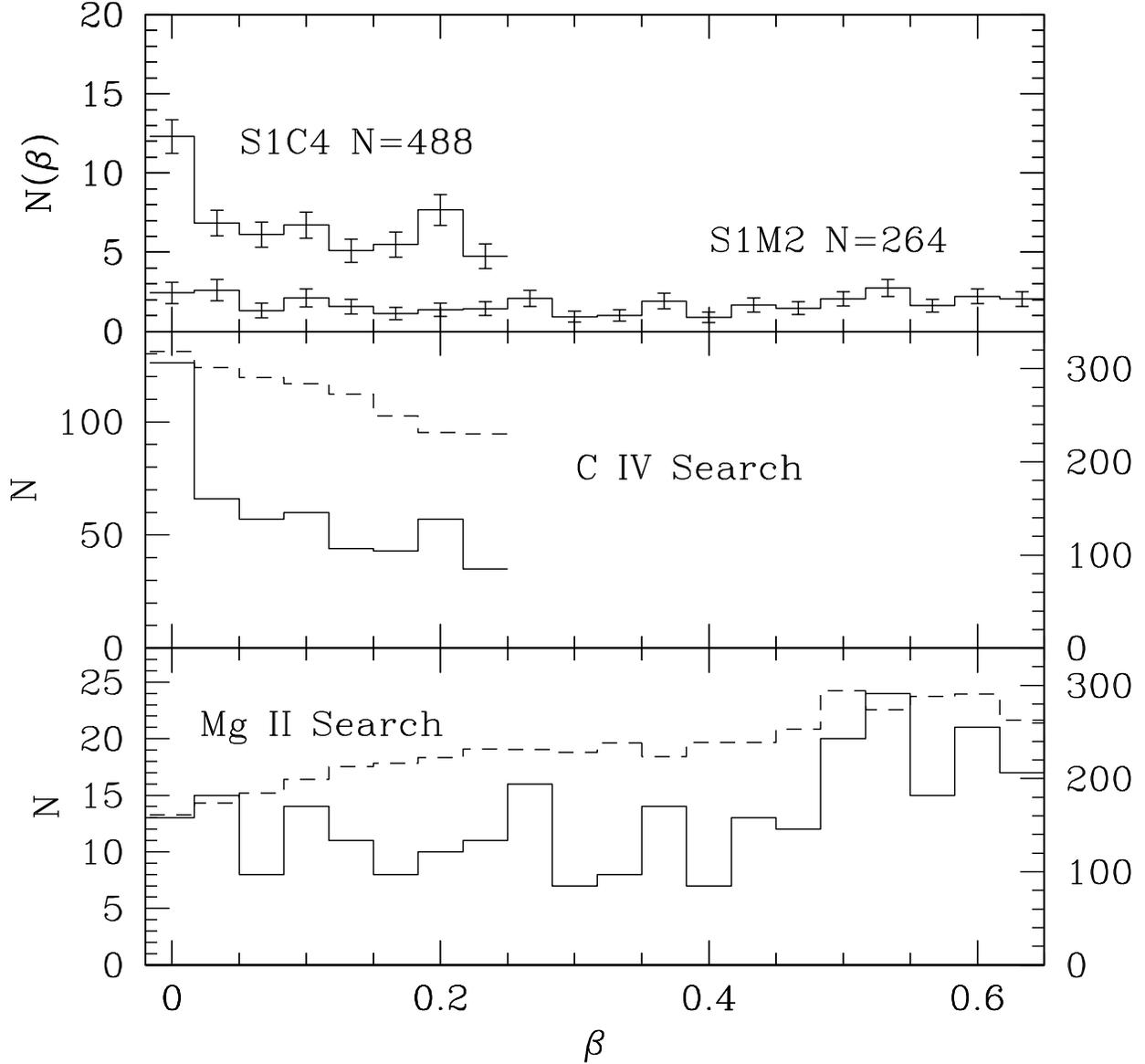}
\caption{Sample S1 $dN/d\beta$.  ({\em Top}) Normalized velocity
distribution of \ion{C}{4} and \ion{Mg}{2} in Sample S1.  ({\em
Middle}) Number of \ion{C}{4} absorbers and number of QSOs searched as
a function of velocity.  See Figure~\ref{fig:fig25} for an explanation
of the lines.  ({\em Bottom}) Number of \ion{Mg}{2} absorbers and
number of QSOs searched as a function of velocity.
\label{fig:fig27}}
\end{figure}

\begin{figure}[htpb]
\plotone{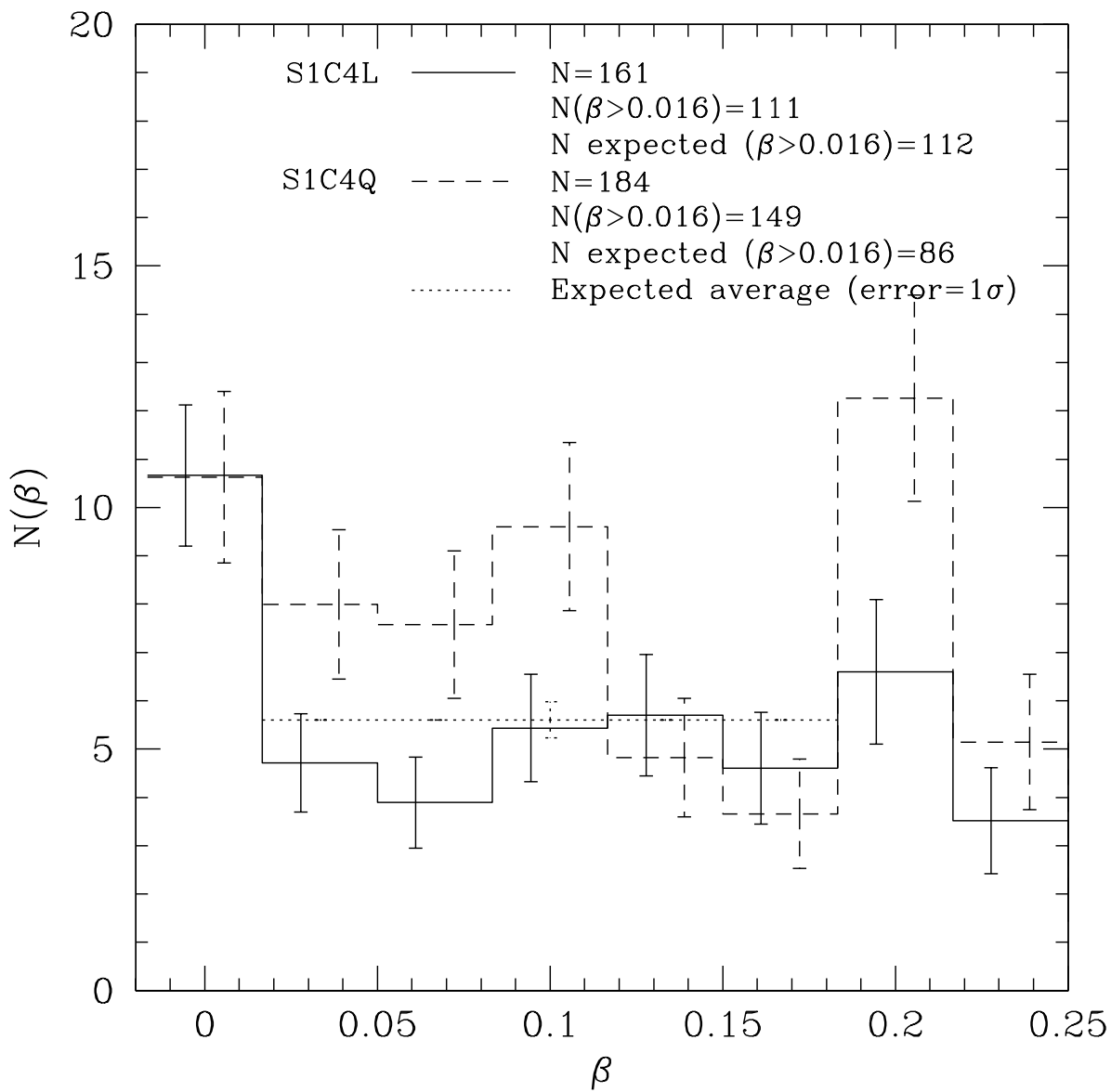}
\caption{Samples S1C4L and S1C4Q $dN/d\beta$.  Normalized velocity
distribution of \ion{C}{4} for radio-loud (solid line) and radio-quiet
(dashed line) quasars in Samples S1C4L and S1C4Q.  In this (and
similar plots) the expected level of $dN/d\beta$ is plotted in the
velocity range $5000$ to $55,000\,{\rm km\,s^{-1}}$ for comparison.
The upper right hand corner gives the number of absorbers, the number
of absorbers with $v>5000\,{\rm km\,s^{-1}}$ and the expected number
of absorbers with $v>5000\,{\rm km\,s^{-1}}$.\label{fig:fig28}}
\end{figure}

\begin{figure}[htpb]
\plotone{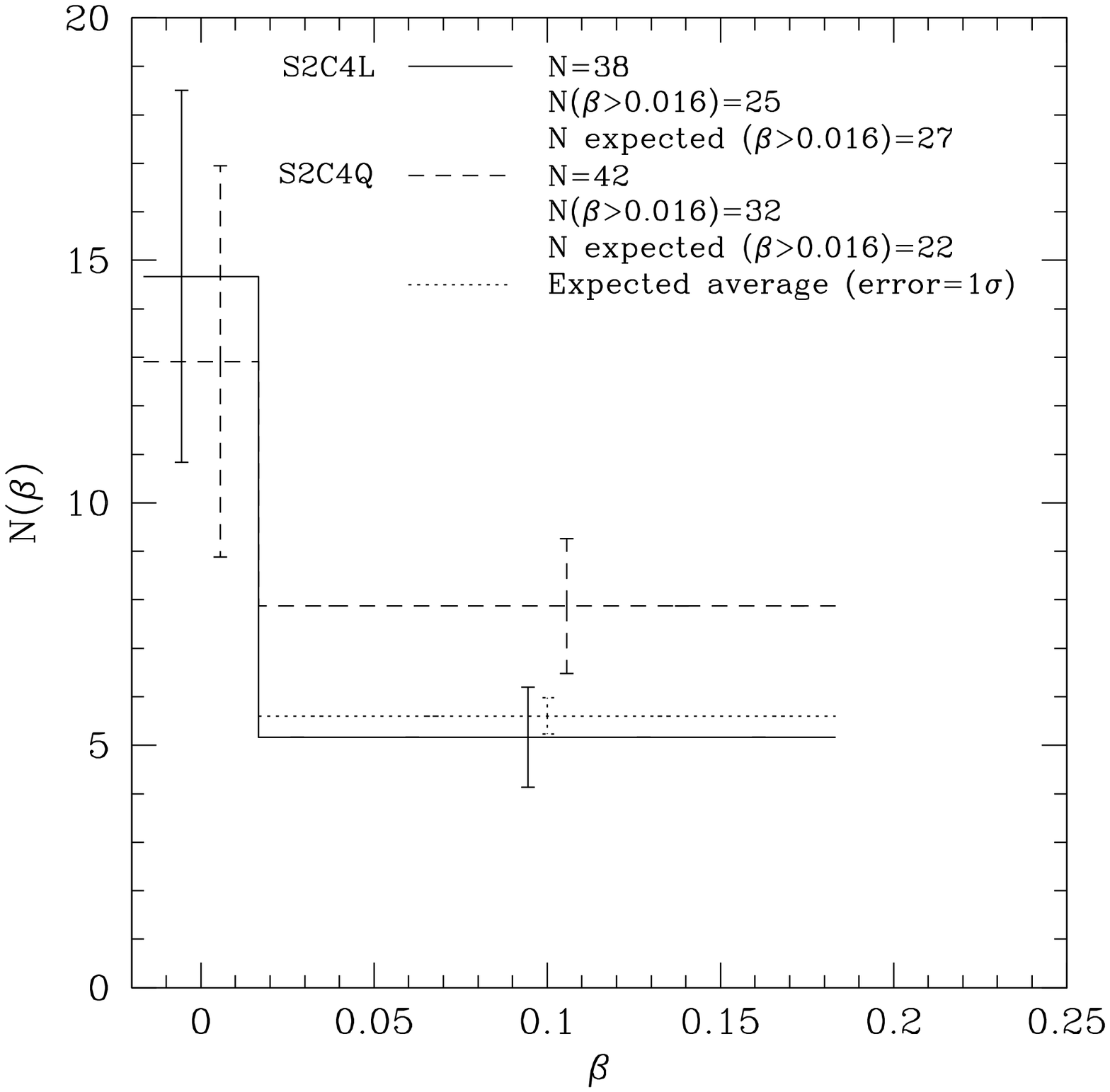}
\caption{Samples S2C4L and S2C4Q $dN/d\beta$.  Normalized velocity
distribution of \ion{C}{4} from radio-loud and radio-quiet quasars in
Samples S2C4L and S2C4Q.\label{fig:fig29}}
\end{figure}

\begin{figure}[htpb]
\plotone{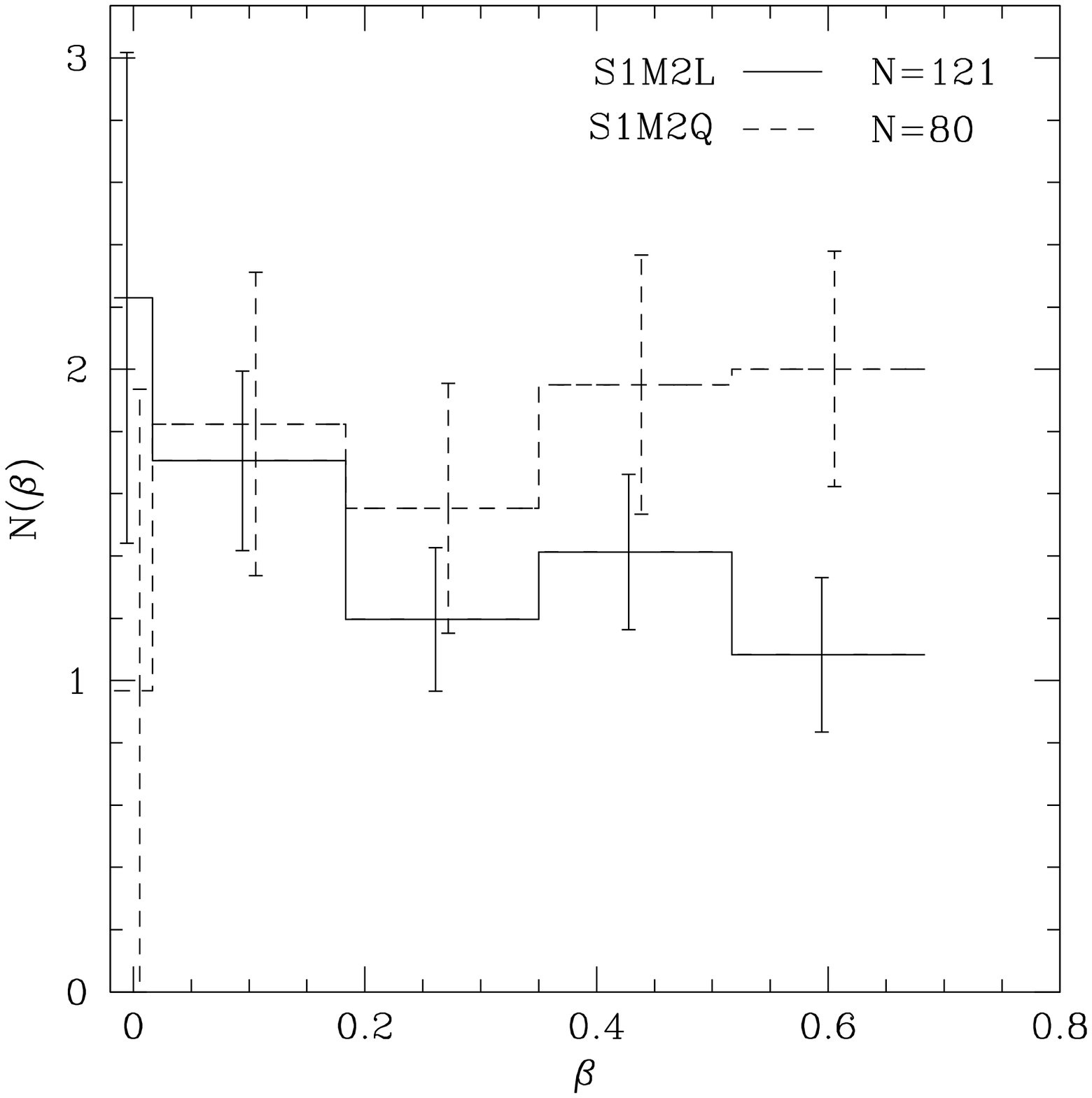}
\caption{Samples S1M2L and S1M2Q $dN/d\beta$.  Normalized velocity
distribution of \ion{Mg}{2} from radio-loud and radio-quiet quasars in
Samples S1M2L and S1M2Q.\label{fig:fig30}}
\end{figure}

\begin{figure}[htpb]
\plotone{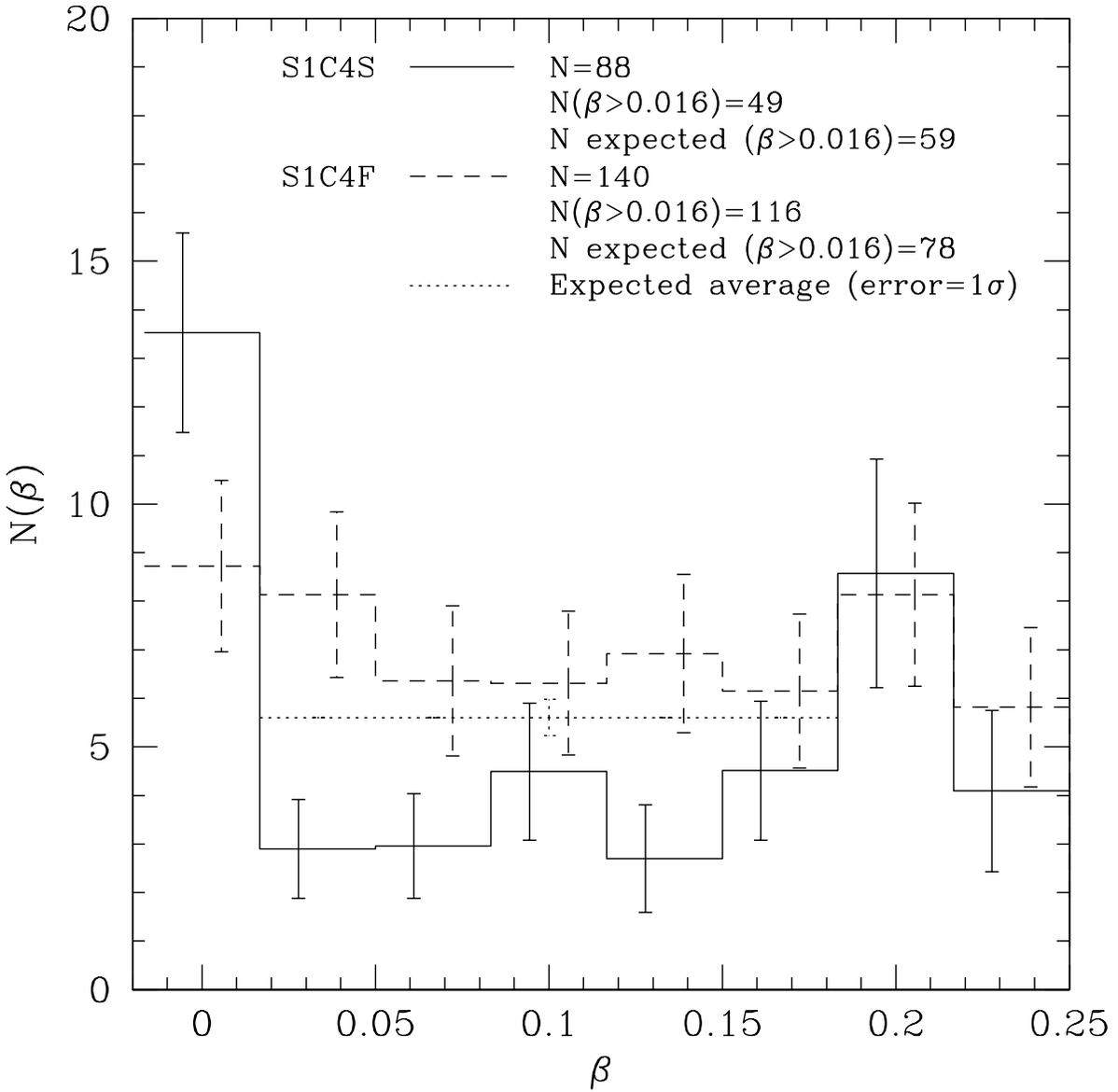}
\caption{Samples S1M2S and S1M2F $dN/d\beta$.  Normalized velocity
distribution of \ion{C}{4} from steep- (solid line) and flat-spectrum
(dashed line) quasars in Samples S1C4S and S1C4F.\label{fig:fig31}}
\end{figure}

\begin{figure}[htpb]
\plotone{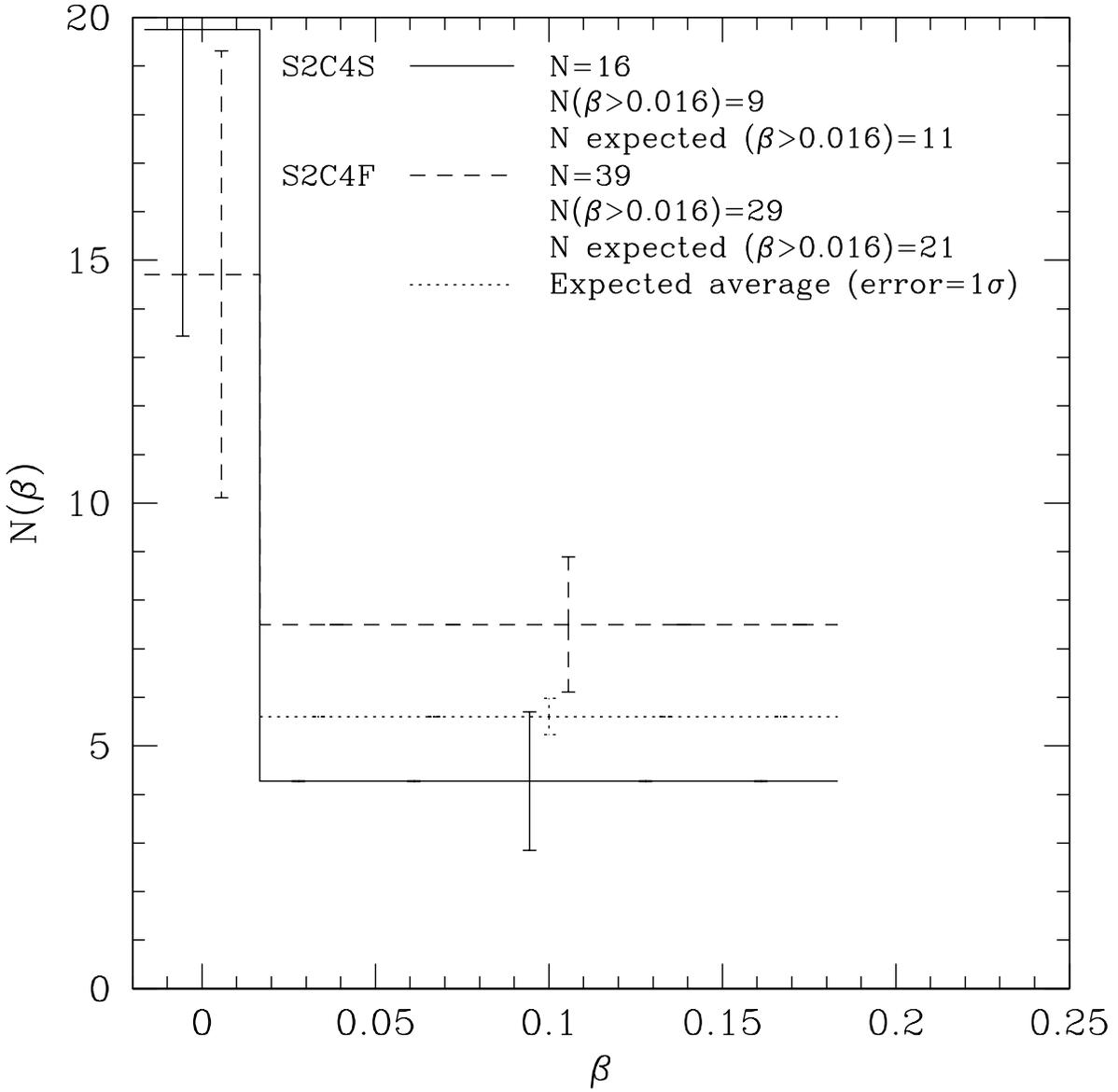}
\caption{Samples S2C4S and S2C4F $dN/db\beta$.  Normalized velocity
distribution of \ion{C}{4} from steep- and flat-spectrum quasars from
Samples S2C4S and S2C4F.\label{fig:fig32}}
\end{figure}

\begin{figure}[htpb]
\plotone{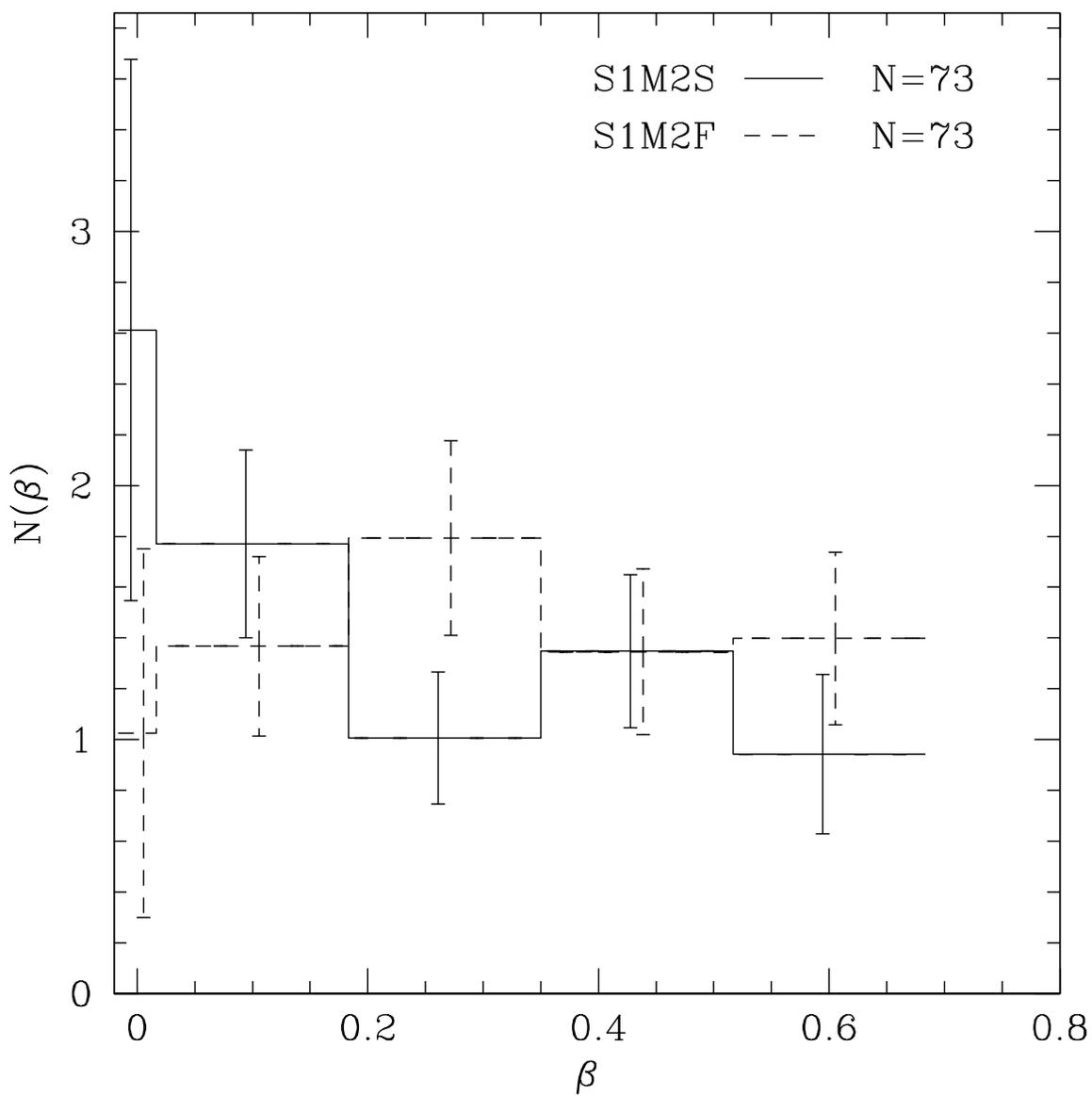}
\caption{Normalized velocity distribution of \ion{Mg}{2} from steep-
and flat-spectrum quasars in Sa mples S1M2S and S1M2F.\label{fig:fig33}}
\end{figure}

\clearpage

\clearpage

\begin{deluxetable}{lcrrccrrcc}
\tabletypesize{\footnotesize}
\rotate
\tablecaption{Observed Targets. \label{tab:tab1}}
\tablewidth{0pt}
\tablecolumns{10}
\tablehead{
\colhead{Object} & \colhead{$z$} & 
\colhead{RA} & \colhead{Dec} &
\colhead{$E$} & \colhead{$O-E$} & 
\colhead{Fpeak} & \colhead{Fint} &
\colhead{Blue} &
\colhead{Red} \\
\colhead{} & \colhead{} & 
\colhead{} & \colhead{} & 
\colhead{} & \colhead{} & 
\colhead{(mJy)} & \colhead{(mJy)} &
\colhead{S/N} &
\colhead{S/N}
}
\startdata
FIRST J$004717.7-015623$ & 2.480 & 00 47 17.799 & $-01$ 56 23.85 & 17.60 & 0.50 & 1.53 & 1.27	& 78.0 & 44.0 \\
FIRST J$021039.8-015213$ & 2.372 & 02 10 39.879 & $-01$ 52 13.81 & 17.97 & 0.44 & 1.51 & 1.76   	& 55.5 & 32.5 \\
FIRST J$025625.6-011911$ & 2.491 & 02 56 25.629 & $-01$ 19 11.91 & 17.99 & 0.16 & 25.78 & 27.56 	& 32.0 & 17.5 \\
FIRST J$072550.5+281906$ & 2.666 & 07 25 50.587 & $+28$ 19 06.12 & 16.98 & 1.18 & 37.99 & 38.85 	& 80.0 & 60.0 \\
FIRST J$072928.4+252451$ & 2.300 & 07 29 28.470 & $+25$ 24 51.80 & 17.22 & 0.91 & 6.83 & 6.47   	& 80.5 & 54.0 \\
FIRST J$080413.6+251633$ & 2.292 & 08 04 13.671 & $+25$ 16 33.63 & 17.49 & 0.58 & 3.48 & 3.68   	& 76.5 & 55.0 \\
FIRST J$082107.6+310751$ & 2.609 & 08 21 07.610 & $+31$ 07 51.41 & 16.40 & 0.45 & 85.41 & 89.98  	& 76.0 & 68.5 \\
FIRST J$085726.9+331316$ & 2.340 & 08 57 26.931 & $+33$ 13 16.86 & 17.15 & 0.55 & 20.69 & 21.29   	& 73.5 & 44.0 \\
\enddata
\end{deluxetable}

\clearpage

\begin{deluxetable}{lll}
\tabletypesize{\footnotesize}
\tablecaption{Observations. \label{tab:tab2}}
\tablewidth{0pt}
\tablehead{
\colhead{QSO} & \colhead{UT Date} & \colhead{UT Time} 
}
\startdata
FBQS0047-0156 & 1998 Aug 19 & 8:55\tablenotemark{1}, 9:38, 10:28 \\
 & 1998 Aug 23 & 9:08, 10:05 \\
 & 1998 Dec 12 & 2:57, 3:32 \\
 & 1998 Dec 13 & 2:12, 2:47 \\
FBQS0210-0152 & 1998 Dec 12 & 4:18, 4:56, 5:31 \\
 & 1998 Dec 13 & 3:41, 4:16 \\
 & 1998 Dec 27 & 2:40, 3:16 \\
FBQS0256-0119 & 1998 Dec 13 & 6:07, 7:10 \\
\enddata
\tablenotetext{1}{1800 second exposure.  Unless otherwise indicated, all exposures are 1800 seconds.}
\tablenotetext{2}{900 second exposure.}
\tablenotetext{3}{803 second exposure.}
\tablenotetext{4}{1497 second exposure.}
\tablenotetext{4}{2400 second exposure.}
\end{deluxetable}

\clearpage

\begin{deluxetable}{llllp{0.5cm}llll}
\tabletypesize{\small}
\tablewidth{0pt}
\tablecolumns{9}
\tablecaption{Search Linelist. \label{tab:tab3}}
\tablehead{
\colhead{Ion} &
\colhead{$\lambda_{\rm vac}$} &
\colhead{f} &
\colhead{IP (eV)} &
\colhead{} &
\colhead{Ion} &
\colhead{$\lambda_{\rm vac}$} &
\colhead{f} &
\colhead{IP (eV)}
}
\startdata
\ion{C}{2} & 1036.3367 & 0.125 & 24.376 &   & \ion{Ni}{2} & 1370.132 & 0.100 & \\        
\ion{C}{2} & 1334.5323 & 0.118 & &	    & \ion{Ni}{2} & 1741.549 & 0.0679 & \\       
\ion{O}{6} & 1031.9265 & 0.130 & 138.080 &  & \ion{Ni}{2} & 1709.600 & 0.0470 & \\       
\ion{O}{6} & 1037.6155 & 0.0648 & &         & \ion{Si}{4} & 1393.755 & 0.528 & 45.13 \\  
\ion{Ar}{1} & 1048.2199 & 0.230 & 15.755 &  & \ion{Si}{4} & 1402.770 & 0.262 & \\        
\ion{Ar}{1} & 1066.6599 & 0.0594 & &	    & \ion{C}{4} & 1548.202 & 0.1940 & 64.476 \\ 
\ion{S}{4} & 1062.671 & 0.0377 & 47.29 &    & \ion{C}{4} & 1550.774 & 0.0970 & \\       
\ion{N}{2} & 1083.990 & 0.101 & 29.593 &    & \ion{C}{1} & 1656.9282 & 0.142 & 11.256 \\
\enddata
\end{deluxetable}

\clearpage

\begin{deluxetable}{rrrrrrrlc}
\tabletypesize{\footnotesize}
\tablecaption{All absorption lines. \label{tab:tab4}}
\tablewidth{0pt}
\tablecolumns{9}
\tablehead{
\colhead{N} & \colhead{$\lambda$} & \colhead{$\sigma_{\lambda}$} &
\colhead{${\rm W_{obs}}$} & \colhead{$\sigma_{\rm W}$} &
\colhead{fwhm} &
\colhead{SL} &
\colhead{ID} &
\colhead{$z$}
}
\startdata
\cutinhead{FBQS0047-0156 ($z=2.479$)}
 1 & 3977.09\tablenotemark{a} & 0.15 & 1.99 & 0.08 & 6.87 & 21.2 & & \\*
 2 & 3984.01\tablenotemark{a} & 0.25 & 1.19 & 0.08 & 6.88 & 13.0 & & \\*
 5 & 4012.43\tablenotemark{a} & 0.41 & 0.94 & 0.11 & 5.01 &  8.3 & & \\
 6 & 4044.96\tablenotemark{a} & 0.11 & 3.14 & 0.10 & 6.85 & 26.4 & & \\
 7 & 4063.58\tablenotemark{a} & 0.46 & 0.68 & 0.09 & 5.99 &  7.6 & & \\
 8 & 4076.98\tablenotemark{a} & 0.23 & 1.28 & 0.08 & 13.19 & 14.6 & & \\
 9 & 4089.69\tablenotemark{a} & 0.30 & 0.91 & 0.07 & 8.83 & 11.3 & & \\
10 & 4094.06\tablenotemark{a} & 0.29 & 0.89 & 0.07 & 5.10 & 11.3 & & \\
\enddata
\tablenotetext{a}{Line in Lyman $\alpha$ forest}
\tablenotetext{b}{Possibly a continuum fitting artifact}
\tablenotetext{c}{Questionable identification}
\tablenotetext{d}{Possibly a telluric residual}
\tablenotetext{e}{Possible night sky artifact}

\end{deluxetable}

\clearpage

\begin{deluxetable}{ll}
\tablewidth{0pt}
\tablecolumns{2}
\tablecaption{Absorption Line Samples.\label{tab:tab5}}
\tablehead{
\colhead{Sample} &
\colhead{Notes}
}
\startdata
S1 & All data from \citet{v+00} and new data presented herein \\
S2 & New data plus \citet{v+00} data with $2.2 \le z_{em} \le 2.8$ \\
S3 & New data presented herein \\
C4 & Only \ion{C}{4} systems with both lines stronger than $0.15{\rm \AA}$ greater than $5\sigma$ \\
M2 & Only \ion{M}{2} systems with both lines stronger than $0.30{\rm \AA}$ greater than $5\sigma$ \\
L & Systems found in quasars with $\log L_{20} \ge 33.70\,{\rm ergs/s/cm^2/Hz}$ \\
Q & Systems found in quasars with $\log L_{20} \le 32.80\,{\rm ergs/s/cm^2/Hz}$ \\
S & Systems found in quasars with $\alpha_6^{20} \le -0.4$ \\
F & Systems found in quasars with $\alpha_6^{20} > -0.4$ \\
\enddata
\end{deluxetable}


\clearpage

\begin{thebibliography}{40}
\expandafter\ifx\csname natexlab\endcsname\relax\def\natexlab#1{#1}\fi

\bibitem[{{Anderson} {et~al.}(1987){Anderson}, {Weymann}, {Foltz}, \& {Chaffee,
  F. H., Jr.}}]{awf+87}
{Anderson}, S.~F., {Weymann}, R.~J., {Foltz}, C.~B., \& {Chaffee, F. H., Jr.}
  1987, \aj, 94, 278

\bibitem[{{Bahcall} \& {Spitzer}(1969)}]{bs69}
{Bahcall}, J.~N. \& {Spitzer}, L., J. 1969, \apjl, 156, L63

\bibitem[{{Bahcall et al.}(1993)}]{bbb+93}
{Bahcall, J. N., et al.} 1993, \apjs, 87, 1

\bibitem[{{Barlow} \& {Sargent}(1997)}]{bs97}
{Barlow}, T.~A. \& {Sargent}, W. L.~W. 1997, \aj, 113, 136

\bibitem[{{Burbidge} {et~al.}(1966){Burbidge}, {Lynds}, \& {Burbidge}}]{blb66}
{Burbidge}, E.~M., {Lynds}, C.~R., \& {Burbidge}, G.~R. 1966, \apj, 144, 447

\bibitem[{{Crotts}(1989)}]{cro89}
{Crotts}, A. P.~S. 1989, \apj, 336, 550

\bibitem[{{Filippenko}(1982)}]{fil82}
{Filippenko}, A.~V. 1982, \pasp, 94, 715

\bibitem[{{Foltz} {et~al.}(1988){Foltz}, {Chaffee, F. H., Jr.}, {Weymann}, \&
  {Anderson}}]{fcw+88}
{Foltz}, C.~B., {Chaffee, F. H., Jr.}, {Weymann}, R.~J., \& {Anderson}, S.~F.
  1988, in Proceedings of the QSO Absorption Line Meeting, ed. J.~C. Blades,
  D.~A. Turnshek, \& C.~A. Norman (Cambridge: Cambridge University Press), 53

\bibitem[{{Foltz} {et~al.}(1986){Foltz}, {Weymann}, {Peterson}, {Sun},
  {Malkan}, \& {Chaffee, F. H., Jr.}}]{fwp+86}
{Foltz}, C.~B., {Weymann}, R.~J., {Peterson}, B.~M., {Sun}, L., {Malkan},
  M.~A., \& {Chaffee, F. H., Jr.} 1986, \apj, 307, 504

\bibitem[{{Gregg} {et~al.}(1996){Gregg}, {Becker}, {White}, {Helfand},
  {McMahon}, \& {Hook}}]{gbw+96}
{Gregg}, M.~D., {Becker}, R.~H., {White}, R.~L., {Helfand}, D.~J., {McMahon},
  R.~G., \& {Hook}, I.~M. 1996, \aj, 112, 407

\bibitem[{{Hamann} {et~al.}(1997){Hamann}, {Barlow}, {Cohen}, {Junkkarinen}, \&
  {Burbidge}}]{hbc+97}
{Hamann}, F., {Barlow}, T., {Cohen}, R.~D., {Junkkarinen}, V., \& {Burbidge},
  E.~M. 1997, in ASP Conf. Ser. 128: Mass Ejection from Active Galactic Nuclei,
  ed. N.~Arav, I.~Shlosman, \& R.~J. Weymann (San Francisco: ASP), 19

\bibitem[{{Horne}(1986)}]{hor86}
{Horne}, K. 1986, \pasp, 98, 609

\bibitem[{{Jannuzi et al.}(1996)}]{jhk+96}
{Jannuzi, B. T., et al.} 1996, \apjl, 470, L11

\bibitem[{{Jannuzi et al.}(1998)}]{jbb+98}
---. 1998, \apjs, 118, 1

\bibitem[{{Moore}(1971)}]{m71}
{Moore}, C.~E. 1971, Atomic Energy Levels (Washington, D.C.: United States
  Department of Commerce, National Bureau of Standards)

\bibitem[{{Morton} {et~al.}(1988){Morton}, {York}, \& {Jenkins}}]{myj88}
{Morton}, D.~C., {York}, D.~G., \& {Jenkins}, E.~B. 1988, \apjs, 68, 449

\bibitem[{{Orr} \& {Browne}(1982)}]{ob82}
{Orr}, M. J.~L. \& {Browne}, I. W.~A. 1982, \mnras, 200, 1067

\bibitem[{{Padovani} \& {Urry}(1992)}]{pu92}
{Padovani}, P. \& {Urry}, C.~M. 1992, \apj, 387, 449

\bibitem[{{Quashnock} {et~al.}(1996){Quashnock}, {Vanden Berk}, \&
  {York}}]{qvy96}
{Quashnock}, J.~M., {Vanden Berk}, D.~E., \& {York}, D.~G. 1996, \apjl, 472,
  L69

\bibitem[{{Richards} {et~al.}(2000){Richards}, {Laurent-Muehleisen}, {Becker},
  \& {York}}]{rlb00}
{Richards}, G.~T., {Laurent-Muehleisen}, S., {Becker}, R., \& {York}, D.~G.
  2000, \apj, submitted

\bibitem[{{Richards} {et~al.}(1999){Richards}, {York}, {Yanny}, {Kollgaard},
  {Laurent-Muehleisen}, \& {Vanden Berk}}]{ryy+99}
{Richards}, G.~T., {York}, D.~G., {Yanny}, B., {Kollgaard}, R.~I.,
  {Laurent-Muehleisen}, S.~A., \& {Vanden Berk}, D.~E. 1999, \apj, 513, 576

\bibitem[{{Rowan-Robinson}(1977)}]{rr77}
{Rowan-Robinson}, M. 1977, \apj, 213, 635

\bibitem[{{Sargent} {et~al.}(1988){Sargent}, {Steidel}, \&
  {Boksenberg}}]{ssb88}
{Sargent}, W. L.~W., {Steidel}, C.~C., \& {Boksenberg}, A. 1988, \apjs, 68, 539

\bibitem[{{Schneider et al.}(1993)}]{shj+93}
{Schneider, D. P., et al.} 1993, \apjs, 87, 45

\bibitem[{{Steidel}(1990)}]{ste90}
{Steidel}, C.~C. 1990, \apjs, 72, 1

\bibitem[{{Steidel} {et~al.}(1997){Steidel}, {Dickinson}, {Meyer},
  {Adelberger}, \& {Sembach}}]{sdm+97}
{Steidel}, C.~C., {Dickinson}, M., {Meyer}, D.~M., {Adelberger}, K.~L., \&
  {Sembach}, K.~R. 1997, \apj, 480, 568

\bibitem[{{Stocke} {et~al.}(1992){Stocke}, {Morris}, {Weymann}, \&
  {Foltz}}]{smw+92}
{Stocke}, J.~T., {Morris}, S.~L., {Weymann}, R.~J., \& {Foltz}, C.~B. 1992,
  \apj, 396, 487

\bibitem[{{Stockton} \& {Lynds}(1966)}]{sl66}
{Stockton}, A.~N. \& {Lynds}, C.~R. 1966, \apj, 144, 451

\bibitem[{{Turnshek}(1988)}]{tur88}
{Turnshek}, D.~A. 1988, in Proceedings of the QSO Absorption Line Meeting, ed.
  J.~C. Blades, D.~A. Turnshek, \& C.~A. Norman (Cambridge: Cambridge
  University Press), 17

\bibitem[{{Tytler} \& {Fan}(1992)}]{tf92}
{Tytler}, D. \& {Fan}, X.~M. 1992, \apjs, 79, 1

\bibitem[{{Vanden Berk}(1997)}]{van97}
{Vanden Berk}, D.~E. 1997, PhD thesis, University of Chicago

\bibitem[{{Vanden Berk} {et~al.}(1996){Vanden Berk}, {Quashnock}, {York}, \&
  {Yanny}}]{vqy+96}
{Vanden Berk}, D.~E., {Quashnock}, J.~M., {York}, D.~G., \& {Yanny}, B. 1996,
  \apj, 469, 78

\bibitem[{{Vanden Berk et al.}(1999)}]{vls+99}
{Vanden Berk, D. E., et al.} 1999, \apjs, 122, 355

\bibitem[{{Vanden Berk et al.}(2001)}]{v+00}
---. 2001, in preparation

\bibitem[{{Weymann}(1997)}]{wey97}
{Weymann}, R. 1997, in ASP Conf. Ser. 128: Mass Ejection from Active Galactic
  Nuclei, ed. N.~Arav, I.~Shlosman, \& R.~J. Weymann (San Francisco: ASP), 3

\bibitem[{{Weymann} {et~al.}(1991){Weymann}, {Morris}, {Foltz}, \&
  {Hewett}}]{wmf+91}
{Weymann}, R.~J., {Morris}, S.~L., {Foltz}, C.~B., \& {Hewett}, P.~C. 1991,
  \apj, 373, 23

\bibitem[{{York} {et~al.}(1986){York}, {Dopita}, {Green}, \&
  {Bechtold}}]{ydg+86}
{York}, D.~G., {Dopita}, M., {Green}, R., \& {Bechtold}, J. 1986, \apj, 311,
  610

\bibitem[{{York} {et~al.}(1991){York}, {Yanny}, {Crotts}, {Carilli},
  {Garrison}, \& {Matheson}}]{yyc+91}
{York}, D.~G., {Yanny}, B., {Crotts}, A., {Carilli}, C., {Garrison}, E., \&
  {Matheson}, L. 1991, \mnras, 250, 24

\bibitem[{{York et al.}(1999)}]{ybl+99}
{York, D. G., et al.} 1999, BAAS, 195, 52.07

\bibitem[{{Young} {et~al.}(1982){Young}, {Sargent}, \& {Boksenberg}}]{ysb82}
{Young}, P., {Sargent}, W. L.~W., \& {Boksenberg}, A. 1982, \apjs, 48, 455

\end{thebibliography}

\end{document}